\documentclass[aps,prl,twocolumn,superscriptaddress,nofootinbib]{revtex4-2}
\usepackage{graphicx}
\usepackage{amsmath,amssymb,amsthm}
\usepackage{bm}
\usepackage[colorlinks=true,linkcolor=blue,citecolor=blue,urlcolor=blue]{hyperref}

\newcommand{\id}{\mathrm{id}}
\newcommand{\Tr}{\mathrm{Tr}}
\newcommand{\cD}{\mathcal D}
\newcommand{\cE}{\mathcal E}
\newcommand{\cL}{\mathcal L}
\newcommand{\cN}{\mathcal N}
\newcommand{\loc}{{\rm loc}}
\newcommand{\cF}{\mathcal F}
\newcommand{\cR}{\mathcal R}
\newcommand{\cC}{\mathcal C}
\newcommand{\cM}{\mathcal M}

\newcommand{\cH}{\mathcal H}
\newcommand{\calI}{\mathcal I}
\newcommand{\cS}{\mathcal S}
\newcommand{\cB}{\mathcal B}
\newcommand{\cJ}{\mathcal J}
\newcommand{\cP}{\mathcal P}
\newcommand{\cQ}{\mathcal Q}
\newcommand{\cU}{\mathcal U}
\newcommand{\cA}{\mathcal A}
\newcommand{\cT}{\mathcal T}
\newcommand{\syn}{\mathrm{syn}}
\newcommand{\supp}{\mathrm{supp}}
\newcommand{\dist}{\mathrm{dist}}
\newcommand{\diam}{\mathrm{diam}}

\newcommand{\F}{\mathbb F}
\newcommand{\Z}{\mathbb Z}
\newcommand{\Zloc}{\mathfrak Z_{\loc}}
\newcommand{\Zfull}{\mathfrak Z}

\theoremstyle{plain}
\newtheorem{theorem}{Theorem}

\newtheorem{lemma}{Lemma}

\newtheorem{proposition}{Proposition}

\newtheorem{corollary}{Corollary}

\theoremstyle{remark}

\makeatletter
\newcommand{\SMhidecontents}{\let\SMsavedcontentsline\contentsline\let\contentsline\@gobblefour}
\newcommand{\SMshowcontents}{\let\contentsline\SMsavedcontentsline}
\makeatother

\begin{document}

\title{Robust Quantum Memory in Thermal Equilibrium}

\author{Jiaxin Liu}
\affiliation{Hangzhou Institute of Extremely-Weak Magnetic Field Major National Science and Technology Infrastructure, Hangzhou 310051, China}
\affiliation{School of Instrumentation and Optoelectronic Engineering, Beihang University, Haidian, Beijing 100191, China}
\author{Danyue Ma}
\email[Corresponding author: ]{madanyue0419@buaa.edu.cn}
\affiliation{Hangzhou Institute of Extremely-Weak Magnetic Field Major National Science and Technology Infrastructure, Hangzhou 310051, China}
\affiliation{School of Instrumentation and Optoelectronic Engineering, Beihang University, Haidian, Beijing 100191, China}
\author{Jiancheng Fang}
\affiliation{Hangzhou Institute of Extremely-Weak Magnetic Field Major National Science and Technology Infrastructure, Hangzhou 310051, China}
\affiliation{School of Instrumentation and Optoelectronic Engineering, Beihang University, Haidian, Beijing 100191, China}

\date{October 17, 2026}

\begin{abstract}
A self-correcting quantum memory stores a qubit robustly in thermal equilibrium, as a magnet stores a bit, but whether it can exist beyond idealized models has been open for two decades. We prove that the four-dimensional toric code stays self-correcting under any weak local perturbation whenever the system evolves locally and stays in thermal equilibrium. The reason is that flipping the qubit requires an excitation spanning the whole system, and the energy cost of such an excitation grows with its length. The quantum fluctuations caused by the perturbation lower this cost but cannot remove it, so such excitations stay exponentially rare at any low temperature. The argument uses only two local properties of the ideal code: every excitation can be erased locally, and the stored information is invisible to local operators. Every stabilizer code with these two properties stays self-correcting under perturbation as well. Self-correction therefore holds for every Hamiltonian near these codes and is a robust collective property of quantum matter. In principle, quantum information can be as robust in thermal equilibrium as classical information.
\end{abstract}

\maketitle
\addtocontents{toc}{\protect\SMhidecontents}

A ferromagnet stores a bit in thermal equilibrium: reversing a region costs an energy proportional to its boundary, so thermal fluctuations cannot erase the magnetization~\cite{Peierls1936}. A self-correcting quantum memory stores a qubit the same way~\cite{DKLP2002}, dispensing with the measurement and feedback on which today's quantum processors spend most of their physical qubits~\cite{Google2025,Bluvstein2024,Brown2016}. The memory requires excitations whose energy grows with their size, like the reversed regions of the magnet. In two and three dimensions the toric code~\cite{Kitaev2003} has point-like excitations and fails this requirement~\cite{AFH2009,BravyiTerhal2009}; in four dimensions all excitations are closed loops whose energy grows with their length, so the Peierls argument of the ferromagnet applies~\cite{DKLP2002,AHHH2010,Hastings2011}. But the argument holds only for the ideal code, the toric code without the perturbation. Its Hamiltonian is a sum of commuting checks and therefore exactly solvable. Every later construction and criterion of self-correction likewise assumes an exactly solvable Hamiltonian~\cite{MazacHamma2012,Haah2011,BravyiHaah2013,HongGuoLucas2025,Placke2024,CLBT2010,BCHM2013,BGL2025}. No real material has exactly such a Hamiltonian, because weak perturbations are always present. Whether self-correction is a property of matter, or only of these models, has therefore remained open~\cite{Brown2016,Hastings2011,MKS2026,YinLucas2026}.

\begin{figure*}[t]
\includegraphics[width=\textwidth]{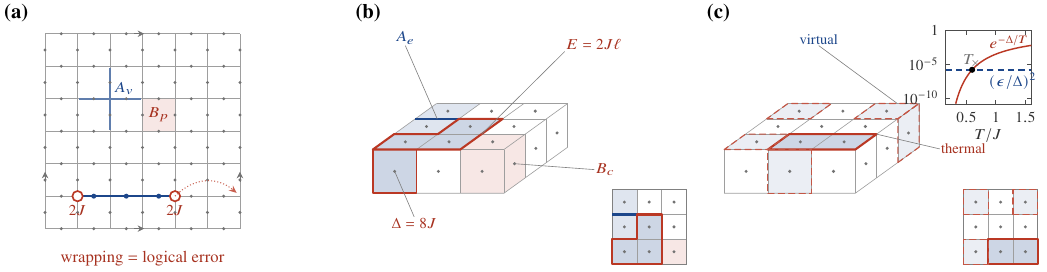}
\caption{Excitations of the ideal and perturbed toric codes. (a)~Two-dimensional toric code: qubits on edges, star checks $A_v$, plaquette checks $B_p$. A $Z$-error string ends on two violated stars that cost $2J$ each at any separation; a string wrapping the torus flips a logical qubit. (b)~Four-dimensional toric code (three-dimensional slab): qubits on plaquettes, checks $A_e$ on edges and $B_c$ on cubes; a membrane of flipped plaquettes is bounded by a loop of violated edge checks of energy $2J\ell$. (c)~Perturbed code: single plaquettes flicker for a time $\sim1/\Delta$ with density $\sim(\epsilon/\Delta)^2$ around a thermal loop of density $\sim e^{-\Delta/T}$; inset, the two densities cross at $T_\times$ ($\epsilon=0.01J$).}
\label{fig:enemies}
\end{figure*}

The difficulty is that the perturbation adds quantum fluctuations to the thermal ones, and in the toric code the two have so far been controlled only separately. Quantum fluctuations alone are controlled at zero temperature, where topological order and the higher-form symmetry behind it survive weak local perturbations~\cite{BHM2010,HastingsWen2005,GKSW2015,McGreevy2023}. Thermal fluctuations alone are controlled too. The Peierls argument protects the ideal code, and classical topological memories survive perturbations~\cite{Wegner1971,Thorngren2026}. The perturbed code faces both kinds at once, and together they defeat the Peierls argument. The argument requires each excitation to have a definite energy, which quantum fluctuations destroy by creating and annihilating excitations. The damage is greatest at low temperature, where the memory should work best. Thermal excitations become exponentially rare, but quantum fluctuations do not. Generic quantum many-body systems share this problem. The theory of mixed-state phases rests on exactly solvable examples, which have no quantum fluctuations~\cite{MKS2026,Sang2025,SangHsieh2025,FanBaoAltmanVishwanath2024,BaoFanVishwanathAltman2023,SangZouHsieh2024,EllisonCheng2025,SohalPrem2025}, and the convergence theory of quantum Gibbs samplers covers only high temperature~\cite{CKG2023,ChenNature2025,RouzeFrancaAlhambra2024,CMTW2025,Bakshi2024,Kochanowski2024,DobrushinShlosman2026} and a few special models~\cite{Rakovszky2026,GamarnikKianiZlokapa2024}.

Here we prove that the two kinds of fluctuation can be controlled together, so that self-correction holds for every Hamiltonian near the ideal code. In imaginary time, quantum fluctuations become short-lived bubbles that a linked-cluster expansion counts alongside the thermal excitations. The bubbles lower the energy cost of large excitations without removing it, so a Peierls argument applies again. The argument needs only two properties of the ideal code: every excitation can be erased locally, and the stored information is invisible to local operators. Both are properties of the checks alone, and any code with both, in any dimension, is self-correcting as well. For any such code, large excitations stay exponentially rare below a fixed temperature whenever the perturbation is weaker than a fixed threshold. The threshold holds down to zero temperature. Topological order at nonzero temperature thereby survives generic perturbations, and quantum Gibbs samplers for these Hamiltonians mix slowly at low temperature. Because the stored information is invisible to local operators, no local term can favor one stored state. Self-correction can thus be a robust collective property of quantum matter. In thermal equilibrium, the system's own interactions protect a qubit the way a magnet protects a bit, without symmetry or fine-tuning.

\emph{Model and main results.}---The four-dimensional toric code has qubits on the plaquettes of a four-dimensional periodic lattice of linear size $L$ [Fig.~\ref{fig:enemies}(b)]. Two kinds of checks act on them: for every edge $e$, the product $A_e$ of Pauli $X$ over the plaquettes containing $e$; for every cube $c$, the product $B_c$ of Pauli $Z$ over the plaquettes bounding $c$. All checks commute. Violated checks form the excitations, which are closed loops. A $Z$ error on one plaquette violates its four edge checks, the smallest loop. More generally, a membrane of $Z$ errors is bounded by a loop of violated edge checks, and a membrane of $X$ errors by a loop of violated cube checks on the dual lattice. The ideal Hamiltonian is a sum of checks, and the states with no excitations form its ground space. A membrane that spans the torus has no boundary and acts within the ground space, which encodes six logical qubits. The perturbed Hamiltonian is
\begin{equation}
H=-\sum_eJ_eA_e-\sum_cJ_cB_c+V,\qquad \|V\|_\loc\le\epsilon,
\label{eq:H}
\end{equation}
with couplings between $J_{\min}$ and $J_{\max}$ and $V$ an arbitrary finite-range sum of Hermitian terms; $\|V\|_\loc$ is the total strength of the terms touching any one qubit. $V$ need not commute with the checks and may break every symmetry of the code.

\begin{figure}[t]
\includegraphics[width=\columnwidth]{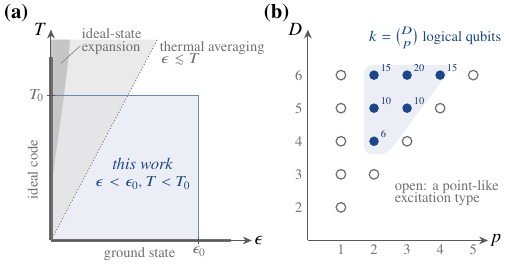}
\caption{Stability region and scope of the condition. (a)~The $(\epsilon,T)$ plane (schematic): earlier results on the axes; an ideal-state expansion controls only a sliver closing as $T\to0$, thermal averaging only the wedge $\epsilon\lesssim T$, this work the rectangle $\epsilon<\epsilon_0$, $T<T_0$, uniform down to $T=0$. (b)~Generalized toric codes with qubits on the $p$-cells of a $D$-dimensional torus: filled, the condition holds and $\binom Dp$ logical qubits are protected; open, one excitation type is point-like.}
\label{fig:results}
\end{figure}

Our first result concerns the thermal state $\rho\propto e^{-H/T}$ itself. Measuring the checks of the ideal code in $\rho$ produces a pattern of excitations. There are $\epsilon_0,T_0>0$, independent of $L$, such that for all $\epsilon\le\epsilon_0$ and $T\le T_0$ [the rectangular stable region of Fig.~\ref{fig:results}(a)]
\begin{equation}
\Pr_\rho[\text{an excitation loop of length}\ge\ell]\le CL^{c}e^{-4\ell},
\label{eq:static}
\end{equation}
with $C,c$ independent of $L$ and $T$. Loops spanning the system are therefore exponentially rare. Our second result turns this into a memory. Encode the logical qubits thermally, by the map $\cE$ that places the logical state in its sector of $\rho$ (Appendix~C). Then let the system evolve under any quasi-local dynamics $\cL$ that preserves $\rho$ and has bounded rates. With the decoder $\cD$ of the ideal code, which measures the checks and erases small loops in place,
\begin{equation}
\|\cD\,e^{t\cL}\,\cE-\id\|_\diamond\le C'L^{c'}\,t\,e^{-aL}
\label{eq:memory}
\end{equation}
for every input, including halves of entangled pairs. The logical qubits thus survive for times exponential in $L$. The decoder uses no knowledge of $V$ or $T$, so the same ideal decoder works throughout the stable region.

The proof uses only two properties of the ideal code. First, every excitation can be erased locally. Second, the stored information is invisible to local operators: every local operator that commutes with all checks is a product of nearby checks. These two properties, together with bounded locality and finitely many global constraints on excitation patterns (Appendix~A), form what we call the topological Peierls condition. The first property protects against thermal fluctuations and the second against the perturbation; Table~\ref{tab:criterion} shows that dropping either costs its protection. The condition holds for the generalized toric codes with qubits on the $p$-cells of a $D$-dimensional torus, $D\ge4$ and $2\le p\le D-2$~\cite{MazacHamma2012}, which protect $\binom Dp$ logical qubits [Fig.~\ref{fig:results}(b)]. In the language of generalized symmetries~\cite{GKSW2015,McGreevy2023}, the theorem states that the finite-temperature breaking of higher-form symmetries of degrees $1\le q\le D-2$ survives generic explicit breaking (Table~\ref{tab:criterion}; Supplemental Material~\cite{SM}, Sec.~S11). Both properties can be checked on the ideal code alone. Robustness over the whole neighborhood is confirmed at a single point.

\begin{table}[b]
\caption{Two local properties of a code and the protection each provides. Rows: whether every excitation can be erased locally. Columns: whether the stored information is invisible to local operators. In the language of higher-form symmetries the rows are $q\le D-2$ and the columns $q\ge1$ for every symmetry of the code.}
\label{tab:criterion}
\begin{ruledtabular}
\begin{tabular}{p{1.45cm}p{3.1cm}p{3.1cm}}
 & visible to local operators & invisible to local operators \\
\colrule
erasable & 2D Ising ($q=0$): thermal memory, erased by a weak field & 4D and generalized toric codes ($1\le q\le D-2$): both protections (this work) \\
not erasable & 1D Ising ($q=0$): no memory & 2D, 3D toric codes (a symmetry with $q=D-1$): stable at $T=0$, no quantum memory at $T>0$ (3D keeps a classical bit) \\
\end{tabular}
\end{ruledtabular}
\end{table}

\begin{figure}[t]
\includegraphics[width=\columnwidth]{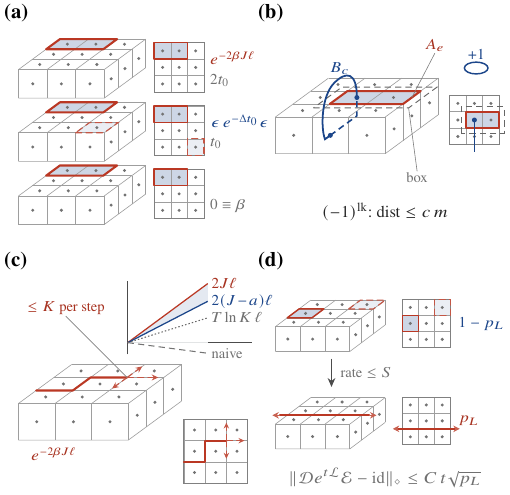}
\caption{Imaginary-time histories of the perturbed thermal state (same slab as Fig.~\ref{fig:enemies}(b)). (a)~Imaginary time as a stack of slices of duration $t_0\sim1/\Delta$, periodic in $\beta$. A virtual loop lives in one slice, with weight $(\epsilon/\Delta)^2$ independent of $\beta$; a thermal loop is present in every slice and winds around the circle at cost $e^{-2\beta J\ell}$. (b)~Loops on the lattice (violated $A_e$) and on the dual lattice (violated $B_c$). When one threads the membrane of the other the amplitude acquires the sign $(-1)^{\rm lk}$; linking requires the two to lie within a box of the smaller one's size, so the signs act at short range, Eq.~\eqref{eq:signs}. (c)~Peierls counting: at most $K^\ell$ loops of length $\ell$ through a point. Inset: virtual loops lower the cost of a loop from $2J\ell$ to at least $2(J-a)\ell$, still above the entropy $T\ell\ln K$. (d)~Patterns with only small loops carry weight $1-p$ and decode correctly; dynamics that preserve the thermal state leak into patterns with a large loop at a rate at most $S\sqrt p$.}
\label{fig:histories}
\end{figure}

\emph{Self-correction of the ideal code.}---The two properties enter the proof exactly where the classical Peierls argument needs them, so we first recall that argument for the ideal code~\cite{DKLP2002,AHHH2010}. At $\epsilon=0$ the checks are conserved and the thermal state is a classical gas of excitation loops, a loop of length $\ell$ carrying weight $e^{-2\beta J\ell}$ (uniform couplings $J$, $\beta=1/T$). A loop of length $\ell$ through a given point can be drawn in at most $K^\ell$ ways, with $K$ a lattice constant [Fig.~\ref{fig:histories}(c)], so
\begin{equation}
\Pr[\text{a loop of length }\ell\text{ through a point}]\le(Ke^{-2\beta J})^\ell .
\label{eq:peierls}
\end{equation}
For $T<2J/\ln K$ the energy of a large loop beats its entropy, and large loops are exponentially rare. A logical error requires a loop stretching across the system at some moment, and dynamics that preserve the thermal state produce one at a rate of order $e^{-cL}$ [Fig.~\ref{fig:histories}(d)].

The weight in Eq.~\eqref{eq:peierls} comes from a comparison: each configuration containing a given loop is mapped to the configuration with that loop erased, which lowers the energy by $2J\ell$. The map requires erasing that loop alone. A small loop bounds a small membrane, and flipping the plaquettes of that membrane removes the loop while leaving every other excitation in place. This is the first property. A single point-like excitation of the two-dimensional toric code has no such membrane [Fig.~\ref{fig:enemies}(a)]. Point-like excitations are created in pairs and can only be moved. The second property enters at readout. The Ising model also erases its domain walls locally and so stores a classical bit, but its magnetization is a local observable that a weak field biases; the toric code has no local operator that distinguishes its logical states.

\emph{Quantum fluctuations in imaginary time.}---A perturbation breaks the classical argument in two ways. The checks are no longer conserved, so the excitations one measures are no longer classical objects with a Boltzmann weight. $V$ also cannot be treated as small as a whole, since its total strength $\epsilon L^4$ exceeds the energy $JL$ of a loop spanning the system, and can only be expanded term by term. Quantum Pirogov--Sinai theory~\cite{Ginibre1969,BKU1996,DFF1996} handles quantum perturbations of classical models, but it expands around product states, which a stabilizer code lacks. Term-by-term expansions around the ideal thermal state converge only for $\epsilon\lesssim e^{-c\Delta/T}$, with $\Delta$ the smallest excitation energy, and averaging the perturbation over thermal fluctuations reaches only $\epsilon\lesssim T$ [Fig.~\ref{fig:results}(a)]; both windows close in the low-temperature corner.

At low temperature, most excitations one would measure are not thermal at all. A perturbation term that flips one plaquette creates the smallest loop, of energy $\Delta=8J$, and second-order perturbation theory gives such virtual loops a density of order $(\epsilon/\Delta)^2$, independent of temperature. Thermal loops have density $e^{-\Delta/T}$. The two cross at $T_\times=\Delta/[2\ln(\Delta/\epsilon)]$ [Fig.~\ref{fig:enemies}(c)]; below $T_\times$ quantum fluctuations outnumber the thermal excitations the memory was designed to resist.

Imaginary time turns these fluctuations into geometric objects. The thermal state is a sum over histories on a circle of circumference $\beta$, which we cut into slices of fixed duration $t_0\sim1/\Delta$ [Fig.~\ref{fig:histories}(a)]. A perturbation term that creates a small loop must annihilate it again. The loop is born with amplitude $\epsilon$, pays $e^{-\Delta t}$ while it lives, and dies with amplitude $\epsilon$. It is a bubble of lifetime $1/\Delta$ and weight $(\epsilon/\Delta)^2$, the virtual density above. A thermal loop is present in every slice and winds once around the circle, a sheet of weight $e^{-2\beta J\ell}$. Only large structures in space-time are dangerous, and the problem becomes a Peierls argument in $4+1$ dimensions. The weight of a bubble is set by $\epsilon$ alone, with no factor of $\beta$, so the threshold survives the limit $T\to0$. The protection is energetic. The energy cost of a loop is already present in the ground state, and thermal fluctuations only add entropy against it.

The central estimate states that quantum fluctuations cannot remove the tension of excitations, their free-energy cost per unit length. Fix a large loop $\Gamma$, let $K_\Gamma$ be the perturbed Hamiltonian with the few terms that could move $\Gamma$ removed, and let $K_\Gamma^F$ be the same Hamiltonian after $\Gamma$ has been erased by flipping its membrane. We prove
\begin{equation}
\frac1\beta\left|\ln\frac{Z(K_\Gamma^F)}{Z(K_\Gamma)}\right|\le a\,|\Gamma|,\qquad a<J,
\label{eq:key}
\end{equation}
uniformly in $L$, $T\le T_0$, and $\epsilon\le\epsilon_0$. The perturbation shifts the free-energy cost of $\Gamma$ by at most $a|\Gamma|$, below its bare cost $2J|\Gamma|$ [Fig.~\ref{fig:histories}(c), inset]. Inserting Eq.~\eqref{eq:key} into the counting of Eq.~\eqref{eq:peierls} gives Eq.~\eqref{eq:static}.

\emph{Locality of the topological signs.}---One feature has no classical counterpart. Histories carry signs~\cite{Klich2010}. The two kinds of loops have mutual statistics. When a loop of violated edge checks threads the membrane of a loop of violated cube checks, the amplitude changes sign, an Aharonov--Bohm phase of $\pi$ [Fig.~\ref{fig:histories}(b)]. In imaginary time a history is a set of bubbles and sheets, its components $C$, and its weight is
\begin{equation}
\begin{gathered}
\prod_C\omega(C)\prod_{C<D}\sigma(C,D),\\
\sigma(C,D)=-1\ \Rightarrow\ {\rm dist}(C,D)\le c\min(m(C),m(D)),
\end{gathered}
\label{eq:signs}
\end{equation}
where $\omega(C)$ is the amplitude of component $C$ on its own, $m(C)$ its size, and $\sigma(C,D)=-1$ when $C$ and $D$ are linked. These signs are the threat. If distant components could pick up signs from one another, the Peierls counting would collapse.

Linking, however, requires closeness. To link a loop of size $m$, another loop must enter a box of size $m$ around it [Fig.~\ref{fig:histories}(b)]. We prove an empty-box lemma: a component acquires no sign from anything outside a box of its own size. Both properties of the code enter here. By the first, each small component can be erased inside its box, so it leaves no trace outside; by the second, a small component carries no logical information, so nothing outside can feel it. This is the second half of Eq.~\eqref{eq:signs}. Signs act only between neighbors. The new case at $T>0$ is a bubble linked with a sheet that winds around the thermal circle. A virtual loop feels a thermal excitation through the Aharonov--Bohm phase at short range only, so the thermal excitation renormalizes only the free energy of its neighborhood. Linked components are counted like connected clusters, the gas stays dilute, and Eq.~\eqref{eq:key} follows from an expansion whose small parameter is $\epsilon$ alone (Appendix~B); at the lowest temperatures, $T\lesssim J/L^4$, a one-line entropy bound replaces this expansion.

\emph{Lifetime of the logical qubits.}---Rare excitations become a long-lived memory through a stationarity argument, a fully quantum version of the stationarity criterion of Ref.~\cite{CLBT2010}. Call an excitation pattern good if all its loops are small. On good patterns every local operation commutes with the logical algebra, because its effect can be erased locally and leaves no logical trace, the two properties once more. The dynamics preserves the thermal state, in which bad patterns have probability $p\le e^{-cL}$ for large $L$ by Eq.~\eqref{eq:static}, so the encoded state leaks into the bad set only through the local moves of the dynamics, at a rate bounded by their total strength $S$ times $\sqrt p$ [Fig.~\ref{fig:histories}(d)]. Integrating over time gives Eq.~\eqref{eq:memory}, for any input and any reference system entangled with it.

\begin{figure}[t]
\includegraphics[width=\columnwidth]{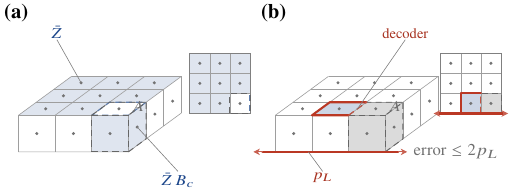}
\caption{Erasure recovery in the perturbed thermal state (same slab as Fig.~\ref{fig:enemies}(b)). (a)~A logical operator $\bar Z$ is a sheet of $Z_p$ spanning the system. Multiplying it by the cube check $B_c$ of a small region $A$ moves it off $A$: a small region carries no logical information. (b)~If $A$ is erased and refilled with a fixed state, the checks outside $A$ show an open string of excitations ending on $A$, which the ideal decoder closes inside $A$. The only failure is a loop threading $A$ and spanning the system (bottom, wrapping the periodic direction), of probability $p$; the recovery error is at most $2p$, with no knowledge of the perturbation or the temperature.}
\label{fig:code}
\end{figure}

\emph{Consequences.}---Topological order survives at nonzero temperature in the perturbed system, in its information-theoretic sense. The perturbed thermal state remains a quantum error-correcting code (Fig.~\ref{fig:code}, Appendix~C). Let $A$ be any region inside a box of side $cL$ and let $\cN_A$ be any physical noise acting on $A$. Decoding after the noise gives
\begin{equation}
\cD\,\cN_A\,\cE=(1-p)\,\id+p\,\Xi_A,
\label{eq:code}
\end{equation}
with $\Xi_A$ a channel. If $A$ is erased, inserting any fixed state and running the ideal decoder restores the logical qubits with error at most $2p$, and the logical information visible inside $A$ is bounded by the same $2p$~\cite{CleveGottesmanLo1999}. In this sense, topological order at nonzero temperature~\cite{Hastings2011} is a property of every Hamiltonian near the ideal code.

For open quantum systems, Eq.~\eqref{eq:memory} establishes a robust steady-state degeneracy. Under every quasi-local dynamics that preserves the thermal state, every thermally encoded state keeps its logical content for times exponential in $L$. Because the argument sees the bath only through the state it preserves, the generator may change from one time interval to the next, and a bath that preserves the thermal state only approximately adds its accumulated residual to the error (Appendix~C). Steady-state degeneracy usually needs a symmetry or engineered nonequilibrium dynamics~\cite{Squeezing2025}; here it comes from the Hamiltonian, and the bath need only be quasi-local and thermal.

For quantum algorithms, Eq.~\eqref{eq:memory} implies an obstruction. Any quasi-local dynamics that preserves the low-temperature thermal state of these noncommuting Hamiltonians needs, from a worst-case initial state, a time exponential in $L$ to reach it (Appendix~C). This extends the slow mixing known for commuting codes~\cite{GamarnikKianiZlokapa2024} to generic noncommuting perturbations. The obstruction is the stored information itself. A sampler that prepared these states quickly would also erase their qubits quickly.

\emph{Discussion.}---We have proved that self-correcting quantum memory is stable under generic weak local perturbations. In an open region of Hamiltonians around the ideal code, the system passively holds a qubit in low-temperature thermal equilibrium, with a storage lifetime exponential in the linear system size and with no symmetry protection or fine-tuning. Self-correction was conceived as a property of a phase of matter~\cite{Kitaev2003}, yet for two decades it could be proved only at exactly solvable points. The circuit definitions of thermal phases~\cite{MKS2026,Sang2025} place thermal states that local channels transform into each other in one phase. Decoding and re-encoding with our decoder already transforms the thermal states of the stable region into each other globally, a concrete starting point for constructing the corresponding local channels. That local transformation would identify the whole stable region as a thermal phase, as quasi-adiabatic continuation~\cite{HastingsWen2005,BHM2010} does for topological order at zero temperature, and self-correction would become the property of a phase it was conceived to be.

Because the two properties leave the dimension of the lattice free, the central question this result raises is three-dimensional: can a local three-dimensional system support a self-correcting quantum memory that is stable under generic weak local perturbations, with a storage lifetime exponential in the linear system size? Any candidate three-dimensional code can be tested against the two properties at its exactly solvable point. Achieving this would let three-dimensional matter hold a qubit robustly in thermal equilibrium, the way a magnet holds a classical bit.

\begin{acknowledgments}
\emph{Acknowledgments.}---This work was supported by the Fundamental and Interdisciplinary Disciplines Breakthrough Plan of the Ministry of Education of China (Grant No.~JYB2025XDXM606).
\end{acknowledgments}

\makeatletter
\def\bibsection{\par\vskip\z@\@plus1fil\onecolumngrid@push\begingroup\baselineskip26\p@\bib@device{\textwidth}{245.5\p@}\endgroup\nobreak\@nobreaktrue\addvspace{19\p@}\par\onecolumngrid@pop}
\makeatother

\clearpage
\onecolumngrid
\section*{End Matter}
\twocolumngrid

\setcounter{equation}{0}\renewcommand{\theequation}{A\arabic{equation}}\renewcommand{\theHequation}{A\arabic{equation}}%
\emph{Appendix A: The topological Peierls condition}---Consider stabilizer codes on $D$-dimensional periodic lattices of linear size $L$, with $H_0=-\sum_aJ_aS_a$, $0<J_{\min}\le J_a\le J_{\max}$, commuting Pauli checks $S_a$ of bounded weight, a bounded number of checks per qubit, and at most $k_{\max}$ logical qubits. Checks come in finitely many types, and excitations of one type are grouped into connected clusters by a fixed finite-range adjacency. The topological Peierls condition has three parts.
(i)~\emph{Local erasure.} Every cluster $C$ of diameter $r<cL$ that occurs in some valid excitation pattern is the full syndrome (the set of violated checks) of a Pauli operator $F_C$ supported within distance $C_0(r+1)$ of $C$, and $F_C$ creates no excitations of the other types.
(ii)~\emph{Global constraints.} Clusters of each type carry a charge $q(C)\in\mathbb F_2^h$, with $h$ fixed, such that valid patterns have zero total charge and every union of non-touching clusters with zero total charge is valid. For the four-dimensional toric code $h=4$ counts the homology classes of loops.
(iii)~\emph{Local generation.} Every Pauli operator $P$ supported in a box of diameter $r<cL$ that commutes with all checks is a product of checks within distance $C_0(r+1)$ of the box,
\begin{equation}
P=\zeta\prod_{a\in Y}S_a,\qquad|\zeta|=1 .
\label{eq:gen}
\end{equation}
Properties 1 and 2 of the main text are (i) and (iii); (ii) is the bookkeeping that closes the Peierls counting when several loops are present.

\emph{Theorem.} For every such family and every finite range of $V$ there are $\epsilon_0,T_0,a,C,L_0>0$, independent of $L$, $T$, and $V$, such that for $L\ge L_0$, $T\le T_0$, and $\|V\|_\loc\le\epsilon_0$: (a)~$\Pr_\rho[\text{a cluster of}\ge u\text{ excitations}]\le CL^{D(h+1)}e^{-4u}$; (b)~the decoder $\cD_L$ built from the checks and the erasure rule (i), and a linear thermal encoding $\cE_{\beta,V}$ with $\cD_L\cE_{\beta,V}=\id$, satisfy $\|\cD_Le^{t\cL}\cE_{\beta,V}-\id\|_\diamond\le CL^{D(h+3)/2}\,t\,e^{-aL}$ for every dynamics of the class of Appendix~C. For the four-dimensional toric code the two exponents are $20$ and $14$. Section~S8 of the Supplemental Material~\cite{SM} proves the theorem by substituting the general constants into the four-dimensional proof of Secs.~S2--S6.

The condition holds for the generalized toric codes with qubits on $p$-cells, $D\ge4$, $2\le p\le D-2$, with $X$ checks on $(p-1)$-cells and $Z$ checks on $(p+1)$-cells, $k=\binom Dp$ and $h=\max\{\binom D{p-1},\binom D{p+1}\}$ (Sec.~S8.3); the two- and three-dimensional toric codes violate (i) for their point-like type. The class is closed under finite stacking, pinned ancillas, and conjugation with a finite-depth local unitary, which admits non-Pauli checks and shrinks the allowed regions of Appendix~C by the light-cone radius (Sec.~S8.4).

\setcounter{equation}{0}\renewcommand{\theequation}{B\arabic{equation}}\renewcommand{\theHequation}{B\arabic{equation}}%
\emph{Appendix B: Histories, signs, and the central estimate}---Fix $\Gamma$, $K_\Gamma$, and $K_\Gamma^F$ as in the main text, and cut $\beta$ into $N=\lfloor\beta/t_0\rfloor$ slices of duration $t_{\rm s}\in[t_0,2t_0]$. Within each slice the interaction picture around $H_0$ expands $e^{-t_{\rm s}K_\Gamma}$ exactly into connected groups $g$ of perturbation terms, with operator weights of norm at most $\prod_{i\in g}(e^{C(t_0)\|v_i\|}-1)$; $C(t_0)$ depends only on $t_0$, $J_{\max}$, and the range of $V$. Expanding each weight in the Pauli basis and inserting the check projectors between slices writes $Z(K_\Gamma)$ as an exact sum over histories, periodic excitation patterns $s_0,\dots,s_N=s_0$ with Pauli insertions $P_k$, of weight
\begin{equation}
\Bigl(\prod_kc_k\Bigr)\,e^{-t_{\rm s}\sum_kE(s_k)}\,\frac1d\,\Tr_{s_0}(P_N\cdots P_1),\;\; E(s)=2\!\sum_{a\in s}\!J_a ,
\label{eq:history}
\end{equation}
where $c_k$ is the product of the Pauli coefficients of the groups in slice $k$ and the trace vanishes unless the net insertion is a check product times a phase (Sec.~S3).

A history splits into connected space-time components; the components of size at least $c_0L$ form the macroscopic background. By (i) and (iii), a component $C$ of size $m(C)<c_0L$ has a valid pattern in every slice, a net insertion $\zeta_CS_C$ generated inside a box of diameter $O(m(C))$, and a weight $|\omega(C)|\le(e^{C(t_0)\epsilon}-1)^{a(C)}e^{-2t_0J_{\min}d(C)}$, with $a(C)$ insertions and $d(C)$ excitation slices. The weight of a history with only local components is exactly $\prod_C\omega(C)\prod_{C<D}\sigma(C,D)$ with
\begin{equation}
\sigma(C,D)=(-1)^{\kappa(C,D)}\,\chi_{S_C}(s_D)\,\chi_{S_D}(s_C),
\label{eq:sigma}
\end{equation}
where $\kappa(C,D)$ counts the anticommuting pairs of insertions in which the insertion of $D$ precedes that of $C$, $\chi_S(s)=\pm1$ is the eigenvalue of the check product $S$ on the pattern $s$, and $s_C,s_D$ are the patterns of $C$ and $D$ in the slice where the circle is cut. This is Eq.~\eqref{eq:signs} (Sec.~S4.1).

\emph{Empty-box lemma.} Let $\mathcal B_C$ be the space-time box of diameter $cm(C)$ around a local component $C$. If $D$ has no insertion and no excitation in $\mathcal B_C$, then $\sigma(C,D)=+1$; the same holds with $D$ replaced by the whole macroscopic background. The lemma uses (i) to erase $C$ inside its box and (iii) to show that nothing outside the box sees the net insertion of $C$ (Sec.~S4.2).

\emph{Linked clusters.} Components joined by $\sigma=-1$ form linked clusters, and at most $K_{\rm link}^m$ linked clusters of size $m$ are rooted at a given point, with $K_{\rm link}$ a geometric constant (Sec.~S4.4). For $\epsilon\le\bar\epsilon$ the gas of linked clusters converges, and $\ln Z_\loc$ changes by at most $\sum_\gamma|\Delta z_\gamma|e^{m(\gamma)}$ under any change of the cluster activities (Sec.~S4.5). Erasing $\Gamma$ flips the signs of the insertions in the components touching $\Gamma$ and nothing else, so $|\ln[Z_\loc(K_\Gamma^F)/Z_\loc(K_\Gamma)]|\le C_1\epsilon N|\Gamma|$ (Sec.~S4.6). The macroscopic background keeps its logical trace as a whole and changes $\ln Z$, relative to the local gas, by $O(NL^Dr^{c_0L})$ with $r<1$, which is negligible for $\beta\le A_*L^D$ (Secs.~S4.7 and S4.8).

\emph{Entropy bridge.} For Hermitian $A,B$ on a space of dimension $2^n$, $f_\beta=-\beta^{-1}\ln Z_\beta$ lies between $E_0-n\ln2/\beta$ and the ground energy $E_0$, so a bound $|f_{\beta_*}(A)-f_{\beta_*}(B)|\le a$ at one $\beta_*$ gives
\begin{equation}
\frac1\beta\Bigl|\ln\frac{Z_\beta(A)}{Z_\beta(B)}\Bigr|\le a+\frac{2n\ln2}{\beta_*}\qquad(\beta\ge\beta_*).
\label{eq:bridge}
\end{equation}
With $\beta_*=A_*L^D$ and $n\le C_{\rm vol}L^D$ the correction is a fixed margin $\delta<J_{\min}$, paid by the bare tension, and Eq.~\eqref{eq:key} follows for all $T\le T_0$ with $a<J_{\min}$ (Sec.~S5). Section~S2 of the Supplemental Material~\cite{SM} reduces the probability that a given loop $\Gamma$ is fully violated to Eq.~\eqref{eq:key} through a Dyson expansion of the few terms that move $\Gamma$ within a fixed time window. The union over loops with the grouping (ii) then gives (a) and Eq.~\eqref{eq:static}.

\setcounter{equation}{0}\renewcommand{\theequation}{C\arabic{equation}}\renewcommand{\theHequation}{C\arabic{equation}}%
\emph{Appendix C: Encoding, decoder, and operational statements}---Let $W:\mathcal H_{\rm log}\otimes\mathcal H_S\to\mathcal H_{\rm phys}$ be the unitary identification defined by the decoder, which erases small clusters by (i) and completes the remaining patterns by a fixed global rule; let $G=W(I\otimes G_S)W^\dagger$ project on good patterns and $p=\Tr[\rho(I-G)]$. In the four-dimensional proof, good patterns have all clusters of diameter below $10^{-3}L$, and allowed regions are boxes of that side. The thermal encoding and decoder are
\begin{equation}
\begin{gathered}
\cE(X)=W(X\otimes\rho_S)W^\dagger,\quad \rho_S=\Tr_{\rm log}(W^\dagger\rho W),\\
\cD(Y)=\Tr_S(W^\dagger YW),
\end{gathered}
\label{eq:encoding}
\end{equation}
with $\cD\cE=\id$ (Sec.~S6.1).

\emph{One-sided identity.} Every operator $O$ on an allowed region satisfies $[O,\bar M]G=0$ for every logical operator $\bar M=W(M\otimes I)W^\dagger$ (Sec.~S1.4). For a channel $\cN_A$ on an allowed region the identity gives $\cD\cN_A\cE=(1-p)\id+p\,\Xi_A$ with $\Xi_A$ a channel, which is Eq.~\eqref{eq:code}. Erasing $A$ and inserting a fixed state is the special case with recovery error at most $2p$, and the reduced channel of $\cE$ on $A$ is within $2p$ of a constant channel, which bounds the logical information visible in $A$ (Sec.~S7).

\emph{Stationarity bound.} Let $\cL_s$ be a time-dependent generator with $\cL_s(\rho)=0$, split into local terms of total strength $S(s)$ and a tail of strength $u(s)$ beyond radius $r$, and let $\Phi(t,0)$ be its evolution. Then
\begin{equation}
\|\Phi(t,0)\cE-\cE\|_\diamond\le\min\Bigl\{2,\;2d\!\int_0^t\!\bigl[4S(s)\sqrt p+2u(s)\bigr]ds\Bigr\}.
\label{eq:dyn}
\end{equation}
The bound rests on the one-sided identity. On good inputs the logical replacement map commutes with every local map, and $\rho$ is within $2\sqrt p$ of its good part (Secs.~S6.2 and S6.3). With $S\le CL^D$, $u\le CL^De^{-\mu r}$, and $r=cL$ this is Eq.~\eqref{eq:memory}. A generator that preserves $\rho$ only approximately adds $2d\int_0^t\|\cL_s(\rho)\|_1ds$ to the bound.

\emph{Mixing time.} For every dynamics of this class, the worst-case trace distance to $\rho$ after time $t$ is at least $1-1/d-\mathfrak B(t)/2$, with $\mathfrak B(t)$ the integral in Eq.~\eqref{eq:dyn}, so mixing to accuracy $\eta<1-1/d$ takes a time of at least $cL^{-D(h+3)/2}e^{aL}$ (Sec.~S7.4).

\clearpage
\onecolumngrid
\makeatletter
\def\normalsize{\@setfontsize\normalsize\@xpt{11.5}\abovedisplayskip 10\p@ \@plus2\p@ \@minus5\p@
  \belowdisplayskip\abovedisplayskip\abovedisplayshortskip\abovedisplayskip\belowdisplayshortskip\abovedisplayskip\let\@listi\@listI}
\makeatother
\normalsize
\setcounter{page}{1}\renewcommand{\thepage}{S\arabic{page}}
\setcounter{equation}{0}\setcounter{figure}{0}\setcounter{table}{0}\setcounter{section}{0}
\renewcommand{\theequation}{S\arabic{equation}}\renewcommand{\theHequation}{S\arabic{equation}}
\renewcommand{\thesection}{S\arabic{section}}\renewcommand{\theHsection}{S\arabic{section}}
\renewcommand{\thesubsection}{\thesection.\arabic{subsection}}\renewcommand{\theHsubsection}{S\arabic{section}.\arabic{subsection}}
\makeatletter\renewcommand{\p@subsection}{}\makeatother
\renewcommand{\thetable}{S\Roman{table}}\renewcommand{\theHtable}{S\Roman{table}}
\renewcommand{\thefigure}{S\arabic{figure}}\renewcommand{\theHfigure}{S\arabic{figure}}
\renewcommand{\Pr}{\mathrm{Pr}}
\emergencystretch=1.5em
\setcounter{secnumdepth}{4}
\addtocontents{toc}{\protect\SMshowcontents}

\begin{center}
{\large\bfseries Supplemental Material for ``Robust Quantum Memory in Thermal Equilibrium''\par}
\vspace{0.8em}
Jiaxin Liu, Danyue Ma, and Jiancheng Fang
\end{center}
\begin{quote}
\small This Supplemental Material contains the complete proofs of the results stated in the main text, the verification of the topological Peierls condition for the generalized toric codes, an exactly solvable sector that isolates the quantum origin of low-temperature excitations, a comparison with earlier methods, the dictionary between the two local properties and higher-form symmetries, and the extensions that the proof admits as it stands.  Equations of the main text are cited as Eq.~(1)--(7) and (A1)--(C2); equations here are numbered (S1), (S2), \ldots
\end{quote}
\vspace{0.5em}
\setcounter{tocdepth}{2}
\tableofcontents
\clearpage

\section*{Overview}
\label{SM:sec:overview}

The main text proves two statements about the thermal state $\rho\propto e^{-\beta H}$ of the perturbed four-dimensional toric code, Eq.~(1): large excitation loops are exponentially rare, Eq.~(2), and thermally encoded qubits survive for a time exponential in $L$, Eq.~(3).  Both follow from the central estimate of the main text, the tension bound Eq.~(5), which states that quantum fluctuations shift the free-energy cost of a large excitation loop by less than its bare cost.  This document proves the tension bound and everything that follows from it.

Section~\ref{SM:sec:geometry} fixes the lattice conventions and proves the two local properties of the ideal code, which every later step uses.  The proof is then a chain of five steps, each of which occupies one section.  A fixed-window expansion turns the probability of a given large loop $\Gamma$ into the free-energy cost of erasing $\Gamma$, Sec.~\ref{SM:sec:window}.  That cost is written as an exact sum over histories in imaginary time, cut into slices of fixed duration, Sec.~\ref{SM:sec:histories}.  Histories with only small components form a gas of linked clusters whose signs act at short range, and a finite deletion induction controls the gas for $\beta$ up to a power of $L$, Sec.~\ref{SM:sec:gas}.  An entropy bound extends the control to all lower temperatures, which yields Eq.~(5) and then Eq.~(2), Sec.~\ref{SM:sec:bridge}.  A stationarity argument turns Eq.~(2) into the memory bound Eq.~(3) for every quasi-local dynamics that preserves $\rho$, Sec.~\ref{SM:sec:memory}.  The remaining sections collect the operational consequences, Sec.~\ref{SM:sec:operational}; the general theorem of Appendix~A and the model family that satisfies it, Sec.~\ref{SM:sec:general}; the exactly solvable sector that isolates the crossover temperature $T_\times$, Sec.~\ref{SM:sec:checks}; the comparison with earlier methods, Sec.~\ref{SM:sec:comparison}; the higher-form-symmetry dictionary for Table~I, Sec.~\ref{SM:sec:symmetry}; and the extensions within the hypotheses, Sec.~\ref{SM:sec:extensions}.  Table~\ref{SM:tab:roadmap} maps each main-text statement to the section that proves it.

\begin{table}[b]
\caption{Location of the proofs of the main-text statements.}
\label{SM:tab:roadmap}
\renewcommand{\baselinestretch}{1}\small
\begin{ruledtabular}
\begin{tabular}{lll}
Main-text statement & Content & Section \\
\colrule
Properties 1, 2; decoder & local erasure, local generation, decoder identity, grouping of loops & \ref{SM:sec:geometry} \\
Eq.~(4) and the Peierls counting & fixed-window reduction & \ref{SM:sec:window} \\
Eqs.~(B1), (B2) & exact history representation, pairwise signs & \ref{SM:sec:histories}, \ref{SM:sec:gas} \\
Eq.~(6), empty-box lemma, linked clusters & linking locality, cluster counting, deletion induction, background & \ref{SM:sec:gas} \\
Eqs.~(5), (B3), (2) & entropy bridge, static theorem & \ref{SM:sec:bridge} \\
Eqs.~(3), (C1), (C2) & encoding, replacement map, stationarity bound, dynamics class & \ref{SM:sec:memory} \\
Eq.~(7), mixing time & local noise, erasure recovery, leakage, mixing-time bound & \ref{SM:sec:operational} \\
Appendix~A theorem & topological Peierls condition, generalized toric codes, closure & \ref{SM:sec:general} \\
$T_\times$, Fig.~1(c) & exactly solvable single-plaquette sector & \ref{SM:sec:checks} \\
Table~I & higher-form symmetries & \ref{SM:sec:symmetry} \\
\end{tabular}
\end{ruledtabular}
\end{table}

Throughout, $\|\cT\|_{1\to1}=\sup_X\|\cT(X)\|_1/\|X\|_1$ is the induced trace norm of a superoperator on the full physical matrix space, a map supported in a region $Q$ means $\cT_Q\otimes\id_{Q^c}$, and $\|\cdot\|_\diamond$ is the diamond norm.  A \emph{pattern} is a set of violated checks; $\syn(P)$ denotes the pattern created by a Pauli operator $P$ acting on a state with no excitations.  Constants named $C,c,K$ with subscripts are independent of $L$, $\beta$, and $V$ unless their dependence is displayed.

\section{Geometry, local generation, and the decoder}
\label{SM:sec:geometry}

This section fixes the lattice conventions and proves the two properties of the ideal code that the main text uses: every small excitation cluster is erased by a nearby Pauli operator, and every local Pauli operator commuting with all checks is a product of nearby checks.  Both follow from one explicit chain contraction on a box.  The same contraction defines the decoder and yields the one-sided identity, Lemma~\ref{SM:lem:decoder}, on which the memory bound and all operational results rest.  The section ends with the grouping of large clusters into unions that bound, which the static bound of Sec.~\ref{SM:sec:window} counts.

\subsection{Lattice, checks, and excitation patterns}
\label{SM:sec:lattice}

The lattice is the periodic cubical complex $T_L^4=(\Z/L\Z)^4$ with $\F_2$ coefficients: $k$-cells are products of $k$ unit intervals and $4-k$ points, and $\partial$ is the cellular boundary.  Qubits sit on the plaquettes (2-cells).  For every edge $e$ and every cube $c$,
\begin{equation}
A_e=\prod_{p\supset e}X_p,\qquad B_c=\prod_{p\subset\partial c}Z_p,\qquad
H_0=-\sum_eJ_eA_e-\sum_cJ_cB_c,
\label{SM:eq:H0}
\end{equation}
with $0<J_{\min}\le J_e,J_c\le J_{\max}$.  For a $2$-chain $z$ (a set of plaquettes) write $Z(z)=\prod_{p\in z}Z_p$ and $X(z)=\prod_{p\in z}X_p$, so that $A_e=X(\delta e)$ with $\delta e$ the set of plaquettes containing $e$, and $B_c=Z(\partial c)$.

\begin{lemma}[elementary lattice properties]
\label{SM:lem:lattice}
\begin{enumerate}
\item[(i)] There are $n=6L^4$ qubits, $4L^4$ edge checks, and $4L^4$ cube checks.  Each check acts on $w=6$ qubits, and each qubit lies in four edge checks and four cube checks.  All checks commute, and the group they generate does not contain $-I$.
\item[(ii)] The edge-check pattern of $Z(z)$ is $\partial z$, a closed $1$-chain, and $Z(z)$ commutes with every cube check; the cube-check pattern of $X(x)$ is the set of cubes $c$ with $|\partial c\cap x|$ odd, which under the cellular duality $*$ of $T^4_L$ is the closed dual $1$-chain $\partial(*x)$, and $X(x)$ commutes with every edge check.  A pattern is valid if and only if its edge part is a $1$-chain that bounds and its cube part is a dual $1$-chain that bounds; the obstruction to bounding is the class in $H_1(T^4;\F_2)=\F_2^4$, so there are $h=4$ global constraints per type, and the code space has dimension $d=2^6$ by $H_2(T^4;\F_2)=\F_2^6$.
\item[(iii)] The energy of a pattern $s$ above the ground energy $E_{\rm ref}=-\sum_aJ_a$ is $E(s)=2\sum_{a\in s}J_a$; the smallest excitation, a single flipped plaquette, violates four edge checks and costs $\Delta=8J$ for uniform couplings.
\end{enumerate}
\end{lemma}

\begin{proof}
(i) A plaquette is fixed by a vertex and two of the four directions, an edge or a cube by a vertex and one or three directions.  An edge in direction $i$ lies in the plaquettes spanned by $i$ and one of the other three directions, two per direction; a cube has six faces; a plaquette has four edges and lies in the cubes obtained by adding one of the two remaining directions in either orientation.  Two $X$-type or two $Z$-type checks commute.  An edge check $A_e$ and a cube check $B_c$ share the plaquettes that contain $e$ and lie in $\partial c$: two if $e$ is an edge of $c$ and none otherwise, so they commute.  A product of checks is $X(\delta y)Z(\partial u)$ with coefficient $+1$ for a $1$-chain $y$ and a $3$-chain $u$, hence never $-I$.
(ii) $A_e$ anticommutes with $Z(z)$ if and only if $|\delta e\cap z|$, the number of plaquettes of $z$ containing $e$, is odd, that is, $e\in\partial z$; $\partial\partial=0$ makes $\partial z$ closed.  $B_c$ anticommutes with $X(x)$ if and only if $|\partial c\cap x|$ is odd; the duality $*$ sends cubes to dual edges, plaquettes to dual plaquettes, and incidence to incidence, so this set is $\partial(*x)$ on the dual lattice.  A general Pauli operator is a phase times $X(x)Z(z)$ and produces both patterns.  Conversely a closed $1$-chain $\gamma$ with $\gamma=\partial z$ is the edge pattern of $Z(z)$, and a closed $1$-chain that does not bound is not a boundary of any $2$-chain, hence not the pattern of any $Z$-type operator; the same holds on the dual lattice.  The remaining statements are the homology of the torus with $\F_2$ coefficients: by the K\"unneth formula $H_k(T^4;\F_2)=\F_2^{\binom4k}$, with $H_1$ generated by the four coordinate circles and $H_2$ by the six coordinate planes; a closed $1$-chain bounds if and only if its class in $H_1$ vanishes; and the code space, the joint $+1$ eigenspace of all checks, has dimension $2^{\dim H_2}=2^6$, by the dimension count $k=n-{\rm rank}(\delta)-{\rm rank}(\partial)=\dim H_2$ for a CSS code built from a chain complex.
(iii) $H_0=-\sum_aJ_aS_a=E_{\rm ref}+2\sum_aJ_an_a$ with $n_a=\tfrac12(I-S_a)$ the projector on $S_a=-1$, and $Z_p$ anticommutes with exactly the four edge checks of $p$.
\end{proof}

The perturbation is $V=\sum_iv_i$ with a fixed range $R$, terms of equal support merged, pure scalars removed, and each term Hermitian; replacing $v_i$ by $(v_i+v_i^\dagger)/2$ changes neither $V$ nor the support, because $V$ is Hermitian, and does not increase $\|V\|_\loc=\sup_p\sum_{i:p\in\supp v_i}\|v_i\|$.  Two geometric constants recur: $k_R$, the largest number of qubits in one term, and $b=4k_R$, the largest number of checks of one type that fail to commute with one term, since a term fails to commute with a check only if the check touches its support.

Excitations of one type are grouped into connected \emph{clusters} by adjacency in the line graph of the lattice: two edges are adjacent when they share a vertex, two dual edges when they share a dual vertex.

\begin{lemma}[cluster enumeration]
\label{SM:lem:clustercount}
The line graph has degree at most $14$, and the number of connected sets of $m$ edges containing a given edge is at most $14^{2(m-1)}\le196^m$.  This is the constant $K$ of Eq.~(4).
\end{lemma}

\begin{proof}
Each vertex has eight edges, so an edge is adjacent to seven others at each of its two endpoints, fourteen in all.  Let $\mathcal C$ be a connected set of $m$ edges containing $e$, and let $T$ be a spanning tree of the subgraph of the line graph induced on $\mathcal C$, chosen by a fixed rule.  A depth-first traversal of $T$ from $e$ is a closed walk $e=e_0,e_1,\dots,e_{2(m-1)}=e$ in the line graph that traverses every edge of $T$ once in each direction, so it visits every element of $\mathcal C$, and $\mathcal C$ is the set of edges the walk visits.  Distinct sets have distinct walks, and a walk is determined by its $2(m-1)$ steps, each with at most $14$ choices; hence there are at most $14^{2(m-1)}$ such sets.
\end{proof}

\subsection{The cubical contraction}
\label{SM:sec:contraction}

Fix an integer $r\ge1$ and the interval complex $[0,r]$ with vertices $\mathbf v_0,\dots,\mathbf v_r$ and edges $\mathbf e_j=[j,j+1]$, $0\le j<r$, over $\F_2$.  Define the chain map $\pi_I$ and the degree-raising map $H_I$ by
\begin{equation}
\pi_I(\mathbf v_j)=\mathbf v_0,\quad\pi_I(\mathbf e_j)=0,\qquad H_I(\mathbf v_j)=\sum_{k=0}^{j-1}\mathbf e_k,\quad H_I(\mathbf e_j)=0 .
\label{SM:eq:intervalmaps}
\end{equation}
A box $Q=[0,r_1]\times\cdots\times[0,r_D]$ has as cellular chain complex the tensor product of the $D$ interval complexes, without signs over $\F_2$, with $\partial=\sum_{j=1}^D\id\otimes\cdots\otimes\partial_j\otimes\cdots\otimes\id$, where $\partial_j$ acts on the $j$-th factor.  Define
\begin{equation}
H_Q=\sum_{j=1}^D\pi_{I_1}\otimes\cdots\otimes\pi_{I_{j-1}}\otimes H_{I_j}\otimes\id\otimes\cdots\otimes\id,\qquad
\Pi_Q=\pi_{I_1}\otimes\cdots\otimes\pi_{I_D} ,
\label{SM:eq:HQ}
\end{equation}
where $\pi_{I_j}$ and $H_{I_j}$ act on the $j$-th factor.

\begin{lemma}[contraction]
\label{SM:lem:contraction}
On the chain complex of every box $Q$,
\begin{equation}
\partial H_Q+H_Q\partial=\id+\Pi_Q .
\label{SM:eq:contraction}
\end{equation}
Consequently every closed $k$-chain $z$ with $k\ge1$ inside $Q$ satisfies
\begin{equation}
\partial(H_Qz)=z,
\label{SM:eq:filling}
\end{equation}
with $H_Qz$ a $(k+1)$-chain inside $Q$ determined by $Q$, the coordinate order, and $z$.
\end{lemma}

\begin{proof}
On one interval, $\partial H_I+H_I\partial=\id+\pi_I$: for a vertex, $\partial H_I(\mathbf v_j)=\partial\sum_{k=0}^{j-1}\mathbf e_k=\mathbf v_0+\mathbf v_j$ by telescoping modulo $2$, and $H_I\partial(\mathbf v_j)=0$; for an edge, $\partial H_I(\mathbf e_j)=0$ and $H_I\partial(\mathbf e_j)=H_I(\mathbf v_j)+H_I(\mathbf v_{j+1})=\mathbf e_j$.  Also $\pi_I$ is a chain map, $\partial\pi_I=\pi_I\partial=0$ on edges and vertices alike.  On the box, let $P_j=\pi_{I_1}\otimes\cdots\otimes\pi_{I_j}\otimes\id\otimes\cdots\otimes\id$ with $P_0=\id$, and let $H^{(j)}$ be the $j$-th summand of $H_Q$.  In $\partial H^{(j)}+H^{(j)}\partial$, expand $\partial$ into its $D$ terms: the term with $\partial_i$, $i<j$, gives $\pi_{I_1}\otimes\cdots\otimes(\partial_i\pi_{I_i}+\pi_{I_i}\partial_i)\otimes\cdots$, which vanishes because $\pi_{I_i}$ is a chain map and $2=0$; the term with $i>j$ gives $\cdots\otimes H_{I_j}\otimes\cdots\otimes(\partial_i+\partial_i)\otimes\cdots=0$; and the term $i=j$ gives $\pi_{I_1}\otimes\cdots\otimes\pi_{I_{j-1}}\otimes(\partial_jH_{I_j}+H_{I_j}\partial_j)\otimes\id\otimes\cdots\otimes\id=P_{j-1}+P_j$.  Summing over $j$ telescopes to $P_0+P_D=\id+\Pi_Q$.  For Eq.~\eqref{SM:eq:filling}: $\Pi_Q$ vanishes on every cell with at least one edge factor, that is, in every positive degree, so for a closed $z$ of degree $k\ge1$, $\partial H_Qz=z+H_Q\partial z+\Pi_Qz=z$, and $H_Q$ maps cells of $Q$ to chains of cells of $Q$.
\end{proof}

Equation~\eqref{SM:eq:filling} is the whole content of the first property: a small closed loop bounds a membrane inside its own box, and flipping that membrane erases it [Fig.~\ref{SM:fig:S1}(a)].  For an isolated point, a $0$-chain, $\Pi_Q$ gives the corner $\mathbf v_0$ of $Q$ rather than zero, so $\partial H_Qz=z+\Pi_Qz\ne z$ and no local filling exists [Fig.~\ref{SM:fig:S1}(b)].  This is where the two- and three-dimensional toric codes fail the first property.

A \emph{non-winding box} is a product of coordinate intervals each shorter than $L/4$; it lifts uniquely to a box in $\Z^4$ once an anchor is chosen, and the lemma applies to the lift.  Every connected set of diameter below $L/8$ fits into a non-winding box of comparable diameter.  The \emph{dual box} $Q^*$ of a box $Q$ is the box of dual cells one layer larger than $Q$, which contains the dual of every cell of $Q$; applying Lemma~\ref{SM:lem:contraction} on $Q^*$ fills closed dual chains inside $Q^*$.

\subsection{Local generation}
\label{SM:sec:generation}

\begin{lemma}[local generation]
\label{SM:lem:generation}
Let $P$ be a Pauli operator supported in a non-winding box $Q$ that commutes with every check.  Then $P=\zeta\prod_{a\in Y}S_a$ with $|\zeta|=1$, where $Y$ consists of checks whose supports lie within the two-layer collar $Q^{+2}$ of $Q$.
\end{lemma}

\begin{proof}
Write $P=\zeta X(x)Z(z)$ with $x,z$ sets of plaquettes in $Q$ and $|\zeta|=1$.  By Lemma~\ref{SM:lem:lattice}(ii), commutation with all edge checks means $\partial z=0$ and commutation with all cube checks means $\partial(*x)=0$.  By Eq.~\eqref{SM:eq:filling} in $Q$, $z=\partial(H_Qz)$ with $H_Qz$ a $3$-chain in $Q$, and $Z(z)=\prod_{c\in H_Qz}Z(\partial c)=\prod_{c\in H_Qz}B_c$, since $Z$ is multiplicative in the chain.  For the $X$ part, $*x$ is a closed dual $2$-chain in $Q^*$; by Eq.~\eqref{SM:eq:filling} in $Q^*$, $*x=\partial(H_{Q^*}(*x))$ with $H_{Q^*}(*x)$ a dual $3$-chain, whose cells are the duals of a set $y$ of edges within one layer of $Q$; mapping back, $x=\delta y$, the set of plaquettes containing an odd number of edges of $y$, so $X(x)=\prod_{e\in y}X(\delta e)=\prod_{e\in y}A_e$.  The cube checks used lie in $Q$ and the edge checks used have supports within $Q^{+2}$.
\end{proof}

This is condition (iii) of Appendix~A, Eq.~(A1), with $C_0=2$.  The proof uses the location of the generating checks only, which is what the history expansion of Sec.~\ref{SM:sec:histories} needs; every constant is set by that location and is independent of the volume of $Q$.

\subsection{The decoder and its one-sided identity}
\label{SM:sec:decoder}

Set $\ell_0=\lfloor L/1000\rfloor$ and take $L$ large enough that $\ell_0\ge10$.  The decoder assigns to every valid pattern $s$ a Pauli operator $R_s$ with $\syn(R_s)=s$, as follows.  If every cluster of $s$, of both types, has diameter below $8\ell_0$, choose for each cluster the lexicographically first vertex as anchor, lift the cluster to the short coordinate interval around the anchor, take its bounding box, and fill it by Eq.~\eqref{SM:eq:filling}, on the box for an edge-type cluster and on the dual box for a cube-type cluster; $R_s$ is the product of the fillings, $X$-type first.  Otherwise $R_s$ is the output of a fixed Gaussian elimination over $\F_2$ on the full system.  The map $s\mapsto R_s$ depends only on the ideal checks, and every filling creates its own cluster and no excitation of the other type (Lemma~\ref{SM:lem:lattice}(ii)).

Let $\cH_{\log}$ be a space of dimension $d=64$ with basis $|\mu\rangle$, let $|\bar\mu\rangle$ be an orthonormal basis of the code space (the sector $\Pi_\emptyset$), and let $\cH_S$ have the orthonormal basis $|s\rangle$ indexed by the valid patterns.

\begin{lemma}[decoder unitary]
\label{SM:lem:W}
The map $W:\cH_{\log}\otimes\cH_S\to\cH_{\rm phys}$, $W(|\mu\rangle\otimes|s\rangle)=R_s|\bar\mu\rangle$, is unitary.  For every valid $s$, $W$ maps $\cH_{\log}\otimes|s\rangle$ onto the range of $\Pi_s$; $H_0$ acts on it as $E_{\rm ref}+E(s)$; and with $G_S$ the projector on the span of the \emph{good} patterns, those all of whose clusters have diameter below $\ell_0$, the \emph{good} projector is $G=W(I\otimes G_S)W^\dagger$, with $B=I-G$.
\end{lemma}

\begin{proof}
$R_s$ is unitary and $R_s\Pi_\emptyset R_s^\dagger=\Pi_s$ by Lemma~\ref{SM:lem:sectors}(iii), a statement about the sectors of the stabilizer group whose proof does not use this section, so for each valid $s$ the vectors $R_s|\bar\mu\rangle$, $\mu=1,\dots,d$, form an orthonormal basis of the range of $\Pi_s$; the sectors of distinct valid patterns are orthogonal, and their ranges sum to the whole space since $\sum_s\Pi_s=I$ and only valid patterns contribute.  The energy statement is Lemma~\ref{SM:lem:lattice}(iii).
\end{proof}

Logical operators are $\bar M=W(M\otimes I)W^\dagger$ for operators $M$ on $\cH_{\log}$; they form the logical algebra $\mathfrak A_{\log}$ of Sec.~\ref{SM:sec:memory}.  An \emph{allowed region} is any set, connected or not, whose union with a fixed collar fits inside a non-winding box of side at most $\ell_0$; in four dimensions every set inside a box of side $L/4000$ qualifies for large $L$.

\begin{lemma}[one-sided decoder identity]
\label{SM:lem:decoder}
For $L$ large enough, every operator $O$ supported in an allowed region satisfies
\begin{equation}
[O,\bar M]\,G=0\qquad\text{for every logical operator }\bar M .
\label{SM:eq:onesided}
\end{equation}
Moreover $[H_0,\bar M]=[G,\bar M]=0$.
\end{lemma}

\begin{figure}[t]
\centering
\includegraphics{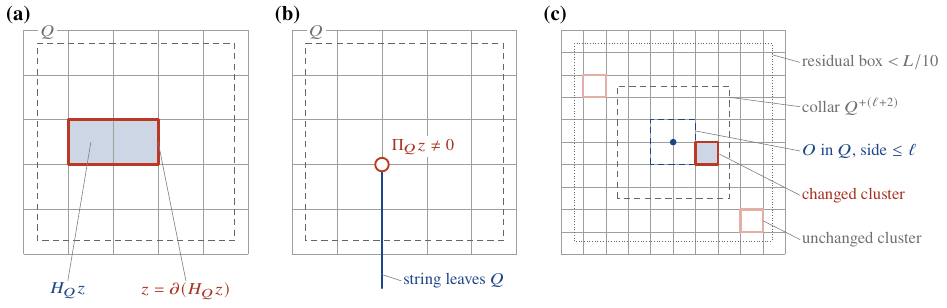}
\caption{Local filling and the decoder identity.  (a)~A closed loop $z$ inside a non-winding box $Q$ bounds the membrane $H_Qz$ inside $Q$, Eq.~\eqref{SM:eq:filling}; flipping the membrane erases the loop.  (b)~An isolated point-like excitation is a $0$-chain, which $\Pi_Q$ maps to the corner of $Q$ rather than to zero: no filling exists inside $Q$, and the string ending on the point leaves the box.  This is the two-dimensional toric code.  (c)~The one-sided identity of Lemma~\ref{SM:lem:decoder}: a local operator $O$ in a box $Q$ of side at most $\ell_0$ changes only the clusters touching the collar $Q^{+2}$, all of which lie in $Q^{+(\ell_0+2)}$; unchanged clusters carry identical fillings, which cancel, and the residual operator has zero pattern inside a box of side below $L/10$, so it is a check product by Lemma~\ref{SM:lem:generation}.}
\label{SM:fig:S1}
\end{figure}

\begin{proof}
By linearity in $O$ it suffices to take $O$ a Pauli operator supported in a box $Q$ of side at most $\ell_0$, and by linearity in the input it suffices to show $O\bar M\,W(|\mu\rangle\otimes|s\rangle)=\bar MO\,W(|\mu\rangle\otimes|s\rangle)$ for every good pattern $s$ and every $\mu$.

\emph{Step 1: the operator $R_{s'}^\dagger OR_s$ is a check product times a phase, for $s'=s\oplus\syn(O)$.}  The checks flipped by $O$ touch $Q$, so $s'$ differs from $s$ only within the collar $Q^{+2}$.  Every cluster of $s$ that changes touches $Q^{+2}$ and has diameter below $\ell_0$, so every changed, merged, or split cluster of $s'$ lies in $Q^{+(\ell_0+2)}$ [Fig.~\ref{SM:fig:S1}(c)] and has diameter at most $\ell_0+2(\ell_0+2)<4\ell_0$; all other clusters of $s'$ are clusters of $s$.  Hence every cluster of $s'$ has diameter below $8\ell_0$ and $R_{s'}$ is computed by the local-filling branch, even if $s'$ is not good.  A cluster common to $s$ and $s'$ receives the same anchor, box, and filling in $R_s$ and $R_{s'}$.  Pauli factors commute up to sign, so $R_{s'}^\dagger OR_s=\pm \tilde R_{s'}^\dagger O\tilde R_s$, where $\tilde R_s$ and $\tilde R_{s'}$ are the products of the fillings of the changed clusters of $s$ and $s'$ and the sign does not depend on $\mu$.  The operator $\tilde R_{s'}^\dagger O\tilde R_s$ has syndrome $s'\oplus\syn(O)\oplus s=\emptyset$ and is supported in the union of $Q$ and the bounding boxes of the changed clusters, which lies in a box of side at most $\ell_0+2(\ell_0+2)+2\cdot4\ell_0<12\ell_0+20<L/10$, a non-winding box.  By Lemma~\ref{SM:lem:generation} it is a check product times a phase.

\emph{Step 2.}  A check product acts on the code space, the sector $\Pi_\emptyset$, as the identity.  Hence $R_{s'}^\dagger OR_s|\bar\mu\rangle=\zeta_{s,O}|\bar\mu\rangle$ with a phase $\zeta_{s,O}$ independent of $\mu$, that is,
\begin{equation}
O\,W(|\mu\rangle\otimes|s\rangle)=\zeta_{s,O}\,W(|\mu\rangle\otimes|s'\rangle)\qquad\text{for every good }s\text{ and every }\mu .
\end{equation}
Therefore, with $M|\mu\rangle=\sum_\nu M_{\nu\mu}|\nu\rangle$,
\begin{equation}
\bar MO\,W(|\mu\rangle\otimes|s\rangle)=\zeta_{s,O}W(M|\mu\rangle\otimes|s'\rangle)=\sum_\nu M_{\nu\mu}\zeta_{s,O}W(|\nu\rangle\otimes|s'\rangle)=\sum_\nu M_{\nu\mu}OW(|\nu\rangle\otimes|s\rangle)=O\bar M\,W(|\mu\rangle\otimes|s\rangle),
\end{equation}
which is Eq.~\eqref{SM:eq:onesided} on the range of $G$.  Finally $W^\dagger H_0W=I\otimes\sum_s(E_{\rm ref}+E(s))|s\rangle\langle s|$ and $W^\dagger GW=I\otimes G_S$ by Lemma~\ref{SM:lem:W}, and both commute with $M\otimes I$.
\end{proof}

The identity holds on the output side without any projection, which is stronger than commutation after compression to the good subspace, and this is what Secs.~\ref{SM:sec:memory} and \ref{SM:sec:operational} use.  The decoder has a polynomial-time implementation: measure the checks, identify clusters, fill them by Eq.~\eqref{SM:eq:filling} or run the global elimination on a matrix of size $O(n)$, apply $R_s^\dagger$, and read the logical qubits.  None of these steps depends on $V$ or $\beta$.

\subsection{Grouping of loops}
\label{SM:sec:grouping}

A cluster too large for a non-winding box need not bound on its own.

\begin{lemma}[charges and grouping]
\label{SM:lem:grouping}
For a $1$-chain $z$ and $i=1,\dots,4$ let
\begin{equation}
\lambda_i(z)=\sum_{(x,\{i\}):\,x_i=L-1}z_{(x,\{i\})}\pmod2 ,
\label{SM:eq:charges}
\end{equation}
the parity of the number of edges of $z$ in direction $i$ that cross the hyperplane $x_i=L-1$.  Then (i) $\lambda_i$ vanishes on boundaries, and a closed $1$-chain bounds if and only if $\lambda_1(z)=\cdots=\lambda_4(z)=0$; (ii) in every valid pattern, every cluster $C_0$ of one type belongs to a union $\Gamma$ of at most $h+1=5$ clusters of that type that bounds, hence is erased by a Pauli operator $F$ with $\syn(F)=\Gamma$ creating no excitation of the other type.  The same holds on the dual lattice.
\end{lemma}

\begin{proof}
(i) The boundary of a plaquette spanned by directions $i,j$ at corner $x$ contains two edges in direction $i$, at $x_j$ and $x_j+1$, which cross the hyperplane $x_i=L-1$ together or not at all; so $\lambda_i$ vanishes on $\partial$ of every plaquette, hence on every boundary.  The coordinate circle in direction $j$, the closed $1$-chain $\mathsf c_j$ formed by the $L$ edges in direction $j$ through a fixed base point, has $\lambda_i(\mathsf c_j)=[i=j]$, so the $\lambda_i$ are linearly independent functionals on $H_1(T^4;\F_2)=\F_2^4$, whose basis the four circles form; a class on which all four vanish is zero.  (ii) The clusters of a valid pattern $s$ of one type are closed $1$-chains, because distinct clusters share no vertex and $\partial s=0$.  Let $c_j\in\F_2^4$ be the charge vectors $(\lambda_i(C_j))_i$ of the clusters $C_0,C_1,\dots,C_r$ of $s$.  By (i), $\sum_{j=0}^rc_j=0$, so $c_0$ lies in the span of $c_1,\dots,c_r$, a subspace of $\F_2^4$ with a basis of at most four of these vectors; hence $c_0=\sum_{j\in J}c_j$ for a set $J\subseteq\{1,\dots,r\}$ with $|J|\le4$.  The union $\Gamma=C_0\cup\bigcup_{j\in J}C_j$ is a closed $1$-chain with all charges zero, so it bounds, $\Gamma=\partial y$, and $F=Z(y)$ has $\syn(F)=\Gamma$ and commutes with every cube check.  On the dual lattice the charges are the dual parities and $F$ is $X$-type.
\end{proof}

This grouping is condition (ii) of Appendix~A and the source of the factor $L^{D(h+1)}$ in the static bound; on the four coordinate circles the $\lambda_i$ form the Kronecker pairing.

\section{Fixed-window reduction}
\label{SM:sec:window}

This section proves that the probability of finding every check of a large loop $\Gamma$ violated in $\rho$ is controlled by the free-energy cost of erasing $\Gamma$, the quantity bounded in Eq.~(5).  The perturbation is expanded in a time window of fixed length $\tau$ and only for the few terms that touch $\Gamma$; all other terms stay inside the exponential.  Combined with the Peierls counting of Sec.~\ref{SM:sec:geometry}, the result is Eq.~(2) once Eq.~(5) is available.  Two standard facts are used.  H\"older's inequality for Schatten norms: for operators $X_1,\dots,X_m$ on a finite-dimensional space and exponents $p_j\in[1,\infty]$ with $\sum_j1/p_j=1$,
\begin{equation}
|\Tr(X_1\cdots X_m)|\le\prod_{j=1}^m\|X_j\|_{p_j},\qquad\|X\|_p=\bigl(\Tr|X|^p\bigr)^{1/p},\quad\|X\|_\infty=\|X\| ,
\label{SM:eq:hoelder}
\end{equation}
Ref.~\cite{Simon2005}, Theorem~2.8.  Monotonicity of the partition function: if $A\le B$ are Hermitian, then $\Tr e^{-\beta B}\le\Tr e^{-\beta A}$, because by the min--max principle the $j$-th eigenvalue of $B$ is at least the $j$-th eigenvalue of $A$ for every $j$.

\subsection{The frozen and erased Hamiltonians}
\label{SM:sec:frozen}

Fix a valid pattern $\Gamma$ of one type, a closed $1$-chain that bounds (Sec.~\ref{SM:sec:geometry}), and a Pauli operator $F$ with $\syn(F)=\Gamma$ that creates no excitation of the other type: the operator $Z(y)$ of a $2$-chain $y$ with $\partial y=\Gamma$ for the edge type, and the operator $X(x)$ of a dual $2$-chain bounded by $\Gamma$ for the cube type.  Let
\begin{equation}
I_\Gamma=\{i:[v_i,S_a]\ne0\text{ for some }a\in\Gamma\},\qquad
V_\Gamma=\sum_{i\in I_\Gamma}v_i,\qquad
\Lambda_\Gamma=\sum_{i\in I_\Gamma}\|v_i\|,\qquad
V_Q=V-V_\Gamma ,
\label{SM:eq:IGamma}
\end{equation}
and define the frozen Hamiltonian, the erased Hamiltonian, and the flipped-check energy [Fig.~\ref{SM:fig:S2}(b)]
\begin{equation}
K_\Gamma=H_0+V_Q,\qquad K_\Gamma^F=H_0+FV_QF^\dagger,\qquad E_\Gamma=2\sum_{a\in\Gamma}J_aS_a .
\label{SM:eq:KKF}
\end{equation}

\begin{lemma}[frozen and erased Hamiltonians]
\label{SM:lem:frozen}
\begin{enumerate}
\item[(i)] $K_\Gamma$ and $K_\Gamma^F$ commute with every check of $\Gamma$; $FK_\Gamma F^\dagger=K_\Gamma^F+E_\Gamma$; and $E_\Gamma$ commutes with $K_\Gamma^F$.
\item[(ii)] $\Lambda_\Gamma\le w\epsilon|\Gamma|$ with $w=6$, and $-\Lambda_\Gamma I\le H-K_\Gamma\le\Lambda_\Gamma I$; hence $Z(K_\Gamma)\le e^{\beta\Lambda_\Gamma}Z_H$ and $Z_H\le e^{\beta\Lambda_\Gamma}Z(K_\Gamma)$.
\item[(iii)] The ratio $Z(K_\Gamma^F)/Z(K_\Gamma)$ does not depend on the choice of $F$.
\end{enumerate}
\end{lemma}

\begin{proof}
(i) $H_0$ and every $v_i$ with $i\notin I_\Gamma$ commute with every check of $\Gamma$, hence so does $K_\Gamma$.  $F$ anticommutes with the checks of $\Gamma$ and commutes with all other checks, so $FS_aF^\dagger=-S_a$ for $a\in\Gamma$, $FH_0F^\dagger=H_0+E_\Gamma$, and $FK_\Gamma F^\dagger=H_0+E_\Gamma+FV_QF^\dagger=K_\Gamma^F+E_\Gamma$.  Since $V_Q$ commutes with every $S_a$, $a\in\Gamma$, the operator $FV_QF^\dagger$ commutes with $FS_aF^\dagger=-S_a$; so $K_\Gamma^F$ commutes with every check of $\Gamma$ and therefore with $E_\Gamma$.

(ii) A term fails to commute with $S_a$ only if its support meets $\supp S_a$, a set of $w=6$ qubits, so
\begin{equation}
\Lambda_\Gamma\le\sum_{a\in\Gamma}\ \sum_{p\in\supp S_a}\ \sum_{i:\,p\in\supp v_i}\|v_i\|\le\sum_{a\in\Gamma}w\|V\|_\loc\le w\epsilon|\Gamma| .
\end{equation}
$H-K_\Gamma=V_\Gamma$ is Hermitian with $\|V_\Gamma\|\le\Lambda_\Gamma$, so $H-\Lambda_\Gamma I\le K_\Gamma\le H+\Lambda_\Gamma I$, and monotonicity of the partition function gives $\Tr e^{-\beta K_\Gamma}\le\Tr e^{-\beta(H-\Lambda_\Gamma)}=e^{\beta\Lambda_\Gamma}Z_H$ and $Z_H\le\Tr e^{-\beta(K_\Gamma-\Lambda_\Gamma)}=e^{\beta\Lambda_\Gamma}Z(K_\Gamma)$.

(iii) Two erasing operators $F_1,F_2$ for the same $\Gamma$ differ by $U=F_2F_1^\dagger$ with $\syn(U)=\emptyset$, which commutes with $H_0$; hence $K_\Gamma^{F_2}=H_0+UF_1V_QF_1^\dagger U^\dagger=UK_\Gamma^{F_1}U^\dagger$, and the two partition functions agree.
\end{proof}

Everything below is organized around the free-energy ratio $Z(K_\Gamma^F)/Z(K_\Gamma)$: Secs.~\ref{SM:sec:histories}--\ref{SM:sec:bridge} bound the free-energy difference $\beta^{-1}|\ln[Z(K_\Gamma^F)/Z(K_\Gamma)]|$ by a constant times the length $|\Gamma|$, Eq.~\eqref{SM:eq:tension}.  The operator norm of $FV_QF^\dagger-V_Q$ grows with the area of the membrane: for a flat membrane of side $r$ with $F$ its product of $X_p$ and $V_Q=-\epsilon\sum_pZ_p$ on the membrane, that norm is $2\epsilon r^2$ while $|\Gamma|=O(r)$.

\subsection{Expansion in the window and the H\"older bound}
\label{SM:sec:hoelder}

Let $\Pi^-_\Gamma=\prod_{a\in\Gamma}n_a$ project on the \emph{marked event} that every check of $\Gamma$ is violated, and fix $0<\tau\le\beta$.  For $n\ge0$ let $\Delta_n(\tau)=\{(t_0,\dots,t_n):t_j\ge0,\ \sum_jt_j=\tau\}$, with $n$-dimensional volume $\tau^n/n!$.

\begin{lemma}[expansion in the window]
\label{SM:lem:duhamel}
\begin{equation}
e^{-\beta H}=\sum_{n\ge0}(-1)^n\sum_{i_1,\dots,i_n\in I_\Gamma}\int_{\Delta_n(\tau)}e^{-t_0K_\Gamma}v_{i_1}e^{-t_1K_\Gamma}v_{i_2}\cdots v_{i_n}e^{-t_nK_\Gamma}\,e^{-(\beta-\tau)H}\,dt ,
\label{SM:eq:duhamel}
\end{equation}
where the term $n=0$ is $e^{-\tau K_\Gamma}e^{-(\beta-\tau)H}$ and the series converges absolutely in norm.
\end{lemma}

\begin{proof}
Write $e^{-\beta H}=e^{-\tau H}e^{-(\beta-\tau)H}$ and $H=K_\Gamma+V_\Gamma$.  The operator $U(t)=e^{tK_\Gamma}e^{-tH}$ satisfies $U(0)=I$ and $U'(t)=-V_\Gamma(t)U(t)$ with $V_\Gamma(t)=e^{tK_\Gamma}V_\Gamma e^{-tK_\Gamma}$, so Picard iteration gives $e^{-\tau H}=e^{-\tau K_\Gamma}\sum_n(-1)^n\int_{0\le u_1\le\cdots\le u_n\le\tau}V_\Gamma(u_n)\cdots V_\Gamma(u_1)\,du$, a norm-convergent series since $\|V_\Gamma(u)\|\le e^{2u\|K_\Gamma\|}\Lambda_\Gamma$ on $[0,\tau]$.  Inserting $V_\Gamma(u)=e^{uK_\Gamma}V_\Gamma e^{-uK_\Gamma}$ turns the $n$-th term into $\int e^{-(\tau-u_n)K_\Gamma}V_\Gamma e^{-(u_n-u_{n-1})K_\Gamma}V_\Gamma\cdots V_\Gamma e^{-u_1K_\Gamma}\,du$; the substitution $t_0=\tau-u_n$, $t_j=u_{n+1-j}-u_{n-j}$ for $1\le j<n$, $t_n=u_1$ maps the ordered simplex bijectively and with unit Jacobian onto $\Delta_n(\tau)$, and expanding every $V_\Gamma$ into its terms gives Eq.~\eqref{SM:eq:duhamel}.
\end{proof}

\begin{figure}[t]
\centering
\includegraphics{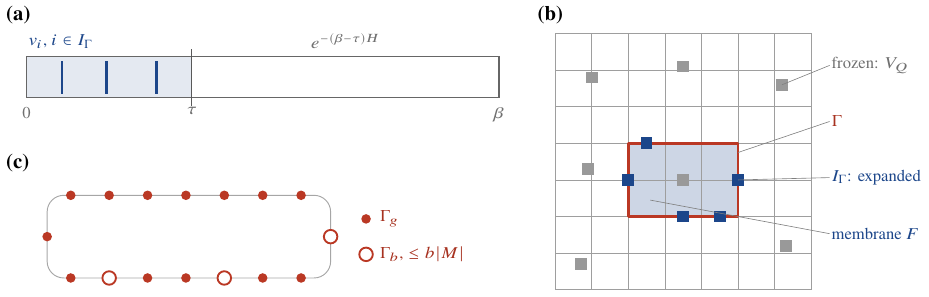}
\caption{Fixed-window reduction.  (a)~The thermal circle of length $\beta$ with the window $[0,\tau]$: only the terms $v_i$ with $i\in I_\Gamma$, which fail to commute with a check of $\Gamma$, are expanded in the window, with the frozen Hamiltonian $K_\Gamma$ between insertions; the factor $e^{-(\beta-\tau)H}$ stays exact.  Every exponent of the resulting bound carries $\tau$, Eq.~\eqref{SM:eq:marked2}.  (b)~The loop $\Gamma$ and its membrane.  Terms touching $\Gamma$ (blue) are expanded; all others (gray) form $V_Q$ and are frozen into $K_\Gamma$; conjugation by the membrane operator $F$ turns $K_\Gamma$ into the erased Hamiltonian $K_\Gamma^F$, Eq.~\eqref{SM:eq:KKF}.  (c)~A word of the expansion disturbs at most $b|\mathsf M|$ checks of $\Gamma$ (open circles).  Low-activity words, $|\Gamma_b|\le\vartheta|\Gamma|$, pay the energy $J_{\min}|\Gamma|$ after erasure, Eq.~\eqref{SM:eq:energycompare}; high-activity words have at least $k_0$ distinct labels and are rare by Eq.~\eqref{SM:eq:highactivity}.}
\label{SM:fig:S2}
\end{figure}

A \emph{word} is a term of Eq.~\eqref{SM:eq:duhamel}: a label sequence $(i_1,\dots,i_n)$ together with a point of $\Delta_n(\tau)$.  Let $\mathsf M\subseteq I_\Gamma$ be its set of distinct labels, let $\Gamma_b(\mathsf M)$ be the set of checks of $\Gamma$ that fail to commute with some $v_i$, $i\in \mathsf M$, and let $\Gamma_g(\mathsf M)=\Gamma\setminus\Gamma_b(\mathsf M)$ [Fig.~\ref{SM:fig:S2}(c)].  A term fails to commute with a check only if the check touches its support, and a qubit lies in four checks of each type, so
\begin{equation}
|\Gamma_b(\mathsf M)|\le b\,|\mathsf M|,\qquad b=4k_R .
\label{SM:eq:Gammab}
\end{equation}
Let $\Pi^-_{\Gamma_g}=\prod_{a\in\Gamma_g(\mathsf M)}n_a$ project on the event that every check of $\Gamma_g(\mathsf M)$ is violated.

\begin{lemma}[H\"older bound for a word]
\label{SM:lem:hoelder}
For every word,
\begin{equation}
\frac{\bigl|\Tr\bigl(\Pi^-_\Gamma\,e^{-t_0K_\Gamma}v_{i_1}e^{-t_1K_\Gamma}\cdots v_{i_n}e^{-t_nK_\Gamma}e^{-(\beta-\tau)H}\bigr)\bigr|}{Z_H}
\le\Bigl(\frac{\Tr \Pi^-_{\Gamma_g}e^{-\beta K_\Gamma}}{Z_H}\Bigr)^{\tau/\beta}\ \prod_{j=1}^n\|v_{i_j}\| .
\label{SM:eq:holder}
\end{equation}
\end{lemma}

\begin{proof}
$\Pi^-_{\Gamma_g}$ is a function of the checks of $\Gamma_g(\mathsf M)$, which commute with $K_\Gamma$ (Lemma~\ref{SM:lem:frozen}(i)) and with every $v_{i_j}$ of the word (by the definition of $\Gamma_g$); hence $\Pi^-_{\Gamma_g}$ commutes with every factor $e^{-t_jK_\Gamma}$ and $v_{i_j}$.  Since $\Pi^-_\Gamma\le \Pi^-_{\Gamma_g}$, $\Pi^-_\Gamma=\Pi^-_\Gamma \Pi^-_{\Gamma_g}=\Pi^-_\Gamma (\Pi^-_{\Gamma_g})^{n+1}$, and moving one factor $\Pi^-_{\Gamma_g}$ to the right of each $e^{-t_jK_\Gamma}$ writes the trace as
\begin{equation}
\Tr\bigl(\Pi^-_\Gamma\,A_0\,v_{i_1}\,A_1\cdots v_{i_n}\,A_n\,D\bigr),\qquad A_j=\Pi^-_{\Gamma_g}e^{-t_jK_\Gamma},\quad D=e^{-(\beta-\tau)H} .
\end{equation}
Apply Eq.~\eqref{SM:eq:hoelder} with exponent $\infty$ for $\Pi^-_\Gamma$ and for every $v_{i_j}$, exponent $\beta/t_j$ for $A_j$, and exponent $\beta/(\beta-\tau)$ for $D$; the exponents satisfy $\sum_jt_j/\beta+(\beta-\tau)/\beta=1$.  Each $A_j$ is a positive operator, because $\Pi^-_{\Gamma_g}$ and $K_\Gamma$ commute, so $\|A_j\|_{\beta/t_j}=\bigl(\Tr A_j^{\beta/t_j}\bigr)^{t_j/\beta}=\bigl(\Tr \Pi^-_{\Gamma_g}e^{-\beta K_\Gamma}\bigr)^{t_j/\beta}$, with the value $1$ when $t_j=0$; likewise $\|D\|_{\beta/(\beta-\tau)}=Z_H^{(\beta-\tau)/\beta}$; and $\|\Pi^-_\Gamma\|_\infty\le1$, $\|v_{i_j}\|_\infty=\|v_{i_j}\|$.  Multiplying the factors and dividing by $Z_H=Z_H^{\tau/\beta}Z_H^{(\beta-\tau)/\beta}$ gives Eq.~\eqref{SM:eq:holder}.
\end{proof}

Only the norms of the insertions enter Eq.~\eqref{SM:eq:holder}.

\subsection{Energy comparison for low-activity words}
\label{SM:sec:lowactivity}

Set
\begin{equation}
\vartheta=\frac{J_{\min}}{2(J_{\min}+J_{\max})}\in(0,\tfrac14] .
\label{SM:eq:chi}
\end{equation}
A word is \emph{low-activity} if $|\Gamma_b(\mathsf M)|\le\vartheta|\Gamma|$ and \emph{high-activity} otherwise.

\begin{lemma}[energy comparison]
\label{SM:lem:energy}
For every label set $\mathsf M$ with $|\Gamma_b(\mathsf M)|\le\vartheta|\Gamma|$,
\begin{equation}
\Tr \Pi^-_{\Gamma_g}e^{-\beta K_\Gamma}\le e^{-\beta J_{\min}|\Gamma|}\,Z(K_\Gamma^F),
\label{SM:eq:energycompare}
\end{equation}
and consequently
\begin{equation}
\frac{\Tr \Pi^-_{\Gamma_g}e^{-\beta K_\Gamma}}{Z_H}\le e^{-\beta J_{\min}|\Gamma|}\,\frac{Z(K_\Gamma^F)}{Z(K_\Gamma)}\,e^{\beta w\epsilon|\Gamma|} .
\label{SM:eq:lowactivity}
\end{equation}
\end{lemma}

\begin{proof}
Conjugate by $F$.  Since $F$ flips every check of $\Gamma\supseteq\Gamma_g$, $F\Pi^-_{\Gamma_g}F^\dagger=\prod_{a\in\Gamma_g}(I-n_a)=:\Pi^+_{\Gamma_g}$, the projector on $S_a=+1$ for all $a\in\Gamma_g$, and $FK_\Gamma F^\dagger=K_\Gamma^F+E_\Gamma$ by Lemma~\ref{SM:lem:frozen}(i).  The three operators $\Pi^+_{\Gamma_g}$, $E_\Gamma$, and $K_\Gamma^F$ commute pairwise: the first two are functions of the checks of $\Gamma$, with which $K_\Gamma^F$ commutes.  Hence
\begin{equation}
\Tr \Pi^-_{\Gamma_g}e^{-\beta K_\Gamma}=\Tr\bigl(\Pi^+_{\Gamma_g}e^{-\beta K_\Gamma^F}e^{-\beta E_\Gamma}\bigr)=\Tr\bigl(e^{-\beta K_\Gamma^F/2}\,\Pi^+_{\Gamma_g}e^{-\beta E_\Gamma}\Pi^+_{\Gamma_g}\,e^{-\beta K_\Gamma^F/2}\bigr) .
\end{equation}
On the range of $\Pi^+_{\Gamma_g}$ every $S_a$ with $a\in\Gamma_g$ equals $+1$ and every $S_a$ with $a\in\Gamma_b$ is at least $-1$, so there
\begin{equation}
E_\Gamma\ \ge\ 2\Bigl[\sum_{a\in\Gamma_g}J_a-\sum_{a\in\Gamma_b}J_a\Bigr]\ \ge\ 2\bigl[J_{\min}(|\Gamma|-|\Gamma_b|)-J_{\max}|\Gamma_b|\bigr]\ \ge\ 2\bigl[J_{\min}-(J_{\min}+J_{\max})\vartheta\bigr]|\Gamma|=J_{\min}|\Gamma| ,
\end{equation}
using $|\Gamma_b|\le\vartheta|\Gamma|$ and the value of $\vartheta$.  Thus $\Pi^+_{\Gamma_g}e^{-\beta E_\Gamma}\Pi^+_{\Gamma_g}\le e^{-\beta J_{\min}|\Gamma|}\Pi^+_{\Gamma_g}$, and since $X\mapsto\Tr(e^{-\beta K_\Gamma^F/2}Xe^{-\beta K_\Gamma^F/2})$ preserves operator inequalities,
\begin{equation}
\Tr \Pi^-_{\Gamma_g}e^{-\beta K_\Gamma}\le e^{-\beta J_{\min}|\Gamma|}\Tr\bigl(\Pi^+_{\Gamma_g}e^{-\beta K_\Gamma^F}\bigr)\le e^{-\beta J_{\min}|\Gamma|}Z(K_\Gamma^F) ,
\end{equation}
which is Eq.~\eqref{SM:eq:energycompare}.  Dividing by $Z_H$ and using $Z(K_\Gamma)\le e^{\beta\Lambda_\Gamma}Z_H\le e^{\beta w\epsilon|\Gamma|}Z_H$ from Lemma~\ref{SM:lem:frozen}(ii) gives Eq.~\eqref{SM:eq:lowactivity}.
\end{proof}

The sum over all words, low- and high-activity alike, of $\prod_j\|v_{i_j}\|$ times the simplex volume is
\begin{equation}
\sum_{n\ge0}\frac{\tau^n}{n!}\sum_{i_1,\dots,i_n\in I_\Gamma}\prod_{j=1}^n\|v_{i_j}\|=\sum_{n\ge0}\frac{(\tau\Lambda_\Gamma)^n}{n!}=e^{\tau\Lambda_\Gamma} .
\label{SM:eq:allwords}
\end{equation}

\subsection{Combinatorial bound for high-activity words}
\label{SM:sec:highactivity}

\begin{lemma}[high-activity words]
\label{SM:lem:high}
Let $k_0=\lfloor\vartheta|\Gamma|/b\rfloor+1$ and $C_{\rm act}=2ewb/\vartheta$.  Every high-activity word has at least $k_0$ distinct labels, and if $\tau\epsilon\le1$ and $C_{\rm act}\tau\epsilon\le1$, then
\begin{equation}
\sum_{\text{high-activity words}}\frac{\tau^n}{n!}\prod_{j=1}^n\|v_{i_j}\|\ \le\ \sum_{\substack{\mathsf M\subseteq I_\Gamma\\|\mathsf M|\ge k_0}}\ \prod_{i\in \mathsf M}z_i\ \le\ e^{\tau\Lambda_\Gamma}\Bigl(\frac{2e\tau\Lambda_\Gamma}{k_0}\Bigr)^{k_0}\ \le\ e^{\tau\Lambda_\Gamma}\bigl(C_{\rm act}\tau\epsilon\bigr)^{\vartheta|\Gamma|/b},\qquad z_i=e^{\tau\|v_i\|}-1 .
\label{SM:eq:highactivity}
\end{equation}
\end{lemma}

\begin{proof}
By Eq.~\eqref{SM:eq:Gammab}, $\vartheta|\Gamma|<|\Gamma_b(\mathsf M)|\le b|\mathsf M|$ gives $|\mathsf M|>\vartheta|\Gamma|/b$, hence $|\mathsf M|\ge k_0$.  Group the words by their label set $\mathsf M$.  For a fixed $\mathsf M$, the sum over all $n$ and all sequences in $\mathsf M^n$ of $(\tau^n/n!)\prod_j\|v_{i_j}\|$ is $\prod_{i\in \mathsf M}e^{\tau\|v_i\|}$, and inclusion--exclusion over the label set, as in Lemma~\ref{SM:lem:coefficients}, shows that the sequences whose label set is exactly $\mathsf M$ contribute $\prod_{i\in \mathsf M}(e^{\tau\|v_i\|}-1)=\prod_{i\in \mathsf M}z_i$; summing over $|\mathsf M|\ge k_0$ gives the first inequality.  For the second, every $\mathsf M$ with $|\mathsf M|\ge k_0$ is counted at least once in $\sum_{|\mathsf M_0|=k_0}\prod_{i\in \mathsf M_0}z_i\sum_{\mathsf M'\subseteq I_\Gamma\setminus \mathsf M_0}\prod_{i\in \mathsf M'}z_i$, and the inner sum is at most $\prod_{i\in I_\Gamma}(1+z_i)=e^{\tau\Lambda_\Gamma}$.  Since $\tau\|v_i\|\le\tau\epsilon\le1$ and $e^x-1\le2x$ for $0\le x\le1$, $\sum_iz_i\le2\tau\Lambda_\Gamma$; the elementary symmetric polynomial satisfies $\sum_{|\mathsf M_0|=k_0}\prod_{i\in \mathsf M_0}z_i\le(\sum_iz_i)^{k_0}/k_0!$, because the expansion of $(\sum_iz_i)^{k_0}$ contains every product over a $k_0$-subset $k_0!$ times; and $k_0!\ge(k_0/e)^{k_0}$.  For the third, $\Lambda_\Gamma\le w\epsilon|\Gamma|$ and $k_0>\vartheta|\Gamma|/b$ give $2e\tau\Lambda_\Gamma/k_0\le(2ewb/\vartheta)\tau\epsilon=C_{\rm act}\tau\epsilon\le1$, and a base at most $1$ raised to the power $k_0\ge\vartheta|\Gamma|/b$ is at most the same base raised to $\vartheta|\Gamma|/b$.
\end{proof}

For a high-activity word the factor in Eq.~\eqref{SM:eq:holder} is bounded without the energy comparison: $\Pi^-_{\Gamma_g}\le I$ and Lemma~\ref{SM:lem:frozen}(ii) give
\begin{equation}
\Bigl(\frac{\Tr \Pi^-_{\Gamma_g}e^{-\beta K_\Gamma}}{Z_H}\Bigr)^{\tau/\beta}\le\Bigl(\frac{Z(K_\Gamma)}{Z_H}\Bigr)^{\tau/\beta}\le e^{\tau\Lambda_\Gamma} .
\label{SM:eq:highfactor}
\end{equation}

\subsection{The marked-event bound}
\label{SM:sec:marked}

\begin{proposition}[marked event]
\label{SM:prop:marked}
Let $0<\tau\le\beta$ with $\tau\epsilon\le1$ and $C_{\rm act}\tau\epsilon\le1$.  Then
\begin{equation}
\Tr(\rho \Pi^-_\Gamma)\le e^{\tau\Lambda_\Gamma}\Bigl[e^{-\tau J_{\min}|\Gamma|}\,\Theta_\Gamma^{\tau/\beta}+e^{\tau\Lambda_\Gamma}(C_{\rm act}\tau\epsilon)^{\vartheta|\Gamma|/b}\Bigr],\qquad
\Theta_\Gamma=\frac{Z(K_\Gamma^F)}{Z(K_\Gamma)}\,e^{\beta w\epsilon|\Gamma|} .
\label{SM:eq:marked}
\end{equation}
If moreover the tension bound holds in the form
\begin{equation}
\frac1\beta\Bigl|\ln\frac{Z(K_\Gamma^F)}{Z(K_\Gamma)}\Bigr|\le a|\Gamma|,\qquad a=C_{\rm tw}\epsilon+2\delta ,
\label{SM:eq:tension}
\end{equation}
which is Eq.~(5) with the constants of Sec.~\ref{SM:sec:bridge}, then
\begin{equation}
\Tr(\rho \Pi^-_\Gamma)\le e^{-\tau[J_{\min}-(C_{\rm tw}+2w)\epsilon-2\delta]\,|\Gamma|}
+e^{2w\tau\epsilon|\Gamma|}\,(C_{\rm act}\tau\epsilon)^{\vartheta|\Gamma|/b} .
\label{SM:eq:marked2}
\end{equation}
\end{proposition}

\begin{proof}
$\Tr(\rho \Pi^-_\Gamma)=\Tr(\Pi^-_\Gamma e^{-\beta H})/Z_H$.  Insert Eq.~\eqref{SM:eq:duhamel}, bound the absolute value of the series by the sum of the absolute values of its words, and bound each word by Lemma~\ref{SM:lem:hoelder}.  For a low-activity word, Lemma~\ref{SM:lem:energy} bounds the first factor of Eq.~\eqref{SM:eq:holder} by $\bigl(e^{-\beta J_{\min}|\Gamma|}\Theta_\Gamma\bigr)^{\tau/\beta}=e^{-\tau J_{\min}|\Gamma|}\Theta_\Gamma^{\tau/\beta}$, uniformly in the word, and the sum of the remaining factors $(\tau^n/n!)\prod_j\|v_{i_j}\|$ over low-activity words is at most $e^{\tau\Lambda_\Gamma}$ by Eq.~\eqref{SM:eq:allwords}.  For a high-activity word, Eq.~\eqref{SM:eq:highfactor} bounds the first factor by $e^{\tau\Lambda_\Gamma}$, and Lemma~\ref{SM:lem:high} bounds the sum of the remaining factors by $e^{\tau\Lambda_\Gamma}(C_{\rm act}\tau\epsilon)^{\vartheta|\Gamma|/b}$.  Adding the two contributions gives Eq.~\eqref{SM:eq:marked}.  Under Eq.~\eqref{SM:eq:tension}, $\Theta_\Gamma^{\tau/\beta}\le e^{\tau(a+w\epsilon)|\Gamma|}$, and with $\Lambda_\Gamma\le w\epsilon|\Gamma|$ the first term of Eq.~\eqref{SM:eq:marked} is at most $e^{-\tau[J_{\min}-a-2w\epsilon]|\Gamma|}$ and the second at most $e^{2w\tau\epsilon|\Gamma|}(C_{\rm act}\tau\epsilon)^{\vartheta|\Gamma|/b}$.
\end{proof}

Every exponent in Eq.~\eqref{SM:eq:marked2} carries the window length $\tau$, which is fixed, in place of $\beta$.  This is the reason the final threshold is uniform in temperature.

\subsection{The static bound in four dimensions}
\label{SM:sec:static4d}

The constants are chosen in the order of Sec.~\ref{SM:sec:constants}: $\delta=J_{\min}/64$, then
\begin{equation}
\tau=\beta_0=\max\Bigl\{t_0,\ \frac{2(\ln392+4)}{J_{\min}}\Bigr\},\qquad
\epsilon_0=\min\Bigl\{\bar\epsilon,\ \frac{J_{\min}}{8(C_{\rm tw}+2w)},\ \frac1{4w\tau},\ \frac1{C_{\rm act}\tau}\Bigl(\frac{e^{-4}}{392}\Bigr)^{b/\vartheta}\Bigr\} ,
\label{SM:eq:eps0}
\end{equation}
where $t_0$, $\bar\epsilon$, and $C_{\rm tw}$ come from Secs.~\ref{SM:sec:histories}--\ref{SM:sec:bridge}; Sec.~\ref{SM:sec:bridge} proves Eq.~\eqref{SM:eq:tension} for all $\beta\ge\beta_0$, $\epsilon\le\epsilon_0$, and all valid $\Gamma$ of one type.

\begin{proposition}[static bound in four dimensions]
\label{SM:prop:static4d}
For $\epsilon\le\epsilon_0$ and $\beta\ge\beta_0$,
\begin{equation}
\Pr_\rho[\text{a cluster of one type with}\ge u\text{ checks}]
\le\frac{4q^u}{1-q}\sum_{r=1}^{5}M_{\rm ch}^r\Bigl(\frac q{1-q}\Bigr)^{r-1}
\le CL^{20}e^{-4u},\qquad q=e^{-4},\quad M_{\rm ch}=4L^4 .
\label{SM:eq:static4d}
\end{equation}
\end{proposition}

\begin{proof}
\emph{The two bases.}  With $\tau=\beta_0$ and $\epsilon\le\epsilon_0$, the hypotheses of Proposition~\ref{SM:prop:marked} hold: $\tau\epsilon\le1/(4w)\le1$ and $C_{\rm act}\tau\epsilon\le(e^{-4}/392)^{b/\vartheta}\le1$.  The bracket in the first exponent of Eq.~\eqref{SM:eq:marked2} is at least $J_{\min}-J_{\min}/8-J_{\min}/32\ge J_{\min}/2$, so the first term is at most $(e^{-\tau J_{\min}/2})^{|\Gamma|}\le(e^{-4}/392)^{|\Gamma|}$ by the choice of $\tau$.  In the second term, $e^{2w\tau\epsilon}\le e^{1/2}$ and $(C_{\rm act}\tau\epsilon)^{\vartheta/b}\le e^{-4}/392$, so it is at most $(e^{1/2}e^{-4}/392)^{|\Gamma|}$.  Both bases, multiplied by the counting constant $196$ of Sec.~\ref{SM:sec:geometry}, are at most $e^{-4}=q$:
\begin{equation}
\Tr(\rho \Pi^-_\Gamma)\le2\Bigl(\frac q{196}\Bigr)^{|\Gamma|}\qquad\text{for every valid }\Gamma\text{ of one type}.
\label{SM:eq:PGammafinal}
\end{equation}

\emph{Union bound.}  Fix a pattern measured in $\rho$ and suppose it contains a cluster $C_0$ of one type with at least $u$ checks.  By Sec.~\ref{SM:sec:grouping}, $C_0$ belongs to a union $\Gamma=C_0\cup C_1\cup\cdots\cup C_{r-1}$ of $r\le5$ clusters of the pattern that bounds, and every check of $\Gamma$ is violated in this pattern.  Hence the event in Eq.~\eqref{SM:eq:static4d} is contained in the union, over all sets $\Gamma$ that are valid patterns of one type and consist of $r\le5$ connected sets of checks, one of which has at least $u$ checks, of the marked events $\Pi^-_\Gamma$, and the probability is at most $\sum_\Gamma\Tr(\rho \Pi^-_\Gamma)$.  Enumerate $\Gamma$ by the type, the number $r$ of its connected sets, and for each connected set its root check, its size $m_j$, and its shape: there are $M_{\rm ch}=4L^4$ checks of each type and at most $196^{m_j}$ connected sets of size $m_j$ containing a given check (Sec.~\ref{SM:sec:geometry}); the enumeration overcounts, which only enlarges the bound.  With Eq.~\eqref{SM:eq:PGammafinal} and $|\Gamma|=\sum_jm_j$,
\begin{equation}
\sum_\Gamma\Tr(\rho \Pi^-_\Gamma)\le2\cdot2\sum_{r=1}^{5}M_{\rm ch}^r\Bigl(\sum_{m_1\ge u}q^{m_1}\Bigr)\Bigl(\sum_{m\ge1}q^{m}\Bigr)^{r-1}
=\frac{4q^u}{1-q}\sum_{r=1}^{5}M_{\rm ch}^r\Bigl(\frac q{1-q}\Bigr)^{r-1},
\end{equation}
where the factor $2$ outside counts the two excitation types.  The last sum is at most $5M_{\rm ch}^5$ times a constant, which gives $CL^{20}e^{-4u}$.
\end{proof}

A loop of length $u$ is a cluster of $u$ checks, so Eq.~\eqref{SM:eq:static4d} is Eq.~(2) with $c=20$ and $\ell$ in place of $u$.  The counting is a union bound over the joint measurement statistics of both excitation types in the Gibbs state $\rho$ itself.  In general dimension the same steps give $\Pr\le CL^{D(h+1)}e^{-4u}$ with at most $h+1$ clusters per union, Sec.~\ref{SM:sec:general}.

It remains to prove the tension bound Eq.~\eqref{SM:eq:tension}.  That is the task of the next three sections.

\section{Exact history representation in fixed imaginary-time slices}
\label{SM:sec:histories}

This section writes $Z(K_\Gamma)$ as an exact finite sum over histories: periodic sequences of excitation patterns with Pauli insertions between them, Eq.~(B1).  The representation keeps every sign, every logical trace, and the true physical normalization.  Its one organizing choice is to cut the thermal circle into slices of a fixed duration $t_0$, so that the weight of a small history is set by $\epsilon$ and $t_0$ alone.  This is the mathematical form of the statement in the main text that a bubble costs $(\epsilon/\Delta)^2$ independently of temperature.  Throughout, $\Gamma$, $F$, and $K_\Gamma=H_0+V_Q$ are as in Sec.~\ref{SM:sec:window}: $V_Q=\sum_{i\in\calI}v_i$ is the frozen perturbation, $\calI=\{i:i\notin I_\Gamma\}$ is its label set, $\|V_Q\|_\loc\le\|V\|_\loc\le\epsilon$, and every $v_i$ with $i\in\calI$ commutes with every check of $\Gamma$.  Every statement holds for every finite $L$ and every finite number of slices.

\subsection{Sectors, patterns, and characters}
\label{SM:sec:sectors}

Write $H_0=E_{\rm ref}+H_+$ with
\begin{equation}
E_{\rm ref}=-\sum_aJ_a,\qquad H_+=2\sum_aJ_an_a,\qquad n_a=\tfrac12(I-S_a),
\label{SM:eq:Hplus}
\end{equation}
where $a$ runs over the checks of both types, $J_a$ is the coupling of check $a$, and $n_a$ is the projector on $S_a=-1$.  For a set $s$ of checks let
\begin{equation}
\Pi_s=\prod_{a\in s}n_a\prod_{a\notin s}(I-n_a) .
\label{SM:eq:sectorproj}
\end{equation}
The checks commute, so the $\Pi_s$ are mutually orthogonal projectors with $\sum_s\Pi_s=I$, and
\begin{equation}
H_+\Pi_s=E(s)\Pi_s,\qquad E(s)=2\sum_{a\in s}J_a .
\label{SM:eq:Esector}
\end{equation}
Let $\mathfrak S$ be the group generated by the checks.  It is abelian, every element is a Pauli operator $X(\delta y)Z(\partial u)$ with sign $+1$, and $-I\notin\mathfrak S$.  Pauli operators modulo phases form the symplectic $\F_2$-space $\F_2^{2n}$, with $\langle P,P'\rangle=1$ when $P$ and $P'$ anticommute; $\mathfrak S$ is an isotropic subspace of dimension $r=n-6$, and $\mathfrak S^\perp$, the space of Pauli operators commuting with every check, has dimension $2n-r=n+6$.

\begin{lemma}[sectors]
\label{SM:lem:sectors}
(i) $\Pi_s\ne0$ if and only if $s=\syn(P)$ for some Pauli operator $P$.  These are the \emph{valid} patterns; they form an $\F_2$-linear space, $\syn(PP')=\syn(P)\oplus\syn(P')$.  (ii) Every nonzero $\Pi_s$ has rank $d=2^6=64$.  (iii) $P\Pi_sP^\dagger=\Pi_{s\oplus\syn(P)}$ for every Pauli operator $P$ and every set $s$.
\end{lemma}

\begin{proof}
(iii) $PS_aP^\dagger=S_a$ when $P$ commutes with $S_a$ and $-S_a$ otherwise, and $\syn(P)$ is the set of checks that anticommute with $P$.  Hence $Pn_aP^\dagger=n_a$ for $a\notin\syn(P)$ and $Pn_aP^\dagger=I-n_a$ for $a\in\syn(P)$; conjugating Eq.~\eqref{SM:eq:sectorproj} exchanges the two kinds of factors exactly for $a\in\syn(P)$, which gives $\Pi_{s\oplus\syn(P)}$.  $PP'$ anticommutes with $S_a$ if and only if exactly one of $P,P'$ does, which is the linearity of $\syn$.

(i) and (ii).  For a set $s$ write $\lambda_s(S_a)=(-1)^{[a\in s]}$.  Suppose first that $\lambda_s$ extends to a character of $\mathfrak S$, that is, $\prod_{a\in Y}\lambda_s(S_a)=1$ for every set $Y$ of checks with $\prod_{a\in Y}S_a=I$.  Choose checks $S_{a_1},\dots,S_{a_r}$ that form a basis of $\mathfrak S$.  Expanding the product $\prod_{j=1}^r\tfrac12(I+\lambda_s(S_{a_j})S_{a_j})$ gives $2^{-r}\sum_{S\in\mathfrak S}\lambda_s(S)S$, because every $S\in\mathfrak S$ is a unique product of basis elements and $\lambda_s$ is multiplicative on such products; this is the projector on the subspace where every $S\in\mathfrak S$ has eigenvalue $\lambda_s(S)$; on that subspace every remaining factor $\tfrac12(I+\lambda_s(S_a)S_a)$ of Eq.~\eqref{SM:eq:sectorproj} is the identity, and off it the basis product vanishes.  Hence $\Pi_s=2^{-r}\sum_{S\in\mathfrak S}\lambda_s(S)S$, and since only $S=I$ has nonzero trace, ${\rm rank}\,\Pi_s=2^{n-r}=64$.  Suppose next that $\lambda_s$ does not extend: some relation $\prod_{a\in Y}S_a=I$ has $\prod_{a\in Y}\lambda_s(S_a)=-1$.  On the range of $\Pi_s$ the operator $\prod_{a\in Y}S_a$ equals $\prod_{a\in Y}\lambda_s(S_a)=-1$, which contradicts $\prod_{a\in Y}S_a=I$ unless $\Pi_s=0$.  Finally, the map $P\mapsto\langle\,\cdot\,,P\rangle$ from $\F_2^{2n}$ to the dual space of $\mathfrak S$ is linear with kernel $\mathfrak S^\perp$ of dimension $2n-r$, so its image has dimension $r$ and is the whole dual space: every character of $\mathfrak S$ has the form $S\mapsto(-1)^{\langle S,P\rangle}$ for some Pauli operator $P$.  Since $(-1)^{\langle S_a,P\rangle}=(-1)^{[a\in\syn(P)]}$, the map $\lambda_s$ extends to a character if and only if $s=\syn(P)$ for some $P$.
\end{proof}

\begin{lemma}[characters]
\label{SM:lem:characters}
For a check product $S=\prod_{a\in Y}S_a$ and a valid pattern $s$ define $\chi_S(s)\in\{\pm1\}$ by $S\Pi_s=\chi_S(s)\Pi_s$.  Then $\chi_S(s)=(-1)^{|Y\cap s|}$ for every set $Y$ of checks with $\prod_{a\in Y}S_a=S$, and
\begin{enumerate}
\item[(a)] $\chi_S(s\oplus s')=\chi_S(s)\chi_S(s')$ and $\chi_{SS'}(s)=\chi_S(s)\chi_{S'}(s)$ for valid $s,s'$ and check products $S,S'$;
\item[(b)] $RSR^\dagger=\chi_S(\syn(R))\,S$ for every Pauli operator $R$;
\item[(c)] $S$ commutes with a Pauli operator $P$ if and only if $\chi_S(\syn(P))=1$.
\end{enumerate}
\end{lemma}

\begin{proof}
On the range of $\Pi_s$ every $S_a$ equals $(-1)^{[a\in s]}$, so $S$ equals $\prod_{a\in Y}(-1)^{[a\in s]}=(-1)^{|Y\cap s|}$ there, for any representation $Y$; since $\Pi_s\ne0$, the value is determined by the operators $S$ and $\Pi_s$.  (a) is the parity formula.  (b): $RS_aR^\dagger=(-1)^{[a\in\syn(R)]}S_a$ for every check, and $\syn(R)$ is valid.  (c) is (b) with $R=P$.
\end{proof}

\begin{lemma}[sector trace]
\label{SM:lem:sectortrace}
Let $Q$ be a Pauli operator with $\syn(Q)=\emptyset$.  Then $Q=\zeta S\bar L$ with $\zeta\in\{\pm1,\pm i\}$, $S$ a check product, and $\bar L$ one of the $4^6$ products of fixed logical Pauli representatives.  For every valid $s$,
\begin{equation}
\frac1d\Tr(\Pi_sQ)=\begin{cases}\zeta\,\chi_S(s)&\text{if }\bar L=I,\\[2pt]0&\text{if }\bar L\ne I.\end{cases}
\label{SM:eq:sectortrace}
\end{equation}
\end{lemma}

\begin{proof}
$\syn(Q)=\emptyset$ means that $Q$ commutes with every check, so $Q$ modulo phase lies in $\mathfrak S^\perp$.  Fix Pauli operators $\bar X_1,\dots,\bar X_6,\bar Z_1,\dots,\bar Z_6$ in $\mathfrak S^\perp$ whose classes form a symplectic basis of $\mathfrak S^\perp/\mathfrak S$, for instance the sheet operators of the main text; every element of $\mathfrak S^\perp$ is then $\zeta S\bar L$ with $\bar L=\prod_j\bar X_j^{x_j}\bar Z_j^{z_j}$.  Since $S\Pi_s=\chi_S(s)\Pi_s$ and $\bar L$ commutes with every $\Pi_s$, $\Tr(\Pi_sQ)=\zeta\chi_S(s)\Tr(\Pi_s\bar L)$, and for $\bar L=I$ the trace equals $d$ by Lemma~\ref{SM:lem:sectors}(ii).  For $\bar L\ne I$ the symplectic form on $\mathfrak S^\perp/\mathfrak S$ is nondegenerate, so some representative $\bar L'$ anticommutes with $\bar L$; it commutes with every check, hence $\bar L'\Pi_s\bar L'^\dagger=\Pi_s$, and
$\Tr(\Pi_s\bar L)=\Tr(\bar L'\Pi_s\bar L\bar L'^\dagger)=\Tr(\Pi_s\bar L'\bar L\bar L'^\dagger)=-\Tr(\Pi_s\bar L)$.
\end{proof}

\subsection{Locality in imaginary time}
\label{SM:sec:locality}

For a label $i\in\calI$ let $\mathcal X(i)$ be the set of checks that fail to commute with $v_i$, let
\begin{equation}
{\rm col}(i)=\supp v_i\ \cup\bigcup_{a\in \mathcal X(i)}\supp S_a
\label{SM:eq:collar}
\end{equation}
be the \emph{collar} of $i$, and let $\mathcal K(i)$ be the set of checks whose support meets ${\rm col}(i)$.  A check that fails to commute with $v_i$ meets $\supp v_i$, so $\mathcal X(i)\subseteq\mathcal K(i)$.  Three constants depend on the range $R$ and the lattice only: $c_R=\max_i|{\rm col}(i)|$, $b_R=\max_i\sum_{a\in \mathcal X(i)}J_a\le8k_RJ_{\max}$, and the maximal degree of the \emph{dependency graph} on $\calI$, in which two labels $i\ne j$ are joined when $\mathcal K(i)\cap\mathcal K(j)\ne\emptyset$.  For $A\subseteq\calI$ write $V_A=\sum_{i\in A}v_i$, ${\rm col}(A)=\bigcup_{i\in A}{\rm col}(i)$, $\mathcal K(A)=\bigcup_{i\in A}\mathcal K(i)$, and for real $u$
\begin{equation}
v_i(u)=e^{uH_+}v_ie^{-uH_+},\qquad V_A(u)=\sum_{i\in A}v_i(u) .
\label{SM:eq:intpicture}
\end{equation}

\begin{lemma}[strict locality]
\label{SM:lem:locality}
Let $H_{\mathcal X(i)}=2\sum_{a\in \mathcal X(i)}J_an_a$.  For every real $u$,
\begin{equation}
v_i(u)=e^{uH_{\mathcal X(i)}}\,v_i\,e^{-uH_{\mathcal X(i)}},\qquad\|v_i(u)\|\le e^{2|u|b_R}\|v_i\| ,
\label{SM:eq:vlocal}
\end{equation}
and $v_i(u)$ is a linear combination of the operators $S_Yv_iS_{Y'}$ with $Y,Y'\subseteq \mathcal X(i)$, where $S_Y=\prod_{a\in Y}S_a$; in particular $v_i(u)$ is supported in ${\rm col}(i)$.  If $i$ and $j$ are not joined in the dependency graph, then $[v_i(u),v_j(u')]=0$ for all real $u,u'$.
\end{lemma}

\begin{proof}
Write $H_+=H_{\mathcal X(i)}+H'$ with $H'=2\sum_{a\notin \mathcal X(i)}J_an_a$.  All $n_a$ commute with each other, and $H'$ commutes with $v_i$, so $e^{uH_+}=e^{uH'}e^{uH_{\mathcal X(i)}}$ and $e^{uH'}$ commutes with $e^{uH_{\mathcal X(i)}}v_ie^{-uH_{\mathcal X(i)}}$; this is the first identity.  Since $e^{\pm uH_{\mathcal X(i)}}=\prod_{a\in \mathcal X(i)}\bigl(I+(e^{\pm2uJ_a}-1)n_a\bigr)$ and $n_a=\tfrac12(I-S_a)$, expanding the two products writes $v_i(u)$ as a combination of the operators $S_Yv_iS_{Y'}$, each supported in ${\rm col}(i)$.  For the norm, $0\le H_{\mathcal X(i)}\le2b_RI$, so one of the factors $e^{\pm uH_{\mathcal X(i)}}$ has norm at most $1$ and the other at most $e^{2|u|b_R}$.  If $i$ and $j$ are not joined, then $\supp v_i\cap\supp v_j=\emptyset$, because a common qubit lies in some check, which would belong to $\mathcal K(i)\cap\mathcal K(j)$; so $v_i$ and $v_j$ commute.  Every $S_a$ with $a\in \mathcal X(i)$ commutes with $v_j$, because a check that fails to commute with both $v_i$ and $v_j$ belongs to $\mathcal X(i)\cap \mathcal X(j)\subseteq\mathcal K(i)\cap\mathcal K(j)$; likewise every $S_b$ with $b\in \mathcal X(j)$ commutes with $v_i$; and checks commute with each other.  Hence every $S_Yv_iS_{Y'}$ commutes with every $S_Uv_jS_{U'}$, and so do the combinations $v_i(u)$ and $v_j(u')$.
\end{proof}

\subsection{Slices and the slice operator}
\label{SM:sec:slices}

Fix $t_0>0$ and, for $\beta\ge t_0$, set $N=\lfloor\beta/t_0\rfloor$ and $t_{\rm s}=\beta/N\in[t_0,2t_0)$ [Fig.~\ref{SM:fig:S3}(a)].  For $A\subseteq\calI$ define the symmetrized slice operator
\begin{equation}
\mathsf G(A)=e^{t_{\rm s}H_+/2}\,e^{-t_{\rm s}(H_++V_A)}\,e^{t_{\rm s}H_+/2},
\label{SM:eq:Gdef}
\end{equation}
a Hermitian operator with $\mathsf G(\emptyset)=I$.

\begin{lemma}[Dyson form]
\label{SM:lem:dyson}
For every $A\subseteq\calI$,
\begin{equation}
\mathsf G(A)=\sum_{n\ge0}(-1)^n\int_{-t_{\rm s}/2\le u_1\le\cdots\le u_n\le t_{\rm s}/2}V_A(u_n)\cdots V_A(u_1)\,du_1\cdots du_n ,
\label{SM:eq:dyson}
\end{equation}
where the term $n=0$ is $I$ and the series converges absolutely in norm.
\end{lemma}

\begin{proof}
Let $U(t)=e^{tH_+}e^{-t(H_++V_A)}$.  Then $U(0)=I$ and $U'(t)=-V_A(t)U(t)$, and Picard iteration, which converges in norm because $\|V_A(t)\|\le e^{2|t|b_R}\sum_{i\in A}\|v_i\|$ is bounded on bounded intervals, gives $U(t_{\rm s})=\sum_n(-1)^n\int_{0\le u_1\le\cdots\le u_n\le t_{\rm s}}V_A(u_n)\cdots V_A(u_1)\,du$.  Now $\mathsf G(A)=e^{-t_{\rm s}H_+/2}U(t_{\rm s})e^{t_{\rm s}H_+/2}$ and $e^{-t_{\rm s}H_+/2}V_A(u)e^{t_{\rm s}H_+/2}=V_A(u-t_{\rm s}/2)$; the substitution $u_j\mapsto u_j+t_{\rm s}/2$ gives Eq.~\eqref{SM:eq:dyson}.
\end{proof}

\begin{lemma}[factorization]
\label{SM:lem:factor}
If $A=A_1\sqcup A_2$ and no label of $A_1$ is joined to a label of $A_2$ in the dependency graph, then $\mathsf G(A)=\mathsf G(A_1)\mathsf G(A_2)=\mathsf G(A_2)\mathsf G(A_1)$.
\end{lemma}

\begin{proof}
By Lemma~\ref{SM:lem:locality} every $V_{A_1}(u)$ commutes with every $V_{A_2}(u')$.  Let $U_B(t)=e^{tH_+}e^{-t(H_++V_B)}$ for $B\in\{A,A_1,A_2\}$.  The product $U_{A_1}(t)U_{A_2}(t)$ equals $I$ at $t=0$, and $U_{A_1}(t)$, a norm limit of polynomials in the operators $V_{A_1}(u)$, commutes with $V_{A_2}(t)$, so
\begin{equation}
\frac{d}{dt}\bigl[U_{A_1}(t)U_{A_2}(t)\bigr]=-V_{A_1}(t)U_{A_1}U_{A_2}-U_{A_1}V_{A_2}(t)U_{A_2}=-V_A(t)\,U_{A_1}(t)U_{A_2}(t) .
\end{equation}
By uniqueness of the solution of $U'=-V_A(t)U$ with $U(0)=I$, $U_{A_1}(t)U_{A_2}(t)=U_A(t)$; the two factors commute for the same reason.  Conjugating with $e^{-t_{\rm s}H_+/2}$ as in Lemma~\ref{SM:lem:dyson} turns $U_B(t_{\rm s})$ into $\mathsf G(B)$ and gives the claim.
\end{proof}

A \emph{compatible family} is a set of connected label sets of the dependency graph that are pairwise disjoint and pairwise not joined by any edge.  For a connected label set $g$ let
\begin{equation}
w_g=\sum_{A\subseteq g}(-1)^{|g\setminus A|}\mathsf G(A) .
\label{SM:eq:wdef}
\end{equation}

\begin{lemma}[M\"obius inversion]
\label{SM:lem:mobius}
(i) The slice operator is the exact finite sum
\begin{equation}
\mathsf G(\calI)=\sum_{\mathfrak F\ \text{compatible}}\ \prod_{g\in\mathfrak F}w_g ,
\label{SM:eq:slicegas}
\end{equation}
where the empty family contributes $I$ and the factors of one family commute.  (ii) $w_g$ is the part of the series \eqref{SM:eq:dyson} for $A=g$ in which every label of $g$ occurs at least once:
\begin{equation}
w_g=\sum_{n\ge|g|}(-1)^n\int_{-t_{\rm s}/2\le u_1\le\cdots\le u_n\le t_{\rm s}/2}\ \sum_{\substack{(i_1,\dots,i_n)\in g^n\\ \{i_1,\dots,i_n\}=g}}v_{i_n}(u_n)\cdots v_{i_1}(u_1)\,du_1\cdots du_n .
\label{SM:eq:wwords}
\end{equation}
(iii) $w_g$ is Hermitian and supported in ${\rm col}(g)$.
\end{lemma}

\begin{proof}
For every $B\subseteq\calI$ set $\hat w_B=\sum_{A\subseteq B}(-1)^{|B\setminus A|}\mathsf G(A)$.  M\"obius inversion on the Boolean lattice gives $\mathsf G(B)=\sum_{A\subseteq B}\hat w_A$ for every $B$: inserting the definition, the coefficient of $\mathsf G(A')$ on the right is $\sum_{A'\subseteq A\subseteq B}(-1)^{|A\setminus A'|}$, which is $1$ for $A'=B$ and $0$ otherwise.  Expand every $\mathsf G(A)$ by Eq.~\eqref{SM:eq:dyson} into words $(i_1,\dots,i_n)\in A^n$; the integrand of a word does not depend on $A$, and a word with label set $M$ occurs in $\mathsf G(A)$ exactly when $M\subseteq A$, so its coefficient in $\hat w_B$ is $\sum_{M\subseteq A\subseteq B}(-1)^{|B\setminus A|}=[M=B]$.  Thus $\hat w_B$ is the sum of the words with label set exactly $B$, which is (ii) for $B=g$.  If $B=B_1\sqcup B_2$ with no edge between $B_1$ and $B_2$, Lemma~\ref{SM:lem:factor} gives $\mathsf G(A)=\mathsf G(A\cap B_1)\mathsf G(A\cap B_2)$ for every $A\subseteq B$, and the double sum factorizes: $\hat w_B=\hat w_{B_1}\hat w_{B_2}$.  By induction, $\hat w_B=\prod_g w_g$ over the connected components $g$ of $B$ in the dependency graph, with commuting factors by Lemma~\ref{SM:lem:factor}.  Hence $\mathsf G(\calI)=\sum_{B\subseteq\calI}\hat w_B=\sum_B\prod_{g\in{\rm comp}(B)}w_g$, and $B\mapsto{\rm comp}(B)$ is a bijection from the subsets of $\calI$ onto the compatible families: the components of a set are pairwise disjoint, connected, and not joined, and the union of a compatible family has exactly its members as components.  This is (i).  (iii): each $\mathsf G(A)$ is Hermitian, so $w_g$ is, and every word in Eq.~\eqref{SM:eq:wwords} is supported in ${\rm col}(g)$ by Lemma~\ref{SM:lem:locality}.
\end{proof}

\begin{lemma}[norm and Pauli coefficients]
\label{SM:lem:coefficients}
Let $C(t_0)=2t_0e^{2t_0b_R}$.  For every connected label set $g$,
\begin{equation}
\|w_g\|\le\prod_{i\in g}\bigl(e^{C(t_0)\|v_i\|}-1\bigr) .
\label{SM:eq:slabnorm}
\end{equation}
Write $w_g=\sum_Pc_{g,P}P$, the sum over the Pauli operators $P$ supported in ${\rm col}(g)$, with $c_{g,P}=2^{-|{\rm col}(g)|}\Tr_{{\rm col}(g)}(Pw_g)$.  The coefficients are real, there are at most $4^{c_R|g|}$ of them, and
\begin{equation}
|c_{g,P}|\le\|w_g\|\le q_\epsilon^{|g|},\qquad q_\epsilon=e^{C(t_0)\epsilon}-1 .
\label{SM:eq:eta}
\end{equation}
\end{lemma}

\begin{proof}
In Eq.~\eqref{SM:eq:wwords} every time argument satisfies $|u_j|\le t_{\rm s}/2\le t_0$, so by Eq.~\eqref{SM:eq:vlocal} the norm of the word $(i_1,\dots,i_n)$ is at most $\prod_jx_{i_j}$ with $x_i=e^{2t_0b_R}\|v_i\|$, and the volume of the simplex is $t_{\rm s}^n/n!$.  Summing over all words in $g^n$ and all $n$ gives $\exp(t_{\rm s}\sum_{i\in g}x_i)=\prod_{i\in g}e^{t_{\rm s}x_i}$; the words whose label set is exactly $g$ are obtained by inclusion--exclusion over the label set, $\sum_{B\subseteq g}(-1)^{|g\setminus B|}\prod_{i\in B}e^{t_{\rm s}x_i}=\prod_{i\in g}(e^{t_{\rm s}x_i}-1)$.  Since $t_{\rm s}\le2t_0$, $e^{t_{\rm s}x_i}-1\le e^{C(t_0)\|v_i\|}-1$.  The Pauli operators on $m$ qubits are an orthogonal basis of the $4^m$-dimensional operator space with $\Tr(P^\dagger P')=2^m[P=P']$, which gives the expansion and the formula for $c_{g,P}$; the coefficients are real because $w_g$ and $P$ are Hermitian, and $|c_{g,P}|\le2^{-m}\|P\|_1\|w_g\|=\|w_g\|$.  Finally $\|v_i\|\le\|V_Q\|_\loc\le\epsilon$, since $\|v_i\|$ is one summand of the local norm at any qubit of $\supp v_i$, and $x\mapsto e^{C(t_0)x}-1$ is increasing.
\end{proof}

The constant $C(t_0)$ depends on $t_0$, the range, and $J_{\max}$ only; the support ${\rm col}(g)$ and the Pauli alphabet are fixed by the range and the lattice, and time enters through the coefficients alone.

\subsection{The trace formula and the history expansion}
\label{SM:sec:traceformula}

\begin{lemma}[trace formula]
\label{SM:lem:tracepower}
\begin{equation}
e^{\beta E_{\rm ref}}Z(K_\Gamma)=\Tr\bigl(e^{-t_{\rm s}H_+}\mathsf G(\calI)\bigr)^N .
\label{SM:eq:tracepower}
\end{equation}
\end{lemma}

\begin{proof}
$e^{-t_{\rm s}H_+}\mathsf G(\calI)=e^{-t_{\rm s}H_+/2}e^{-t_{\rm s}(H_++V_Q)}e^{t_{\rm s}H_+/2}$, so its $N$-th power is $e^{-t_{\rm s}H_+/2}e^{-Nt_{\rm s}(H_++V_Q)}e^{t_{\rm s}H_+/2}$, whose trace is $\Tr e^{-\beta(H_++V_Q)}=e^{\beta E_{\rm ref}}\Tr e^{-\beta K_\Gamma}$, because $Nt_{\rm s}=\beta$ and $K_\Gamma=E_{\rm ref}+H_++V_Q$.
\end{proof}

Number the $N$ factors of Eq.~\eqref{SM:eq:tracepower} from right to left, so that slice $1$ acts first, and insert Eq.~\eqref{SM:eq:slicegas} and the Pauli expansion of Lemma~\ref{SM:lem:coefficients} in every slice.  A \emph{history} $\mathfrak h$ consists of
\begin{enumerate}
\item[(a)] for every slice $k\in\{1,\dots,N\}$ a compatible family $\mathfrak F_k$ and, for every $g\in\mathfrak F_k$, a Pauli operator $P_{k,g}$ supported in ${\rm col}(g)$; these are the \emph{insertions} of $\mathfrak h$, and $P_k=\prod_{g\in\mathfrak F_k}P_{k,g}$, whose factors have disjoint supports because a qubit common to two collars lies in a check common to the two sets $\mathcal K(g)$;
\item[(b)] a sequence of valid patterns $s_k$, $k\in\Z_N$, with $s_k=s_{k-1}\oplus\syn(P_k)$ for every $k$; in particular $s_0=s_N$ and $\syn(P_N\cdots P_1)=\emptyset$.
\end{enumerate}
Its \emph{weight} is
\begin{equation}
\Bigl(\prod_{k}\prod_{g\in\mathfrak F_k}c_{g,P_{k,g}}\Bigr)\,\exp\Bigl[-t_{\rm s}\sum_{k=1}^{N}E(s_k)\Bigr]\,\frac1d\Tr_{s_0}\bigl(P_N\cdots P_1\bigr),\qquad\Tr_{s_0}(X)=\Tr(\Pi_{s_0}X) ,
\label{SM:eq:historyweight}
\end{equation}
which is Eq.~(B1).  By Lemma~\ref{SM:lem:sectortrace}, the last factor is $\zeta\chi_S(s_0)$ when $P_N\cdots P_1=\zeta S$ is a check product times a phase, and $0$ when the net Pauli operator carries a nontrivial logical operator.

\begin{lemma}[history expansion]
\label{SM:lem:expansion}
$e^{\beta E_{\rm ref}}Z(K_\Gamma)/d$ is the sum of the weights \eqref{SM:eq:historyweight} over all histories, an exact finite sum.
\end{lemma}

\begin{proof}
Insert $I=\sum_s\Pi_s$ to the right of every factor $e^{-t_{\rm s}H_+}$ in Eq.~\eqref{SM:eq:tracepower}, so that $e^{-t_{\rm s}H_+}\Pi_{s_k}=e^{-t_{\rm s}E(s_k)}\Pi_{s_k}$ by Eq.~\eqref{SM:eq:Esector}, and expand every slice operator by Eq.~\eqref{SM:eq:slicegas} and Lemma~\ref{SM:lem:coefficients}.  The result is a finite sum over choices $(\mathfrak F_k,P_{k,g},s_k)_{k=1}^N$ of
\begin{equation}
\Bigl(\prod_{k,g}c_{g,P_{k,g}}\Bigr)\,e^{-t_{\rm s}\sum_kE(s_k)}\ \Tr\bigl[\Pi_{s_N}P_N\,\Pi_{s_{N-1}}P_{N-1}\cdots\Pi_{s_1}P_1\bigr] .
\end{equation}
By Lemma~\ref{SM:lem:sectors}(iii), $\Pi_{s_k}P_k\Pi_{s_{k-1}}=P_k\Pi_{s_{k-1}\oplus\syn(P_k)}\Pi_{s_{k-1}}$ vanishes unless $s_k=s_{k-1}\oplus\syn(P_k)$, with $s_0=s_N$ for $k=1$ by cyclicity of the trace, and in that case $\Pi_{s_k}P_k=P_k\Pi_{s_{k-1}}$.  Moving every projector to the right in turn, the trace becomes $\Tr[P_N\cdots P_1\Pi_{s_0}]=\Tr_{s_0}(P_N\cdots P_1)$, and the sum over the valid patterns $s_k$ is restricted to the sequences of (b).
\end{proof}

\subsection{The space-time graph and its components}
\label{SM:sec:components}

The \emph{environment graph} $\mathfrak G$ has nodes of two kinds, label nodes $(i,k)$ with $i\in\calI$ and excitation nodes $(a,k)$ with $a$ a check, both for $k\in\Z_N$, and edges
\begin{enumerate}
\item[(1)] $(i,k)\sim(j,k)$ when $i$ and $j$ are joined in the dependency graph;
\item[(2)] $(a,k)\sim(a',k)$ when $a$ and $a'$ are adjacent in the line graph of the lattice (two edges sharing a vertex, or two dual edges sharing a dual vertex);
\item[(3)] $(a,k)\sim(a,k+1)$;
\item[(4)] $(i,k)\sim(a,k-1)$ and $(i,k)\sim(a,k)$ for every $a\in\mathcal K(i)$.
\end{enumerate}
Rule (4) joins a label to every check it could flip, whether or not its Pauli operator flips it.  The degree of $\mathfrak G$ is bounded by a constant that depends on the range and the lattice only.  A history $\mathfrak h$ \emph{selects} the label nodes $(i,k)$ with $i\in g\in\mathfrak F_k$ and the excitation nodes $(a,k)$ with $a\in s_k$; the \emph{components} of $\mathfrak h$ are the connected components of the subgraph of $\mathfrak G$ induced on the selected nodes.  For a component $C$ let $a(C)$ be its number of label nodes, $d(C)$ its number of excitation nodes, and $m(C)=a(C)+d(C)\ge1$ its \emph{cost}; let $s_C(k)=\{a:(a,k)\in C\}$ be its \emph{section} at time $k$; let $P_C(k)$ be the product of the $P_{k,g}$ over the label sets $g\in\mathfrak F_k$ whose label nodes lie in $C$ (a label set lies entirely in one component, by rule (1) and its connectedness), with $P_C(k)=I$ when there is none; let $T(C)=\{k:C\text{ has a node in slice }k\}$ be its \emph{time projection}, where the node $(a,k)$ lies in slice $k$; and let
\begin{equation}
Q_C=P_C(N)\cdots P_C(1)
\label{SM:eq:QC}
\end{equation}
be its \emph{net Pauli operator}.  Two components \emph{touch} when an edge of $\mathfrak G$ joins a node of one to a node of the other; components of one history never touch.

\begin{lemma}[components]
\label{SM:lem:components}
Let $\mathfrak h$ be a history with components $C$.
\begin{enumerate}
\item[(i)] $s_k=\bigsqcup_Cs_C(k)$, $E(s_k)=\sum_CE(s_C(k))$, and $P_k=\prod_CP_C(k)$ with commuting factors, for every $k$.
\item[(ii)] $s_C(k)=s_C(k-1)\oplus\syn(P_C(k))$ for every $C$ and $k\in\Z_N$; consequently $\syn(Q_C)=\emptyset$.
\item[(iii)] Every cluster of $s_k$ lies in a single component, so each $s_C(k)$ is a union of clusters of $s_k$.
\item[(iv)] $T(C)$ is a cyclic interval of $\Z_N$.
\end{enumerate}
\end{lemma}

\begin{proof}
(i) Every selected node lies in exactly one component.  The factors $P_{k,g}$ of one slice have disjoint supports, so they commute and can be grouped by component.

(ii) The sets $\syn(P_{k,g})$, $g\in\mathfrak F_k$, are pairwise disjoint: $\syn(P_{k,g})\subseteq\mathcal K(g)$ because a check flipped by $P_{k,g}$ meets its support, which lies in ${\rm col}(g)$, and $\mathcal K(g)\cap\mathcal K(g')=\emptyset$ for compatible $g\ne g'$ by the definition of the dependency graph.  Hence $\syn(P_k)=\bigsqcup_g\syn(P_{k,g})$, every flipped check $a\in\syn(P_k)$ is flipped by exactly one label set $g_a\in\mathfrak F_k$, and $\syn(P_C(k))=\bigsqcup_{g\subseteq C}\syn(P_{k,g})$.  Fix a check $a$ and a time $k$.  If $a\in\syn(P_k)$, let $C'$ be the component containing $g_a$; some $i\in g_a$ has $a\in\mathcal K(i)$, and rule (4) joins $(i,k)$ to whichever of the nodes $(a,k-1)$, $(a,k)$ exists.  Exactly one of them exists, because $a$ toggles between $s_{k-1}$ and $s_k$, and it lies in $C'$.  So $a$ belongs to exactly one of $s_{C'}(k-1)$, $s_{C'}(k)$, in agreement with $a\in\syn(P_{C'}(k))$, and for every $C\ne C'$ it belongs to neither section and not to $\syn(P_C(k))$.  If $a\notin\syn(P_k)$, then $a\notin\syn(P_C(k))$ for every $C$; either $a$ is violated at both times $k-1$ and $k$, in which case rule (3) puts $(a,k-1)$ and $(a,k)$ into the same component, or at neither.  In all cases $[a\in s_C(k)]=[a\in s_C(k-1)]\oplus[a\in\syn(P_C(k))]$ for every $C$.  Summing the update rule over $k\in\Z_N$ and using $s_C(N)=s_C(0)$ gives $\bigoplus_k\syn(P_C(k))=\emptyset$, and $\syn(Q_C)=\bigoplus_k\syn(P_C(k))$ by linearity of $\syn$.

(iii) The excitation nodes of one cluster of $s_k$ are joined to each other by rule (2).

(iv) Every edge of $\mathfrak G$ joins two nodes whose slice indices differ by at most one, and $C$ is connected.
\end{proof}

A component $C$ is a \emph{bubble} if $T(C)$ together with its two neighbouring slices misses some slice of $\Z_N$, and a \emph{sheet} otherwise.

\begin{figure}[t]
\centering
\includegraphics{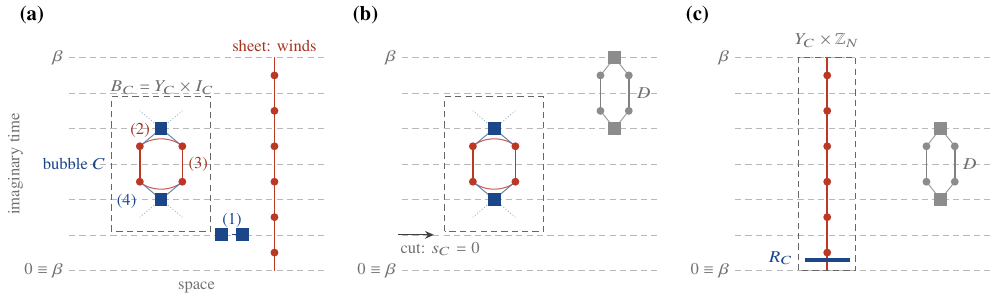}
\caption{Histories in fixed slices and the empty-box lemma, drawn on a one-dimensional cut of space.  (a)~The space-time graph: slices of duration $t_{\rm s}\in[t_0,2t_0)$ (dashed), excitation nodes (red) and active labels (blue squares), joined by the four rules of Sec.~\ref{SM:sec:components}: (1)~dependent labels in one slice, (2)~adjacent excitations in one slice, (3)~the same excitation in consecutive slices, (4)~a label and every check of its collar in the adjacent slices.  A bubble is a component whose time projection misses a slice; a sheet winds around the circle.  $\cB_C=Y_C\times I_C$ is the box of Lemma~\ref{SM:lem:emptybox}.  (b)~Non-winding case: cutting the circle after the first slice of $I_C$, where $s_C=\emptyset$, places every insertion of a component $D$ outside $\cB_C$ after those of $C$ in the cyclic order and separates the two in space, so $\kappa=0$ and both characters equal one.  (c)~Winding case: the repair $R_C$ of $s_C(0)$ lies inside $Y_C$ and commutes with the net Pauli $Q_D$ of any component outside the strip, which forces $\chi_{S_D}(s_C)=1$.  This is a bubble linked with a sheet.}
\label{SM:fig:S3}
\end{figure}

\subsection{Local and macroscopic components}
\label{SM:sec:localcomp}

Attach to every node a set of qubits, its \emph{spatial set}: $\supp S_a$ to $(a,k)$ and ${\rm col}(i)$ to $(i,k)$.  Distances between qubits and between sets of qubits are graph distances in the \emph{plaquette graph} $\mathfrak P$, whose vertices are the plaquettes, two of them adjacent when they share a vertex of the lattice.  The centers of adjacent plaquettes are at $\ell^\infty$ distance at most $1$, and two plaquettes whose centers are at $\ell^\infty$ distance $s$ are joined by a path of at most $12(s+1)$ steps (a unit translation of a plaquette in any lattice direction takes at most three steps); hence a set of $\mathfrak P$-diameter $D$ fits in a coordinate box of side $D+1$, and a coordinate box of side $s$ has $\mathfrak P$-diameter at most $12(s+1)$.  Let $\rho_0$ bound the diameter of every spatial set and $r_{\mathfrak G}$ the distance between the spatial sets of any two adjacent nodes of $\mathfrak G$; both depend on the range and the lattice only.  For a component $C$ let ${\rm sp}(C)$ be the union of the spatial sets of its nodes.  Following a path through the nodes of $C$,
\begin{equation}
\diam{\rm sp}(C)\le C_gm(C),\qquad C_g=\rho_0+r_{\mathfrak G} .
\label{SM:eq:diamPhi}
\end{equation}

Fix $c_0>0$ and set $m_0=\max\{1,\lfloor c_0L\rfloor\}$.  A component is \emph{local} if $m(C)<m_0$ and \emph{macroscopic} otherwise, whether or not it winds.  For a local component let $Q_C^{\rm box}$ be the smallest coordinate box containing ${\rm sp}(C)$, lifted to $\Z^4$ with the lexicographically first cell of ${\rm sp}(C)$ as anchor; its side is at most $C_gm(C)+1$.  Let
\begin{equation}
Y_C=\bigl(Q_C^{\rm box}\bigr)^{+2}
\label{SM:eq:YC}
\end{equation}
be that box enlarged by two layers of cells on every side.  Its side is at most $C_gm(C)+5$, and
\begin{equation}
\diam Y_C\le c_1m(C),\qquad c_1=12C_g+72 .
\label{SM:eq:diamY}
\end{equation}
We take $c_0\le1/(16C_g)$ and $L\ge80$; then every side of $Y_C$ is at most $L/16+5<L/8$, so $Y_C$ is a non-winding box.

\begin{lemma}[local components]
\label{SM:lem:local}
Let $C$ be a local component of a history.
\begin{enumerate}
\item[(i)] Every section $s_C(k)$ is a valid pattern, and there is a Pauli operator $R_C(k)$ supported in $Y_C$ with $\syn(R_C(k))=s_C(k)$.
\item[(ii)] $Q_C=\zeta_CS_C$ with $\zeta_C\in\{\pm1,\pm i\}$ and $S_C$ a product of checks whose supports lie in $Y_C$.
\item[(iii)] The \emph{intrinsic weight}
\begin{equation}
\omega(C)=\Bigl(\prod_{(k,g)\subseteq C}c_{g,P_{k,g}}\Bigr)\,e^{-t_{\rm s}\sum_{k}E(s_C(k))}\,\zeta_C\,\chi_{S_C}(s_C(0))
\qquad\text{satisfies}\qquad
|\omega(C)|\le q_\epsilon^{a(C)}e^{-2t_0J_{\min}d(C)} .
\label{SM:eq:omega}
\end{equation}
\end{enumerate}
\end{lemma}

\begin{proof}
(i) By Lemma~\ref{SM:lem:components}(iii), $s_C(k)$ is a union of clusters of the valid pattern $s_k$.  A valid pattern of one type is a closed $1$-chain, $\partial z$ on the lattice or $\partial(*x)$ on the dual lattice (Sec.~\ref{SM:sec:geometry}); distinct clusters share no vertex, so the boundary of each cluster vanishes separately and each cluster is a closed $1$-chain.  A cluster of $s_C(k)$ lies in ${\rm sp}(C)\subseteq Q_C^{\rm box}$, a non-winding box; by Eq.~\eqref{SM:eq:filling}, applied in $Q_C^{\rm box}$ for the edge type and in the dual box one layer larger for the cube type, it bounds a $2$-chain inside $(Q_C^{\rm box})^{+1}$.  The Pauli operator $Z(\cdot)$ or $X(\cdot)$ of that $2$-chain creates exactly the cluster and no excitation of the other type, because $Z$-type operators commute with the cube checks and $X$-type operators with the edge checks.  The product of these operators over the clusters of $s_C(k)$ is $R_C(k)$: it is supported in $Y_C$, and $\syn(R_C(k))=s_C(k)$ by linearity of $\syn$.  In particular $s_C(k)$ is valid.

(ii) $\syn(Q_C)=\emptyset$ by Lemma~\ref{SM:lem:components}(ii), and $Q_C$ is a product of Pauli operators supported in the collars of the label nodes of $C$, hence in ${\rm sp}(C)\subseteq Q_C^{\rm box}$.  Lemma~\ref{SM:lem:generation} applied to the non-winding box $Q_C^{\rm box}$ gives $Q_C=\zeta_CS_C$ with $S_C$ a product of checks whose supports lie in the two-layer collar $(Q_C^{\rm box})^{+2}=Y_C$.

(iii) $|\prod c|\le\prod_{g\subseteq C}q_\epsilon^{|g|}=q_\epsilon^{a(C)}$ by Eq.~\eqref{SM:eq:eta}; $E(s_C(k))\ge2J_{\min}|s_C(k)|$ and $\sum_k|s_C(k)|=d(C)$, with $t_{\rm s}\ge t_0$; and $|\zeta_C\chi_{S_C}(s_C(0))|=1$.
\end{proof}

\subsection{Sign change under the erasure of $\Gamma$}
\label{SM:sec:signflip}

Let $\mathsf G^F(A)=e^{t_{\rm s}H_+/2}e^{-t_{\rm s}(H_++FV_AF^\dagger)}e^{t_{\rm s}H_+/2}$ be the slice operator of the erased Hamiltonian $K_\Gamma^F=H_0+FV_QF^\dagger$, and for a Pauli operator $P$ write $FPF^\dagger=\theta_F(P)P$ with $\theta_F(P)=\pm1$, a multiplicative function of $P$.

\begin{lemma}[sign change under erasure]
\label{SM:lem:signflip}
(i) $\mathsf G^F(A)=F\mathsf G(A)F^\dagger$ for every $A\subseteq\calI$.  Hence the slice operators of $K_\Gamma^F$ are $w^F_g=Fw_g F^\dagger$, with Pauli coefficients
\begin{equation}
c^F_{g,P}=\theta_F(P)\,c_{g,P} .
\label{SM:eq:signflip}
\end{equation}
(ii) The histories of $K_\Gamma^F$ are the histories of $K_\Gamma$, with the weight of $\mathfrak h$ multiplied by $\prod_{k,g}\theta_F(P_{k,g})=\prod_C\theta_F(Q_C)$, the product over all components of $\mathfrak h$.
(iii) For a local component, $\theta_F(Q_C)=\chi_{S_C}(\Gamma)$, which equals $+1$ whenever no check of $\Gamma$ has its support in $Y_C$; a component without label nodes has $Q_C=I$ and $\theta_F(Q_C)=1$.
\end{lemma}

\begin{proof}
(i) $F$ flips exactly the checks of $\Gamma$: $FS_aF^\dagger=-S_a$ for $a\in\Gamma$ and $FS_aF^\dagger=S_a$ otherwise, so $Fn_aF^\dagger=I-n_a$ for $a\in\Gamma$ and
\begin{equation}
FH_+F^\dagger=H_++E_\Gamma,\qquad E_\Gamma=2\sum_{a\in\Gamma}J_aS_a .
\end{equation}
$E_\Gamma$ commutes with $H_+$, with $V_A$ (every $v_i$, $i\in\calI$, commutes with the checks of $\Gamma$), and with $FV_AF^\dagger$ (which commutes with $FS_aF^\dagger=-S_a$).  Therefore
\begin{equation}
F\mathsf G(A)F^\dagger=e^{t_{\rm s}(H_++E_\Gamma)/2}e^{-t_{\rm s}(H_++E_\Gamma+FV_AF^\dagger)}e^{t_{\rm s}(H_++E_\Gamma)/2}
=\mathsf G^F(A)\,e^{t_{\rm s}E_\Gamma/2}e^{-t_{\rm s}E_\Gamma}e^{t_{\rm s}E_\Gamma/2}=\mathsf G^F(A) .
\end{equation}
M\"obius inversion is linear, so $w^F_g=Fw_g F^\dagger=\sum_Pc_{g,P}\theta_F(P)P$, which is Eq.~\eqref{SM:eq:signflip}.

(ii) Lemmas~\ref{SM:lem:tracepower} and \ref{SM:lem:expansion} apply to $K_\Gamma^F$ with the same $H_+$, $E_{\rm ref}$, slicing, dependency graph, Pauli alphabet, patterns, and energies; only the coefficients change, by Eq.~\eqref{SM:eq:signflip}.  The product of the factors $\theta_F(P_{k,g})$ over the insertions regroups by component because $\theta_F$ is multiplicative, and $\prod_{k,g\subseteq C}\theta_F(P_{k,g})=\theta_F(Q_C)$.

(iii) $\theta_F(Q_C)=\theta_F(S_C)$ since $\zeta_C$ is a scalar, and $FS_CF^\dagger=\chi_{S_C}(\syn(F))S_C=\chi_{S_C}(\Gamma)S_C$ by Lemma~\ref{SM:lem:characters}(b).  With a representation $Y$ of $S_C$ by checks supported in $Y_C$, $\chi_{S_C}(\Gamma)=(-1)^{|Y\cap\Gamma|}=1$ when $Y\cap\Gamma=\emptyset$.
\end{proof}

Patterns, Boltzmann factors, and geometry are unchanged by the erasure; only the signs of insertions near $\Gamma$ change.  This is why erasing $\Gamma$ costs at most $\epsilon$ per unit length of $\Gamma$ per slice, Sec.~\ref{SM:sec:gas}.

\section{The gas of linked clusters and the macroscopic background}
\label{SM:sec:gas}

Histories carry signs from two sources: Pauli insertions of different components may anticommute, and the check product $S_C$ of one component may be odd on the excitations of another.  This section computes the sign exactly, Eq.~(B2), proves that a component acquires no sign from anything outside a box of its own size, groups linked components into clusters, counts the clusters, and controls the resulting gas by a finite deletion induction.  The output is the cost of erasing $\Gamma$ for $\beta\le A_*L^4$, Eq.~\eqref{SM:eq:ratio_bL}.

\subsection{Pairwise signs}
\label{SM:sec:pairwise}

Fix a history $\mathfrak h$ all of whose components are local, and number its components $C_1,\dots,C_M$.  List the insertions $P_{k,g}$ of $\mathfrak h$ as $x_1,\dots,x_n$ in order of increasing $k$, the insertions of one slice in any order (they commute), so that $P_N\cdots P_1=x_n\cdots x_1$; let $\mathrm{cp}(j)$ be the component containing the label set of $x_j$.  Two Pauli operators either commute or anticommute, $xy=\varepsilon(x,y)\,yx$ with $\varepsilon=\pm1$.  For two distinct components $C,D$ define
\begin{equation}
\kappa(C,D)=\#\bigl\{(j,j'):\ \mathrm{cp}(j)=C,\ \mathrm{cp}(j')=D,\ j'<j,\ \varepsilon(x_j,x_{j'})=-1\bigr\},
\label{SM:eq:kappa}
\end{equation}
the number of pairs in which an insertion of $D$ precedes an anticommuting insertion of $C$.

\begin{lemma}[commutation of net Pauli operators]
\label{SM:lem:netcommute}
For distinct local components $C,D$ of a history, $Q_CQ_D=Q_DQ_C$.
\end{lemma}

\begin{proof}
$Q_C=\zeta_CS_C$ by Lemma~\ref{SM:lem:local}(ii), and $\syn(Q_D)=\emptyset$ by Lemma~\ref{SM:lem:components}(ii), so $S_C$ commutes with $Q_D$ by Lemma~\ref{SM:lem:characters}(c).
\end{proof}

\begin{lemma}[regrouping]
\label{SM:lem:regroup}
(i) $x_n\cdots x_1=\Bigl[\prod_{\alpha<\beta}(-1)^{\kappa(C_\alpha,C_\beta)}\Bigr]\,Q_{C_M}\cdots Q_{C_1}$.  (ii) $\kappa(C,D)\equiv\kappa(D,C)\pmod2$ for every pair, so the sign in (i) does not depend on the numbering of the components.
\end{lemma}

\begin{proof}
(i) Move the letters of $C_1$ to the right end of the word, preserving their mutual order and the mutual order of the remaining letters.  Each letter $x_j$ of $C_1$ passes every letter $x_{j'}\notin C_1$ with $j'<j$ exactly once, which produces the factor $\prod_{j'}\varepsilon(x_j,x_{j'})$; over all $j$ this is $\prod_{\beta>1}(-1)^{\kappa(C_1,C_\beta)}$, and the moved letters form $Q_{C_1}$.  Repeat with $C_2$ on the remaining word, and so on.  (ii) Apply (i) to the subword formed by the letters of $C$ and $D$ alone, with either component first: $(-1)^{\kappa(C,D)}Q_DQ_C=(-1)^{\kappa(D,C)}Q_CQ_D$.  By Lemma~\ref{SM:lem:netcommute} and since $Q_CQ_D$ is unitary, $(-1)^{\kappa(C,D)}=(-1)^{\kappa(D,C)}$.
\end{proof}

\begin{lemma}[weight factorization]
\label{SM:lem:factorweight}
The weight \eqref{SM:eq:historyweight} of a history all of whose components are local is exactly
\begin{equation}
\prod_C\omega(C)\prod_{C<D}\sigma(C,D),\qquad
\sigma(C,D)=(-1)^{\kappa(C,D)}\,\chi_{S_C}(s_D(0))\,\chi_{S_D}(s_C(0))\in\{\pm1\},
\label{SM:eq:sigma}
\end{equation}
which is Eq.~(B2) and Eq.~(6).
\end{lemma}

\begin{proof}
By Lemma~\ref{SM:lem:components}(i) the coefficient product and the Boltzmann factor in Eq.~\eqref{SM:eq:historyweight} factorize over components.  By Lemma~\ref{SM:lem:regroup},
\begin{equation}
\frac1d\Tr_{s_0}(x_n\cdots x_1)=\prod_{\alpha<\beta}(-1)^{\kappa(C_\alpha,C_\beta)}\ \prod_C\zeta_C\ \frac1d\Tr\bigl(\Pi_{s_0}S_{C_M}\cdots S_{C_1}\bigr)
=\prod_{\alpha<\beta}(-1)^{\kappa(C_\alpha,C_\beta)}\ \prod_C\zeta_C\,\chi_{S_C}(s_0),
\end{equation}
because every $S_C$ acts on the range of $\Pi_{s_0}$ as the scalar $\chi_{S_C}(s_0)$ and $\Tr\Pi_{s_0}=d$.  Since $s_0=\bigoplus_Ds_D(0)$ with every $s_D(0)$ valid (Lemma~\ref{SM:lem:local}(i)), Lemma~\ref{SM:lem:characters}(a) gives $\chi_{S_C}(s_0)=\prod_D\chi_{S_C}(s_D(0))$.  The factor with $D=C$ completes $\omega(C)$, and the factors with $D\ne C$ regroup into $\prod_{C<D}\chi_{S_C}(s_D(0))\chi_{S_D}(s_C(0))$.
\end{proof}

The sign is a product over pairs because Pauli commutation is bilinear and characters are multiplicative.  The definition of $\kappa$ refers to the cut of the thermal circle between slices $N$ and $1$; the next lemma shows that the factorization does not depend on where the circle is cut.

\begin{lemma}[cut independence]
\label{SM:lem:cut}
For $j\in\Z_N$, relabel the slices cyclically, $k\mapsto k-j$, keeping every family, insertion, and pattern attached to its slice; the initial pattern becomes $s_j$.  This maps histories of $K_\Gamma$ to histories of $K_\Gamma$ and preserves the weight, each $\omega(C)$, and each $\sigma(C,D)$.
\end{lemma}

\begin{proof}
The relabeled sequence satisfies (a) and (b) of Sec.~\ref{SM:sec:traceformula}, and its components, coefficients, and energies are those of $\mathfrak h$.  Its trace factor is $\Tr_{s_j}(P_j\cdots P_1P_N\cdots P_{j+1})$.  Iterating $P_k\Pi_{s_{k-1}}=\Pi_{s_k}P_k$ along the update rule gives $P_j\cdots P_1\Pi_{s_0}=\Pi_{s_j}P_j\cdots P_1$, so by cyclicity of the trace
$\Tr_{s_j}(P_j\cdots P_1P_N\cdots P_{j+1})=\Tr(P_N\cdots P_{j+1}\Pi_{s_j}P_j\cdots P_1)=\Tr(P_N\cdots P_1\Pi_{s_0})=\Tr_{s_0}(P_N\cdots P_1)$;
the weight is preserved.  For a local component $C$ the relabeled net Pauli operator is $Q'_C=A_jQ_CA_j^\dagger$ with $A_j=P_C(j)\cdots P_C(1)$, because $Q'_C=P_C(j)\cdots P_C(1)\,P_C(N)\cdots P_C(j+1)$ and $Q_C=P_C(N)\cdots P_C(j+1)\,A_j$.  By Lemma~\ref{SM:lem:characters}(b), $Q'_C=\zeta_C\chi_{S_C}(\syn(A_j))S_C$ with $\syn(A_j)=s_C(j)\oplus s_C(0)$ by Lemma~\ref{SM:lem:components}(ii), so $S'_C=S_C$, $\zeta'_C=\zeta_C\chi_{S_C}(s_C(j))\chi_{S_C}(s_C(0))$, and the new initial section is $s_C(j)$; hence $\omega'(C)=(\prod c)\,e^{-t_{\rm s}\sum E}\,\zeta'_C\chi_{S_C}(s_C(j))=\omega(C)$.  Finally, the trace identity and the energies do not involve the coefficients, so the same computation with every coefficient replaced by $1$ shows that $e^{-t_{\rm s}\sum_kE(s_k)}\frac1d\Tr_{s_0}(P_N\cdots P_1)=\prod_C\tilde\omega(C)\prod_{C<D}\sigma(C,D)$, with $\tilde\omega(C)=e^{-t_{\rm s}\sum_kE(s_C(k))}\zeta_C\chi_{S_C}(s_C(0))\ne0$, is preserved together with every $\tilde\omega(C)$.  The two components $C$ and $D$ alone form a history whose components are $C$ and $D$ (its patterns $s_C(k)\oplus s_D(k)$ are valid, its label sets are compatible, and no edge joins $C$ to $D$); applied to it, the preserved identity reads $\tilde\omega(C)\tilde\omega(D)\sigma(C,D)$, so $\sigma(C,D)$ is preserved.
\end{proof}

\subsection{The empty-box lemma}
\label{SM:sec:emptybox}

For a local component $C$, the box $Y_C$ of Eq.~\eqref{SM:eq:YC} contains the collars of all its insertions, the supports of the checks generating $S_C$, and the Pauli operators $R_C(k)$ erasing its sections (Lemma~\ref{SM:lem:local}), with $\diam Y_C\le c_1m(C)$.  Its time projection $T(C)$ is a cyclic interval (Lemma~\ref{SM:lem:components}(iv)); let $T^+(C)$ be $T(C)$ together with the slice before it and the slice after it.  If $T^+(C)\ne\Z_N$, let $I_C=T^+(C)$: a cyclic interval of $|T(C)|+2\le m(C)+2$ slices whose first and last slices contain no node of $C$.  Otherwise let $I_C=\Z_N$; in that case every slice lies within one step of $T(C)$, so $|T(C)|\ge N/3$, and $|T(C)|\le m(C)$ gives $N\le3m(C)$.  The space-time box of $C$ is
\begin{equation}
\cB_C=Y_C\times I_C .
\label{SM:eq:BC}
\end{equation}
A node is \emph{inside} $\cB_C$ when its slice lies in $I_C$ and its spatial set meets $Y_C$.  For nodes $u=(\cdot,k)$ and $v=(\cdot,k')$ with spatial sets ${\rm sp}(u),{\rm sp}(v)$ let $\varrho(u,v)=\dist({\rm sp}(u),{\rm sp}(v))+\dist_{\Z_N}(k,k')$, and for components let $\varrho(C,D)=\min_{u\in C,v\in D}\varrho(u,v)$.  A node $v$ inside $\cB_C$ satisfies $\varrho(v,C)\le c_3m(C)$ with $c_3=c_1+2$: its spatial set meets $Y_C$, which contains ${\rm sp}(C)$ and has diameter at most $c_1m(C)$, and its slice lies within $|I_C|-1\le m(C)+1\le2m(C)$ steps of $T(C)$ when $I_C\ne\Z_N$ and within $N/2\le2m(C)$ steps otherwise.

For the union $\cM$ of all macroscopic components of a history $\mathfrak h$, write $s_\cM(k)$ for the union of their sections and $Q_\cM$ for the time-ordered product of their insertions.  Then $s_\cM(k)=s_k\oplus\bigoplus_{C\ \text{local}}s_C(k)$ is valid, and $\syn(Q_\cM)=\emptyset$, because $\syn(P_N\cdots P_1)=\emptyset$ and every local $Q_C$ has empty syndrome.  When $Q_\cM=\zeta_\cM S_\cM$ is a check product times a phase, $Q_\cM$ commutes with every local $Q_C$ (Lemma~\ref{SM:lem:characters}(c)), Lemma~\ref{SM:lem:regroup}(ii) holds for the pair $(C,\cM)$ with $\kappa(C,\cM)$ defined by Eq.~\eqref{SM:eq:kappa}, and $\sigma(C,\cM)$ is defined by Eq.~\eqref{SM:eq:sigma} with $D=\cM$; Sec.~\ref{SM:sec:background} uses it under the name $\xi_C(\cM)$.

\begin{lemma}[empty box]
\label{SM:lem:emptybox}
Let $C$ be a local component of a history $\mathfrak h$, and let $D$ be either another local component of $\mathfrak h$, or the union $\cM$ of all macroscopic components of $\mathfrak h$ when $Q_\cM$ is a check product times a phase.  If no node of $D$ is inside $\cB_C$, then $\sigma(C,D)=+1$.
\end{lemma}

\begin{proof}
In both cases $D$ has these three properties.  (P1)~Every insertion of $D$ in a slice of $I_C$ is supported in a collar that does not meet $Y_C$, hence commutes with every insertion of $C$, all of which are supported in $Y_C$.  (P2)~For every $k\in I_C$, no check of $s_D(k)$ has its support meeting $Y_C$; since $S_C$ is represented by checks supported in $Y_C$, $\chi_{S_C}(s_D(k))=1$.  (P3)~$Q_D=\zeta_DS_D$ is a check product times a phase, and $s_D(0)$ is valid.

\emph{Non-winding case, $I_C\ne\Z_N$} [Fig.~\ref{SM:fig:S3}(b)].  By Lemma~\ref{SM:lem:cut} we may relabel the slices so that the first slice of $I_C$ becomes slice $N$; then $I_C=\{N\}\cup\{1,\dots,|I_C|-1\}$, every node of $C$ lies in a slice of $\{1,\dots,|I_C|-2\}$, and the slices outside $I_C$ are $\{|I_C|,\dots,N-1\}$.  The initial section $s_C(0)=s_C(N)$ is empty, because slice $N$ contains no node of $C$; hence $\chi_{S_D}(s_C(0))=1$.  Slice $N$ lies in $I_C$, so $\chi_{S_C}(s_D(0))=\chi_{S_C}(s_D(N))=1$ by (P2).  An insertion of $D$ that precedes an insertion of $C$ lies in a slice $\le|I_C|-2$, hence in $I_C$, and commutes with every insertion of $C$ by (P1); so no pair is counted in Eq.~\eqref{SM:eq:kappa}, $\kappa(C,D)=0$, and $\sigma(C,D)=1$.

\emph{Winding case, $I_C=\Z_N$} [Fig.~\ref{SM:fig:S3}(c)].  Now every slice lies in $I_C$.  By (P1) every insertion of $D$ commutes with every insertion of $C$, so $\kappa(C,D)=0$, and by (P2) $\chi_{S_C}(s_D(0))=1$.  Let $R_C=R_C(0)$ be the Pauli operator of Lemma~\ref{SM:lem:local}(i), supported in $Y_C$ with $\syn(R_C)=s_C(0)$.  Every insertion of $D$ is supported outside $Y_C$, hence so is $Q_D$, and $[Q_D,R_C]=0$, that is, $[S_D,R_C]=0$.  By Lemma~\ref{SM:lem:characters}(b), $R_CS_DR_C^\dagger=\chi_{S_D}(\syn(R_C))S_D=\chi_{S_D}(s_C(0))S_D$, so $\chi_{S_D}(s_C(0))=1$ and $\sigma(C,D)=1$.

Both cases use (P1), (P2), and (P3) only, which hold for the whole macroscopic background, so the lemma covers $D=\cM$.
\end{proof}

Since $\sigma(C,D)=\sigma(D,C)$ for local components, by Lemma~\ref{SM:lem:regroup}(ii) and the symmetry of the character factors, the contrapositive of Lemma~\ref{SM:lem:emptybox} for $C$ and then for $D$ gives
\begin{equation}
\sigma(C,D)=-1\ \Longrightarrow\ \varrho(C,D)\le c_3\min\{m(C),m(D)\},
\label{SM:eq:linklocal}
\end{equation}
the second line of Eq.~(6): a component with $\sigma(C,D)=-1$ has a node inside $\cB_C$, which lies within $\varrho$-distance $c_3m(C)$ of $C$.  The physical content is the main-text statement that linking requires closeness: a bubble feels a sheet only if it enters a box of its own size around the sheet.  Both properties of the code enter the proof, local erasure through $R_C$ and local generation through $S_C$.

\subsection{Linked clusters and the hard-core gas}
\label{SM:sec:gasdef}

Let $\mathfrak L$ be the set of all local components of all histories of $K_\Gamma$: an element of $\mathfrak L$ is a component $C$ of some history, with its node set, its label sets, its insertions, and its sections, and $m(C)<m_0$.  Such a $C$ is the unique component of the history that consists of $C$ alone: its patterns $s_C(k)$ are valid (Lemma~\ref{SM:lem:local}(i)), its label sets are compatible, and they satisfy the update rule (Lemma~\ref{SM:lem:components}(ii)).  The intrinsic weight $\omega(C)$ of Eq.~\eqref{SM:eq:omega} depends on $C$ alone.  Two elements $C,D\in\mathfrak L$ are \emph{compatible}, $C\sim D$, when their node sets are disjoint and they do not touch; for compatible $C,D$ the sign $\sigma(C,D)$ of Eq.~\eqref{SM:eq:sigma} is defined from the data of $C$ and $D$ alone, and it equals $\sigma(D,C)$.

\begin{lemma}[gas of components]
\label{SM:lem:gascomp}
(i) The components of a history all of whose components are local form a pairwise compatible subset of $\mathfrak L$.  (ii) Every pairwise compatible subset $\{C_1,\dots,C_M\}$ of $\mathfrak L$ is the set of components of exactly one history, whose weight is $\prod_i\omega(C_i)\prod_{i<j}\sigma(C_i,C_j)$.  (iii) Consequently the sum of the weights of all histories all of whose components are local is
\begin{equation}
\Zloc=\sum_{\substack{\mathfrak F\subseteq\mathfrak L\\ \text{pairwise compatible}}}\ \prod_{C\in\mathfrak F}\omega(C)\prod_{\{C,D\}\subseteq\mathfrak F}\sigma(C,D) .
\label{SM:eq:Zloccomp}
\end{equation}
\end{lemma}

\begin{proof}
(i) Distinct components of a history have disjoint node sets and are not joined by any edge of $\mathfrak G$.  (ii) Given pairwise compatible $C_1,\dots,C_M$, define $\mathfrak F_k$ as the union of their label sets in slice $k$, with the given insertions, and $s_k=\bigsqcup_is_{C_i}(k)$.  Label sets of different $C_i$ in one slice are disjoint and not joined in the dependency graph, since a dependency edge is an edge of $\mathfrak G$ by rule (1); so $\mathfrak F_k$ is compatible.  Each $s_k$ is a sum of valid patterns, hence valid, and the update rule for $s_k$ is the sum of the update rules of the $C_i$.  This is a history; its selected nodes are the union of the node sets of the $C_i$, which are connected and pairwise non-adjacent in $\mathfrak G$, so its components are exactly $C_1,\dots,C_M$, and every history with these components has these data.  Its weight is given by Lemma~\ref{SM:lem:factorweight}.  (iii) follows from (i) and (ii).
\end{proof}

Given a pairwise compatible family, join two of its members by an edge when $\sigma=-1$; this is its \emph{linking graph}.  A \emph{linked cluster} $\gamma$ is a pairwise compatible subset of $\mathfrak L$ whose linking graph is connected, with cost $m(\gamma)=\sum_{C\in\gamma}m(C)$, label count $a(\gamma)=\sum_{C\in\gamma}a(C)$, support $\supp\gamma$ the union of the node sets of its members, and \emph{activity}
\begin{equation}
z_\gamma=\prod_{C\in\gamma}\omega(C)\prod_{\{C,D\}\subseteq\gamma}\sigma(C,D) .
\label{SM:eq:activity}
\end{equation}
Two linked clusters are \emph{compatible}, $\gamma\sim\gamma'$, when every member of $\gamma$ is compatible with every member of $\gamma'$ and $\sigma(C,C')=+1$ for all $C\in\gamma$, $C'\in\gamma'$; otherwise they are \emph{incompatible}, $\gamma\not\sim\gamma'$, which includes $\gamma\not\sim\gamma$.  The linking graph and the compatibility relation are defined by node sets and by $\sigma$ alone, so a change of the perturbation coefficients, or setting some of them to zero, changes the activities and nothing else.

\begin{lemma}[polymer representation]
\label{SM:lem:polymer}
Let $\Lambda$ be the finite set of all linked clusters.  Then
\begin{equation}
\Zloc=\sum_{\substack{\mathcal G\subseteq\Lambda\\ \text{pairwise compatible}}}\ \prod_{\gamma\in\mathcal G}z_\gamma ,
\label{SM:eq:Zloc}
\end{equation}
the partition function of a hard-core gas of linked clusters with complex activities.
\end{lemma}

\begin{proof}
Map a pairwise compatible family $\mathfrak F$ of components to the set $\mathcal G$ of the connected components of its linking graph.  Each member of $\mathcal G$ is a linked cluster; two of them are compatible, because their members are compatible and no pair across them has $\sigma=-1$ (such a pair would be an edge of the linking graph joining the two).  Conversely, the union of a pairwise compatible family of linked clusters is a pairwise compatible family of components whose linking graph has exactly those clusters as its connected components.  Under this bijection the summand of Eq.~\eqref{SM:eq:Zloccomp} equals $\prod_{\gamma\in\mathcal G}z_\gamma$: every factor $\sigma(C,D)$ with $C,D$ in different clusters equals $+1$.
\end{proof}

\subsection{Enumeration of linked clusters}
\label{SM:sec:counting}

Three geometric constants enter.  $\Delta_{\mathfrak G}$ is the maximal degree of the environment graph.  $C_{\rm ball}$ bounds the number of nodes within $\varrho$-distance $R$ of any node by $C_{\rm ball}(R+1)^5$: the plaquettes within $\mathfrak P$-distance $R$ of a plaquette have centers within $\ell^\infty$ distance $R$ of its center, at most $6(2R+2)^4$ of them; every plaquette lies in eight checks and in a bounded number of distinct supports of terms; and $2R+1$ slices are involved.  Finally $\Delta_\varrho$ is the maximal degree of the auxiliary graph $\mathfrak G_\varrho$ on the node set in which $u\sim v$ when $\varrho(u,v)\le1$; it is at most $32C_{\rm ball}$.

\begin{lemma}[paths in the auxiliary graph]
\label{SM:lem:paths}
Two nodes with $\varrho(u,v)=R$ are joined in $\mathfrak G_\varrho$ by a path of at most $R+2$ edges.
\end{lemma}

\begin{proof}
Choose plaquettes $p\in {\rm sp}(u)$, $p'\in {\rm sp}(v)$ at $\mathfrak P$-distance $D=\dist({\rm sp}(u),{\rm sp}(v))$ and a $\mathfrak P$-path $p=p_0,p_1,\dots,p_D=p'$, and for each $j$ an edge check $a_j$ whose support contains $p_j$.  The sequence $u,(a_0,k_u),(a_1,k_u),\dots,(a_D,k_u),(a_D,k_u\pm1),\dots,(a_D,k_v),v$ has consecutive $\varrho$-distances at most $1$: ${\rm sp}(u)$ and $\supp S_{a_0}$ share $p_0$, consecutive supports contain plaquettes at distance $1$, the time steps change one slice, and $\supp S_{a_D}$ and ${\rm sp}(v)$ share $p'$.  Its length is $D+\dist_{\Z_N}(k_u,k_v)+2=R+2$.
\end{proof}

\begin{lemma}[enumeration of components]
\label{SM:lem:countcomp}
For every node $x$ and every $m\ge1$, the number of components $C$ of histories, local or macroscopic, with $m(C)=m$ and $x\in C$ is at most $K_{\rm raw}^m$, where $K_{\rm raw}=\Delta_{\mathfrak G}^2\,4^{c_R}$.
\end{lemma}

\begin{proof}
A component is determined by its node set and its insertions.  Given the node set, its sections are the excitation nodes slice by slice; its label nodes in slice $k$ form a set $A_k\subseteq\calI$, and the compatible family of label sets with union $A_k$ is unique, namely the set of connected components of $A_k$ in the dependency graph, as in the proof of Lemma~\ref{SM:lem:mobius}; and $P_C(k)$ is the product of the insertions.  The node set is a connected subset of $\mathfrak G$ with $m$ nodes containing $x$; as in Lemma~\ref{SM:lem:clustercount}, the depth-first traversal of a spanning tree chosen by a fixed rule is a closed walk of length $2(m-1)$ from $x$ that visits every node and determines the set, and each step of the walk has at most $\Delta_{\mathfrak G}$ choices, so there are at most $\Delta_{\mathfrak G}^{2(m-1)}$ such sets.  Given the node set, the component is fixed by the choice of the insertion $P_{k,g}$ of each label set $g$, a Pauli operator supported in ${\rm col}(g)$, one of at most $4^{|{\rm col}(g)|}\le4^{c_R|g|}$, and $\sum_g|g|=a(C)\le m$.  Hence there are at most $\Delta_{\mathfrak G}^{2(m-1)}4^{c_Rm}\le K_{\rm raw}^m$ components.
\end{proof}

\begin{lemma}[tree inequality]
\label{SM:lem:tree}
For every finite tree $T$ with vertex weights $m_i\ge1$,
\begin{equation}
\sum_{\{i,j\}\in T}\min(m_i,m_j)\le\sum_im_i-\max_im_i .
\label{SM:eq:treeineq}
\end{equation}
\end{lemma}

\begin{figure}[t]
\centering
\includegraphics{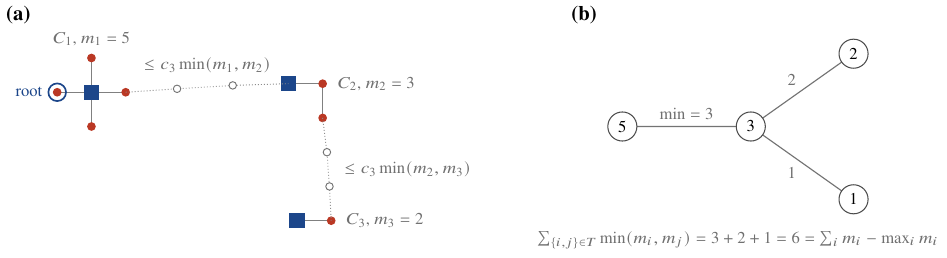}
\caption{Enumeration of linked clusters.  (a)~Two-layer encoding: each component is a rooted connected labeled shape in the space-time graph (layer~1, at most $K_{\rm raw}^m$ shapes of cost $m$); linked pairs, $\sigma=-1$, are joined by deterministic paths of length at most $c_3\min(m_i,m_j)+2$ (layer~2).  The configuration is read back from layer~1 alone, so the paths may cross or reuse nodes.  (b)~The tree inequality \eqref{SM:eq:treeineq} on a tree with weights $5,3,2,1$: the sum of $\min(m_i,m_j)$ over the edges equals the number of edges of the forests induced on $\{i:m_i\ge t\}$, summed over thresholds $t$, and is at most $\sum_im_i-\max_im_i$.}
\label{SM:fig:S4}
\end{figure}

\begin{proof}
For $t\ge1$ let $V_t=\{i:m_i\ge t\}$ and let $T|_{V_t}$ be the forest induced by $T$ on $V_t$.  An edge $\{i,j\}$ satisfies $\min(m_i,m_j)\ge t$ exactly when both ends lie in $V_t$, so $\sum_{\{i,j\}\in T}\min(m_i,m_j)=\sum_{t\ge1}|E(T|_{V_t})|$.  A forest on a nonempty vertex set has fewer edges than vertices, $|E(T|_{V_t})|\le|V_t|-1$ for $1\le t\le\max_im_i$, and $V_t=\emptyset$ beyond.  Summing, $\sum_{t=1}^{\max_im_i}(|V_t|-1)=\sum_im_i-\max_im_i$ [Fig.~\ref{SM:fig:S4}(b)].
\end{proof}

\begin{lemma}[enumeration of linked clusters]
\label{SM:lem:countlink}
For every node $x$ and every $m\ge1$,
\begin{equation}
\#\{\gamma\in\Lambda:\ m(\gamma)=m,\ x\in\supp\gamma\}\le K_{\rm link}^m,\qquad
K_{\rm link}=2\,\Delta_2^{\,2(c_3+4)}\,4^{c_R},\quad\Delta_2=\max\{\Delta_{\mathfrak G},\Delta_\varrho\}+1 .
\label{SM:eq:linkcount}
\end{equation}
\end{lemma}

\begin{proof}
Form the two-layer graph $\mathfrak G^{(2)}$ on the pairs $(u,1)$, $(u,2)$, $u$ a node, with the edges $(u,1)\sim(v,1)$ for $u\sim v$ in $\mathfrak G$, $(u,2)\sim(v,2)$ for $u\sim v$ in $\mathfrak G_\varrho$, and $(u,1)\sim(u,2)$; its degree is at most $\Delta_2$.  Let $\gamma=\{C_1,\dots,C_p\}$ with $x\in C_1$ and $m_i=m(C_i)$.  Choose a spanning tree $T$ of the linking graph of $\gamma$ by a fixed deterministic rule.  For every edge $\{i,j\}$ of $T$, $\sigma(C_i,C_j)=-1$, so by Eq.~\eqref{SM:eq:linklocal} there are nodes $u_{ij}\in C_i$, $v_{ij}\in C_j$ with $\varrho(u_{ij},v_{ij})\le c_3\min(m_i,m_j)$, and by Lemma~\ref{SM:lem:paths} a path $\pi_{ij}$ in $\mathfrak G_\varrho$ from $u_{ij}$ to $v_{ij}$ of at most $c_3\min(m_i,m_j)+2$ edges.  Let
\begin{equation}
A(\gamma)=\bigcup_i\{(u,1):u\in C_i\}\ \cup\ \bigcup_{\{i,j\}\in T}\{(u,2):u\in\pi_{ij}\} .
\end{equation}
$A(\gamma)$ is connected in $\mathfrak G^{(2)}$: each $C_i$ is connected in layer~1, each path is connected in layer~2 and is attached through the vertical edges at $(u_{ij},1)$ and $(v_{ij},1)$, and $T$ is connected.  It contains $(x,1)$, and by Lemma~\ref{SM:lem:tree}, with $p\le m$,
\begin{equation}
|A(\gamma)|\le\sum_im_i+\sum_{\{i,j\}\in T}\bigl(c_3\min(m_i,m_j)+3\bigr)\le m+c_3m+3(p-1)\le(c_3+4)m .
\end{equation}
The cluster is recovered from $A(\gamma)$ and the insertions: the layer-1 nodes of $A(\gamma)$ form $\bigcup_iC_i$, whose connected components in $\mathfrak G$ are exactly the node sets of the $C_i$, because distinct members of $\gamma$ do not touch; with the insertions, each $C_i$ is determined as in Lemma~\ref{SM:lem:countcomp}.  The paths may cross or reuse nodes, since they are never read [Fig.~\ref{SM:fig:S4}(a)].  Hence the number of linked clusters is at most the number of connected subsets of $\mathfrak G^{(2)}$ containing $(x,1)$ with at most $(c_3+4)m$ nodes, times $4^{c_Rm}$ for the insertions: at most $\sum_{s\le(c_3+4)m}\Delta_2^{2s}\,4^{c_Rm}\le2\Delta_2^{2(c_3+4)m}4^{c_Rm}\le K_{\rm link}^m$.
\end{proof}

$K_{\rm raw}$ and $K_{\rm link}$ depend on the range, the collars, the lattice, and the Pauli alphabet only; they are independent of $t_0$, $t_{\rm s}$, $\beta$, $N$, $L$, and the coefficients.

\subsection{Finite deletion induction}
\label{SM:sec:deletion}

Let $c_{\rm adj}=r_{\mathfrak G}+1$ bound the $\varrho$-distance between adjacent nodes of $\mathfrak G$, let $c_5=\max\{c_3,c_{\rm adj}\}$, and let $C_*=C_{\rm ball}(\max\{c_1,c_5\}+1)^5$, so that the number of nodes within $\varrho$-distance $c_1m$ or $c_5m$ of a given node is at most $C_*(m+1)^5$.  Define the geometric function
\begin{equation}
\Upsilon(q)=C_*\sum_{m\ge1}(m+1)^5\,(K_{\rm link}eq)^m .
\label{SM:eq:Sq}
\end{equation}
\emph{Before} choosing $t_0$, fix $q_0>0$ such that
\begin{equation}
K_{\rm link}eq_0<\tfrac12,\qquad \Upsilon(q_0)\le\tfrac1{16},\qquad q_0e^{1/8}\le\tfrac12,\qquad K_{\rm raw}e^{1/4}q_0<\tfrac12,
\label{SM:eq:q0}
\end{equation}
and set $\varkappa=2\Upsilon(q_0)\le\frac18$.  \emph{Then} choose $t_0$ and $\bar\epsilon$ with
\begin{equation}
e^{-2t_0J_{\min}}\le q_0,\qquad e^{C(t_0)\bar\epsilon}-1\le q_0 ,
\label{SM:eq:t0eps}
\end{equation}
for instance $\bar\epsilon=\ln(1+q_0)/C(t_0)$.  Write $\rho_\gamma=q_0^{m(\gamma)}$.

\begin{lemma}[activity bounds]
\label{SM:lem:actbound}
For $\epsilon\le\bar\epsilon$ and every linked cluster $\gamma$,
\begin{equation}
|z_\gamma|\le\Bigl(\frac\epsilon{\bar\epsilon}\Bigr)^{a(\gamma)}\rho_\gamma\le\rho_\gamma .
\label{SM:eq:zsharp}
\end{equation}
\end{lemma}

\begin{proof}
The function $x\mapsto(e^{C(t_0)x}-1)/x$ is increasing on $x>0$, so for $\epsilon\le\bar\epsilon$, $q_\epsilon=e^{C(t_0)\epsilon}-1\le(\epsilon/\bar\epsilon)(e^{C(t_0)\bar\epsilon}-1)\le(\epsilon/\bar\epsilon)q_0$, and $e^{-2t_0J_{\min}}\le q_0$ by Eq.~\eqref{SM:eq:t0eps}.  Equation~\eqref{SM:eq:omega} gives $|\omega(C)|\le(\epsilon/\bar\epsilon)^{a(C)}q_0^{a(C)+d(C)}$ for every member, and $|z_\gamma|=\prod_{C\in\gamma}|\omega(C)|$.
\end{proof}

\begin{lemma}[incompatibility sum]
\label{SM:lem:incomp}
For every linked cluster $\gamma$,
\begin{equation}
\sum_{\gamma'\not\sim\gamma}\rho_{\gamma'}e^{m(\gamma')}\le m(\gamma)\,\Upsilon(q_0)\le\frac{m(\gamma)}{16} .
\label{SM:eq:incompsum}
\end{equation}
\end{lemma}

\begin{proof}
If $\gamma'\not\sim\gamma$, some node of $\gamma'$ lies within $\varrho$-distance $c_5m(\gamma')$ of some node of $\gamma$: a shared node is at distance $0$; two touching members have adjacent nodes, at distance at most $c_{\rm adj}$; and $\sigma(C,C')=-1$ for $C\in\gamma$, $C'\in\gamma'$ gives $\varrho(C,C')\le c_3\min\{m(C),m(C')\}\le c_3m(\gamma')$ by Eq.~\eqref{SM:eq:linklocal}.  Hence, by Lemma~\ref{SM:lem:countlink} and the ball count,
\begin{equation}
\begin{aligned}
\sum_{\gamma'\not\sim\gamma}\rho_{\gamma'}e^{m(\gamma')}&\le\sum_{u\in\supp\gamma}\ \sum_{m'\ge1}\ \sum_{v:\,\varrho(u,v)\le c_5m'}\ \#\{\gamma':m(\gamma')=m',\ v\in\supp\gamma'\}\,q_0^{m'}e^{m'}\\
&\le m(\gamma)\sum_{m'\ge1}C_*(m'+1)^5K_{\rm link}^{m'}(q_0e)^{m'} ,
\end{aligned}
\end{equation}
which is $m(\gamma)\Upsilon(q_0)$.
\end{proof}

\begin{lemma}[finite deletion induction]
\label{SM:lem:deletion}
Let $\Lambda'$ be a finite set of polymers with a symmetric and reflexive incompatibility relation $\not\sim$, costs $m(\gamma)\ge1$, and radii $\rho_\gamma=q^{m(\gamma)}$ with $qe^{1/8}\le\frac12$, such that $\sum_{\gamma'\not\sim\gamma}\rho_{\gamma'}e^{m(\gamma')}\le m(\gamma)/16$ for every $\gamma\in\Lambda'$, and let $b_\gamma=2\rho_\gamma e^{m(\gamma)}$.  For every family of complex activities with $|z_\gamma|\le\rho_\gamma$, the hard-core partition function $Z_{\Lambda'}=\sum_{\mathcal G\subseteq\Lambda'\ \text{pairwise compatible}}\prod_{\gamma\in\mathcal G}z_\gamma$ is nonzero, and for every $\gamma\in\Lambda'$
\begin{equation}
\Bigl|\frac{Z_{\Lambda'\setminus\{\gamma\}}}{Z_{\Lambda'}}\Bigr|\le e^{b_\gamma},\qquad
\Bigl|\frac{\partial\ln Z_{\Lambda'}}{\partial z_\gamma}\Bigr|=\Bigl|\frac{Z_{\Lambda'\setminus N[\gamma]}}{Z_{\Lambda'}}\Bigr|\le\exp\Bigl(\sum_{\gamma'\in N[\gamma]}b_{\gamma'}\Bigr)\le e^{m(\gamma)/8},
\label{SM:eq:deletion}
\end{equation}
where $N[\gamma]=\{\gamma'\in\Lambda':\gamma'\not\sim\gamma\}$, which contains $\gamma$.  The hypotheses hold for every subset $\Lambda'$ of the set $\Lambda$ of linked clusters with $q=q_0$, by Eq.~\eqref{SM:eq:q0} and Lemma~\ref{SM:lem:incomp}.
\end{lemma}

\begin{proof}
A pairwise compatible family either avoids $\gamma$ or contains $\gamma$ together with a pairwise compatible family in $\Lambda'\setminus N[\gamma]$, so
\begin{equation}
Z_{\Lambda'}=Z_{\Lambda'\setminus\{\gamma\}}+z_\gamma Z_{\Lambda'\setminus N[\gamma]},
\label{SM:eq:Zsplit}
\end{equation}
and $\partial Z_{\Lambda'}/\partial z_\gamma=Z_{\Lambda'\setminus N[\gamma]}$.  By the hypothesis and the definition of $b_\gamma$, $\sum_{\gamma'\in N[\gamma]}b_{\gamma'}\le m(\gamma)/8$ for every $\gamma$.  Induct on $|\Lambda'|$; $Z_\emptyset=1$.  Let $|\Lambda'|\ge1$ and assume the statements for all proper subsets.  Fix $\gamma\in\Lambda'$ and list $N[\gamma]\setminus\{\gamma\}=\{\gamma_1,\dots,\gamma_r\}$.  With $\Lambda_0=\Lambda'\setminus\{\gamma\}$ and $\Lambda_j=\Lambda_{j-1}\setminus\{\gamma_j\}$, the ratio $Z_{\Lambda'\setminus N[\gamma]}/Z_{\Lambda'\setminus\{\gamma\}}=\prod_{j=1}^rZ_{\Lambda_j}/Z_{\Lambda_{j-1}}$ is a product of deletion ratios in proper subsets, each of modulus at most $e^{b_{\gamma_j}}$ by the induction hypothesis.  Hence the correction $x_\gamma=z_\gamma Z_{\Lambda'\setminus N[\gamma]}/Z_{\Lambda'\setminus\{\gamma\}}$ in $Z_{\Lambda'}=Z_{\Lambda'\setminus\{\gamma\}}(1+x_\gamma)$ satisfies
\begin{equation}
|x_\gamma|\le\rho_\gamma\exp\Bigl(\sum_{j}b_{\gamma_j}\Bigr)\le\rho_\gamma e^{m(\gamma)/8}=\bigl(qe^{1/8}\bigr)^{m(\gamma)}\le\tfrac12
\end{equation}
by the hypothesis on $q$.  So $Z_{\Lambda'}\ne0$, and $-\ln(1-x)\le2x$ for $0\le x\le\frac12$ gives $|Z_{\Lambda'\setminus\{\gamma\}}/Z_{\Lambda'}|=|1+x_\gamma|^{-1}\le e^{2|x_\gamma|}\le e^{2\rho_\gamma e^{m(\gamma)}}=e^{b_\gamma}$; as $\gamma$ was arbitrary, this is the first bound for every $\gamma\in\Lambda'$.  For the second, $Z_{\Lambda'\setminus N[\gamma]}/Z_{\Lambda'}=(Z_{\Lambda'\setminus\{\gamma\}}/Z_{\Lambda'})\prod_{j=1}^rZ_{\Lambda_j}/Z_{\Lambda_{j-1}}$, a product of $r+1$ deletion ratios, the first in $\Lambda'$ and the others in proper subsets, each of modulus at most $e^{b}$ of the deleted polymer.
\end{proof}

\begin{lemma}[interpolation]
\label{SM:lem:interpolation}
Under the hypotheses of Lemma~\ref{SM:lem:deletion}, let $z^{(0)},z^{(1)}$ be two activity families on $\Lambda'$ with $|z^{(j)}_\gamma|\le\rho_\gamma$, and $Z_0,Z_1$ the corresponding partition functions.  Then
\begin{equation}
\bigl|\ln|Z_1|-\ln|Z_0|\bigr|\le\sum_{\gamma\in\Lambda'}\bigl|z^{(1)}_\gamma-z^{(0)}_\gamma\bigr|\,e^{m(\gamma)} .
\label{SM:eq:interpolation}
\end{equation}
\end{lemma}

\begin{proof}
The family $z(t)=(1-t)z^{(0)}+tz^{(1)}$ satisfies $|z_\gamma(t)|\le\rho_\gamma$ for $0\le t\le1$, so $Z(t)=Z_{\Lambda'}(z(t))$, a polynomial in $t$, has no zero on $[0,1]$ by Lemma~\ref{SM:lem:deletion}, and $\ln|Z(t)|$ is smooth there with $\frac d{dt}\ln|Z(t)|={\rm Re}\sum_\gamma(z^{(1)}_\gamma-z^{(0)}_\gamma)\,\partial\ln Z/\partial z_\gamma$.  Integrating over $[0,1]$ and using $|\partial\ln Z/\partial z_\gamma|\le e^{m(\gamma)/8}\le e^{m(\gamma)}$ from Eq.~\eqref{SM:eq:deletion} gives the claim.
\end{proof}

The statements of this subsection are finite polynomial identities and inequalities proved by induction; the induction plays the role of the Koteck\'y--Preiss condition~\cite{KoteckyPreiss1986} for this finite gas.

\subsection{Cost of erasure in the local gas}
\label{SM:sec:localcost}

Let $\Zloc^F$ be the local gas of the erased Hamiltonian $K_\Gamma^F$.

\begin{lemma}[local cost of erasure]
\label{SM:lem:localtwist}
For $\epsilon\le\bar\epsilon$ and $t_0\le\beta$,
\begin{equation}
\Bigl|\ln\frac{|\Zloc^F|}{|\Zloc|}\Bigr|\le\varkappa\frac\epsilon{\bar\epsilon}N|\Gamma|\le C_1\epsilon N|\Gamma|\le C_2\beta\epsilon|\Gamma|,\qquad
C_1=\frac\varkappa{\bar\epsilon},\quad C_2=\frac{\varkappa}{\bar\epsilon t_0} .
\label{SM:eq:localtwist}
\end{equation}
\end{lemma}

\begin{proof}
By Lemma~\ref{SM:lem:signflip}, the histories of $K_\Gamma^F$ are those of $K_\Gamma$ with every $\omega(C)$ multiplied by $\theta_F(Q_C)=\chi_{S_C}(\Gamma)$; the node sets, the signs $\sigma$, the linking graphs, and the compatibility relation are unchanged.  So $\Zloc^F$ is the gas of Eq.~\eqref{SM:eq:Zloc} with the activities $z^F_\gamma=z_\gamma\prod_{C\in\gamma}\chi_{S_C}(\Gamma)$, which satisfy $|z^F_\gamma|\le\rho_\gamma$, and Lemma~\ref{SM:lem:interpolation} applies.  The difference $z^F_\gamma-z_\gamma$ vanishes unless some $C\in\gamma$ has $\chi_{S_C}(\Gamma)=-1$, which by Lemma~\ref{SM:lem:signflip}(iii) requires a check $a\in\Gamma$ with support in $Y_C$ and a label node in $C$; then the node $(a,k)$, for any slice $k$ of a node of $C$, lies within $\varrho$-distance $\diam Y_C\le c_1m(C)\le c_1m(\gamma)$ of $\supp\gamma$, and $a(\gamma)\ge1$ gives $|z^F_\gamma-z_\gamma|\le2(\epsilon/\bar\epsilon)\rho_\gamma$ by Lemma~\ref{SM:lem:actbound}.  The nodes $(a,k)$ with $a\in\Gamma$ number $N|\Gamma|$, and at most $C_*(m+1)^5K_{\rm link}^m$ linked clusters of cost $m$ have a node within distance $c_1m$ of a given node.  Therefore
\begin{equation}
\sum_\gamma|z^F_\gamma-z_\gamma|e^{m(\gamma)}\le2\frac\epsilon{\bar\epsilon}N|\Gamma|\sum_{m\ge1}C_*(m+1)^5K_{\rm link}^m(q_0e)^m=2\Upsilon(q_0)\frac\epsilon{\bar\epsilon}N|\Gamma|=\varkappa\frac\epsilon{\bar\epsilon}N|\Gamma| ,
\end{equation}
and $N\le\beta/t_0$.
\end{proof}

The number of roots is $N|\Gamma|$, the length of $\Gamma$ times the number of slices, and is independent of the volume.  The constants $C_1,C_2$ are fixed before $\epsilon_0$ is chosen.

\subsection{The local gas with a macroscopic background}
\label{SM:sec:background}

Fix a history $\mathfrak h$ and let $\cM$ be the set of its macroscopic components, its \emph{background}, with total cost $m(\cM)=\sum_{C\in\cM}m(C)$, sections $s_\cM(k)=\bigsqcup_{C\in\cM}s_C(k)$, and net Pauli operator $Q_\cM$, the time-ordered product of the insertions of its members.  As noted before Lemma~\ref{SM:lem:emptybox}, $s_\cM(k)$ is valid and $\syn(Q_\cM)=\emptyset$.  Let $w_{\rm bare}(\cM)$ be the product over the members of $\cM$ of their coefficient products and Boltzmann factors; by Eq.~\eqref{SM:eq:eta}, Eq.~\eqref{SM:eq:t0eps}, and $\epsilon\le\bar\epsilon$,
\begin{equation}
|w_{\rm bare}(\cM)|\le\prod_{C\in\cM}q_\epsilon^{a(C)}e^{-2t_0J_{\min}d(C)}\le q_0^{m(\cM)} .
\label{SM:eq:wbare}
\end{equation}
If $Q_\cM=\zeta_\cM S_\cM$ is a check product times a phase, set $\lambda_\cM=\zeta_\cM\chi_{S_\cM}(s_\cM(0))$ and, for every local component $C$ of $\mathfrak h$,
\begin{equation}
\xi_C(\cM)=(-1)^{\kappa(C,\cM)}\,\chi_{S_C}(s_\cM(0))\,\chi_{S_\cM}(s_C(0)) ,
\label{SM:eq:xi}
\end{equation}
with $\kappa(C,\cM)$ as in Eq.~\eqref{SM:eq:kappa}; otherwise set $\lambda_\cM=0$ and $\xi_C(\cM)=1$.  Individual members of $\cM$ may carry nontrivial logical operators; only the net Pauli operator of the whole background enters.

\begin{lemma}[weight with background]
\label{SM:lem:weightbg}
The weight of $\mathfrak h$ is
\begin{equation}
w_{\rm bare}(\cM)\,\lambda_\cM\ \prod_{C}\omega(C)\prod_{C<D}\sigma(C,D)\prod_C\xi_C(\cM) ,
\label{SM:eq:weightbg}
\end{equation}
the products over the local components of $\mathfrak h$.
\end{lemma}

\begin{proof}
Group the letters $x_1,\dots,x_n$ of Sec.~\ref{SM:sec:pairwise} into the blocks $C_1,\dots,C_M$ (the local components) and $\cM$.  Extracting $C_1,\dots,C_M$ in turn to the right end as in Lemma~\ref{SM:lem:regroup}(i) gives $x_n\cdots x_1=\bigl[\prod_{\alpha<\beta}(-1)^{\kappa(C_\alpha,C_\beta)}\prod_\alpha(-1)^{\kappa(C_\alpha,\cM)}\bigr]Q_\cM Q_{C_M}\cdots Q_{C_1}$; the parity of each $\kappa(C,\cM)$ does not depend on the extraction order, by the argument of Lemma~\ref{SM:lem:regroup}(ii), because $Q_C=\zeta_CS_C$ commutes with $Q_\cM$ (Lemma~\ref{SM:lem:characters}(c), $\syn(Q_\cM)=\emptyset$).  Since every $S_C$ commutes with $Q_\cM$ and acts on the range of $\Pi_{s_0}$ as $\chi_{S_C}(s_0)$,
\begin{equation}
\frac1d\Tr_{s_0}(x_n\cdots x_1)=\Bigl[\prod_{\alpha<\beta}(-1)^{\kappa(C_\alpha,C_\beta)}\prod_\alpha(-1)^{\kappa(C_\alpha,\cM)}\Bigr]\prod_C\zeta_C\chi_{S_C}(s_0)\ \frac1d\Tr(\Pi_{s_0}Q_\cM) ,
\end{equation}
and by Lemma~\ref{SM:lem:sectortrace} the last factor is $\zeta_\cM\chi_{S_\cM}(s_0)$ if $Q_\cM=\zeta_\cM S_\cM$ and $0$ otherwise.  In the first case expand $s_0=s_\cM(0)\oplus\bigoplus_Ds_D(0)$ by Lemma~\ref{SM:lem:characters}(a): $\chi_{S_C}(s_0)=\chi_{S_C}(s_C(0))\prod_{D\ne C}\chi_{S_C}(s_D(0))\,\chi_{S_C}(s_\cM(0))$ and $\chi_{S_\cM}(s_0)=\chi_{S_\cM}(s_\cM(0))\prod_C\chi_{S_\cM}(s_C(0))$.  Together with the coefficient products and Boltzmann factors, which factorize over all components, the factors regroup into Eq.~\eqref{SM:eq:weightbg}.
\end{proof}

Fix now a background $\cM$, that is, a set of pairwise compatible macroscopic components of some history with $Q_\cM=\zeta_\cM S_\cM$.  The histories with background $\cM$ are obtained by adding a pairwise compatible family of local components each of which is compatible with every member of $\cM$, as in Lemma~\ref{SM:lem:gascomp}(ii), and by Lemma~\ref{SM:lem:weightbg} the sum of their weights is $w_{\rm bare}(\cM)\lambda_\cM\Zloc^\cM$, where $\Zloc^\cM$ is the gas of Eq.~\eqref{SM:eq:Zloc} on the same set $\Lambda$ with the activities
\begin{equation}
z_\gamma^{\cM}=\begin{cases}z_\gamma\prod_{C\in\gamma}\xi_C(\cM)&\text{if every member of $\gamma$ is compatible with every member of $\cM$,}\\[2pt]0&\text{otherwise.}\end{cases}
\label{SM:eq:zM}
\end{equation}
Here $\xi_C(\mathcal M)$ for a local component $C$ compatible with $\mathcal M$ is computed in the history formed by $C$ and $\mathcal M$.  The linking graph and the compatibility relation are still those of the original $\sigma$, $|z^\cM_\gamma|\le\rho_\gamma$, and $|z^\cM_\gamma-z_\gamma|\le2\rho_\gamma$.

\begin{lemma}[background comparison]
\label{SM:lem:background}
For every background $\cM$ with $Q_\cM=\zeta_\cM S_\cM$ and $\epsilon\le\bar\epsilon$,
\begin{equation}
\Bigl|\frac{\Zloc^{\cM}}{\Zloc}\Bigr|\le e^{\varkappa\,m(\cM)},\qquad\varkappa=2\Upsilon(q_0)\le\tfrac18 .
\label{SM:eq:background}
\end{equation}
\end{lemma}

\begin{proof}
Apply Lemma~\ref{SM:lem:interpolation} to $z$ and $z^\cM$.  If $z^\cM_\gamma\ne z_\gamma$, then either some member $C$ of $\gamma$ shares a node with a member of $\cM$ or touches one, in which case $\varrho(C,\cM)\le c_{\rm adj}$, or $\xi_C(\cM)=-1$ for some member $C$, in which case Lemma~\ref{SM:lem:emptybox}, applied in the history formed by $C$ and $\cM$ with $D=\cM$, gives a node of $\cM$ inside $\cB_C$, hence $\varrho(C,\cM)\le c_3m(C)$.  Either way $\gamma$ has a node within $\varrho$-distance $c_5m(\gamma)$ of one of the $m(\cM)$ nodes of $\cM$.  By Lemma~\ref{SM:lem:countlink} and the ball count,
\begin{equation}
\sum_\gamma|z^\cM_\gamma-z_\gamma|e^{m(\gamma)}\le2\,m(\cM)\sum_{m\ge1}C_*(m+1)^5K_{\rm link}^m(q_0e)^m=\varkappa\,m(\cM) .
\end{equation}
\end{proof}

The comparison is relative to the full local gas, and the background enters only through its cost $m(\cM)$.

\subsection{The physical partition function}
\label{SM:sec:return}

Write $\Zfull=e^{\beta E_{\rm ref}}Z(K_\Gamma)/d$ for the full history sum of Lemma~\ref{SM:lem:expansion}.  Grouping the histories by their background,
\begin{equation}
\Zfull=\Zloc+\sum_{\cM\ne\emptyset}w_{\rm bare}(\cM)\,\lambda_\cM\,\Zloc^\cM ,
\label{SM:eq:Zfullsplit}
\end{equation}
where $\cM$ runs over the nonempty backgrounds of histories; backgrounds whose net Pauli operator carries a nontrivial logical operator have $\lambda_\cM=0$.  Let $C_{\rm env}$ bound the number of nodes per slice divided by $L^4$: there are $8L^4$ checks and a number of labels proportional to $L^4$.

\begin{lemma}[macroscopic sum]
\label{SM:lem:macro}
Let $r=K_{\rm raw}e^\varkappa q_0$, which is below $\frac12$ by Eq.~\eqref{SM:eq:q0}, and $\delta_{L,N}=C_3NL^4r^{m_0}$ with $C_3=2C_{\rm env}$.  For $\epsilon\le\bar\epsilon$,
\begin{equation}
\Bigl|\frac{\Zfull}{\Zloc}-1\Bigr|\le e^{\delta_{L,N}}-1 ,
\label{SM:eq:macro}
\end{equation}
and if $\delta_{L,N}\le\frac14$ then $\bigl|\ln|\Zfull/\Zloc|\bigr|\le2\delta_{L,N}$.
\end{lemma}

\begin{proof}
By Eqs.~\eqref{SM:eq:Zfullsplit}, \eqref{SM:eq:wbare}, and \eqref{SM:eq:background}, $|\Zfull/\Zloc-1|\le\sum_{\cM\ne\emptyset}\prod_{C\in\cM}(q_0e^\varkappa)^{m(C)}$.  Enlarge the sum to all nonempty sets of macroscopic components, dropping the compatibility between them and the validity of their merged sections: it is then at most $\prod_C\bigl(1+(q_0e^\varkappa)^{m(C)}\bigr)-1\le\exp\bigl(\sum_C(q_0e^\varkappa)^{m(C)}\bigr)-1$, the product and the sum over all components $C$ of histories with $m(C)\ge m_0$.  Each such component contains one of the $C_{\rm env}NL^4$ nodes, and by Lemma~\ref{SM:lem:countcomp} at most $K_{\rm raw}^m$ of them of cost $m$ contain a given node, so $\sum_C(q_0e^\varkappa)^{m(C)}\le C_{\rm env}NL^4\sum_{m\ge m_0}r^m\le2C_{\rm env}NL^4r^{m_0}=\delta_{L,N}$.  For $0\le\delta\le\frac14$, $|x-1|\le e^\delta-1$ implies $-\ln(2-e^\delta)\le\ln|x|\le\delta$, and $-\ln(2-e^\delta)\le2\delta$ on this range because $\phi(\delta)=2\delta+\ln(2-e^\delta)$ vanishes at $0$ and has $\phi'(\delta)=2-e^\delta/(2-e^\delta)>0$ for $\delta\le\frac14$.
\end{proof}

\begin{proposition}[tension bound up to $\beta=A_*L^4$]
\label{SM:prop:ratio_bL}
Let $A_*>0$ and let $L$ be large enough that $C_3A_*L^8r^{m_0}/t_0\le\frac14$.  Then for $\epsilon\le\bar\epsilon$, $t_0\le\beta\le A_*L^4$, and every valid $\Gamma$ of one type,
\begin{equation}
\Bigl|\ln\frac{Z(K_\Gamma^F)}{Z(K_\Gamma)}\Bigr|\le C_2\beta\epsilon|\Gamma|+C_4\beta L^4r^{m_0},\qquad C_4=\frac{4C_3}{t_0} .
\label{SM:eq:ratio_bL}
\end{equation}
\end{proposition}

\begin{proof}
$Z(K_\Gamma)$ and $Z(K_\Gamma^F)$ are positive, and $\Zfull=e^{\beta E_{\rm ref}}Z(K_\Gamma)/d$, $\Zfull^F=e^{\beta E_{\rm ref}}Z(K_\Gamma^F)/d$ with the same prefactor, so
\begin{equation}
\ln\frac{Z(K_\Gamma^F)}{Z(K_\Gamma)}=\Bigl[\ln|\Zfull^F|-\ln|\Zloc^F|\Bigr]+\Bigl[\ln|\Zloc^F|-\ln|\Zloc|\Bigr]-\Bigl[\ln|\Zfull|-\ln|\Zloc|\Bigr] .
\end{equation}
Since $N\le\beta/t_0\le A_*L^4/t_0$, the hypothesis gives $\delta_{L,N}\le\frac14$, and Lemma~\ref{SM:lem:macro}, which applies to $K_\Gamma^F$ as well as to $K_\Gamma$ (the erased Hamiltonian has the same history structure with modified coefficients, Lemma~\ref{SM:lem:signflip}), bounds the first and the third bracket by $2\delta_{L,N}$ each; Lemma~\ref{SM:lem:localtwist} bounds the second by $C_2\beta\epsilon|\Gamma|$.  Finally $4\delta_{L,N}=4C_3NL^4r^{m_0}\le C_4\beta L^4r^{m_0}$.
\end{proof}

The auxiliary gases enter through their moduli only.  Equation~\eqref{SM:eq:ratio_bL} already holds at fixed $\epsilon$ for $\beta$ growing as a power of $L$, which is the regime below $T_\times$ of the main text, where thermal excitations are outnumbered by virtual ones.

\section{Entropy bridge and the static theorem}
\label{SM:sec:bridge}

Equation~\eqref{SM:eq:ratio_bL} covers $\beta\le A_*L^4$.  A one-line inequality between free energies and ground energies extends it to every lower temperature, yielding Eq.~(5) for all $T\le T_0$ at a fixed small cost paid by the bare tension.  This section records the bridge, the order in which all constants are chosen, and the resulting static theorem.

\subsection{The bridge inequality}
\label{SM:sec:bridgeineq}

\begin{lemma}[entropy bridge]
\label{SM:lem:bridge}
Let $A,B$ be Hermitian operators on a space of dimension $\mathsf D$, let $f_\beta(A)=-\beta^{-1}\ln Z_\beta(A)$ with $Z_\beta(A)=\Tr e^{-\beta A}$, and let $E_A=\min\operatorname{spec}A$.  Then
\begin{equation}
E_A-\frac{\ln\mathsf D}{\beta}\le f_\beta(A)\le E_A ,
\label{SM:eq:fsandwich}
\end{equation}
and if $|f_{\beta_*}(A)-f_{\beta_*}(B)|\le a$ for one $\beta_*>0$, then $|E_A-E_B|\le a+\ln\mathsf D/\beta_*$ and, for every $\beta\ge\beta_*$,
\begin{equation}
\frac1\beta\Bigl|\ln\frac{Z_\beta(A)}{Z_\beta(B)}\Bigr|\le a+\frac{2\ln\mathsf D}{\beta_*} ,
\label{SM:eq:bridge}
\end{equation}
which is Eq.~(B3).
\end{lemma}

\begin{proof}
$Z_\beta(A)=\sum_je^{-\beta\lambda_j(A)}$ lies between $e^{-\beta E_A}$ and $\mathcal De^{-\beta E_A}$, which is Eq.~\eqref{SM:eq:fsandwich}.  Then $E_A-E_B\le\bigl(f_{\beta_*}(A)+\ln\mathsf D/\beta_*\bigr)-f_{\beta_*}(B)\le a+\ln\mathsf D/\beta_*$, and the same with $A$ and $B$ exchanged.  For $\beta\ge\beta_*$, $f_\beta(A)-f_\beta(B)\le E_A-\bigl(E_B-\ln\mathsf D/\beta\bigr)\le a+\ln\mathsf D/\beta_*+\ln\mathsf D/\beta\le a+2\ln\mathsf D/\beta_*$, again symmetrically, and $\beta^{-1}\ln[Z_\beta(A)/Z_\beta(B)]=f_\beta(B)-f_\beta(A)$.
\end{proof}

\begin{figure}[t]
\centering
\includegraphics{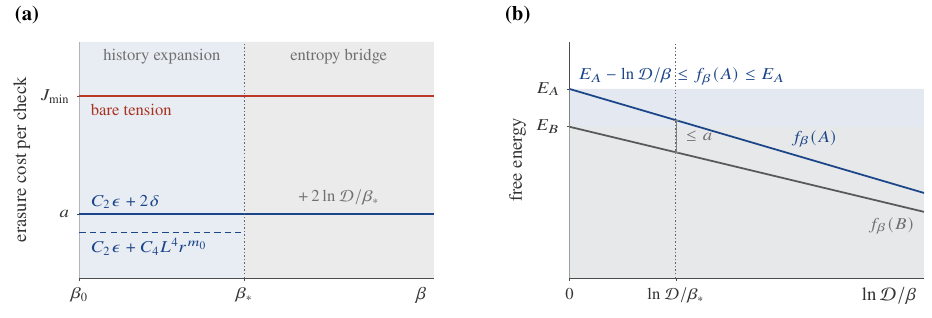}
\caption{The entropy bridge.  (a)~The cost of erasing $\Gamma$ per check as a function of $\beta$: the history expansion controls $\beta\le\beta_*=A_*L^4$ with the bound $C_2\epsilon+C_4L^4r^{m_0}$ per unit length, Eq.~\eqref{SM:eq:ratio_bL}, and the bridge extends the control to every larger $\beta$ at the fixed price $2\ln\mathsf D/\beta_*\le\delta$, Eq.~\eqref{SM:eq:tension_all}.  Both stay below the bare tension $J_{\min}$, so the threshold $\epsilon_0$ is independent of temperature.  (b)~The bridge inequality \eqref{SM:eq:bridge}: each free energy $f_\beta$ lies between the ground energy $E$ and $E-\ln\mathsf D/\beta$; a gap of at most $a$ at $\beta_*$ bounds $|E_A-E_B|$ by $a+\ln\mathsf D/\beta_*$, and hence the gap at every larger $\beta$.}
\label{SM:fig:S5}
\end{figure}

The inequality uses eigenvalue counting only [Fig.~\ref{SM:fig:S5}(b)].  Physically, below the crossover $T_\times$ of the main text the ground-state energy difference already carries the tension of $\Gamma$, and the thermal correction is bounded by the total entropy divided by $\beta_*$.

\subsection{Extension to all low temperatures}
\label{SM:sec:extension}

\begin{proposition}[tension bound, all low temperatures]
\label{SM:prop:tension_all}
Set $\delta=J_{\min}/64$, choose $A_*\ge\beta_0$ with $12\ln2/A_*\le\delta$, and then $L_0$ such that for every $L\ge L_0$: $C_4L^4r^{m_0}\le\delta$ and $C_3A_*L^8r^{m_0}/t_0\le\frac14$.  Then for $L\ge L_0$, $\epsilon\le\bar\epsilon$, $\beta\ge\beta_0$, and every valid $\Gamma$ of one type,
\begin{equation}
\frac1\beta\Bigl|\ln\frac{Z(K_\Gamma^F)}{Z(K_\Gamma)}\Bigr|\le(C_2\epsilon+2\delta)\,|\Gamma| .
\label{SM:eq:tension_all}
\end{equation}
This is Eq.~(5) and Eq.~\eqref{SM:eq:tension} with $C_{\rm tw}=C_2$.
\end{proposition}

\begin{proof}
Let $\beta_*=A_*L^4\ge\beta_0\ge t_0$.  For $\beta_0\le\beta\le\beta_*$, Proposition~\ref{SM:prop:ratio_bL} and $C_4L^4r^{m_0}\le\delta\le\delta|\Gamma|$ give the bound $(C_2\epsilon+\delta)|\Gamma|$.  For $\beta\ge\beta_*$ apply Lemma~\ref{SM:lem:bridge} with $A=K_\Gamma^F$, $B=K_\Gamma$, $\mathsf D=2^{6L^4}$, and $a=(C_2\epsilon+\delta)|\Gamma|$ at $\beta_*$: the correction is $2\ln\mathsf D/\beta_*=12L^4\ln2/(A_*L^4)=12\ln2/A_*\le\delta\le\delta|\Gamma|$ [Fig.~\ref{SM:fig:S5}(a)].
\end{proof}

The splitting point $\beta_*$ grows with $L$ and joins the two regimes; the threshold $\epsilon_0$ and the temperature $T_0=1/\beta_0$ are independent of $L$.  The bound has the form ``$C\epsilon$ plus a fixed margin $2\delta$'', and the margin is what the fixed-window reduction of Sec.~\ref{SM:sec:window} pays from the bare tension $J_{\min}$: with $\epsilon\le\epsilon_0$, $a=C_2\epsilon+2\delta\le J_{\min}/8+J_{\min}/32<J_{\min}$.

\subsection{Order of constants}
\label{SM:sec:constants}

All choices precede the actual $L$, $\beta$, and $V$, and none feeds back into an earlier one.
\begin{enumerate}
\item Fix the range, the label list, the collars, the dependency graph, the environment graph, the Pauli alphabet, and $c_0\le1/(16C_g)$.  These determine $c_1,c_3,c_5$, $C_{\rm ball}$, $C_*$, $K_{\rm raw}$, $K_{\rm link}$, and $C_{\rm env}$, all independent of $t_0$, $\beta$, and $L$.
\item Choose $q_0$ by Eq.~\eqref{SM:eq:q0}; this fixes $\varkappa=2\Upsilon(q_0)\le\frac18$ and $r=K_{\rm raw}e^\varkappa q_0<\frac12$.
\item Choose $t_0$ with $e^{-2t_0J_{\min}}\le q_0$, compute $C(t_0)=2t_0e^{2t_0b_R}$, and set $\bar\epsilon=\ln(1+q_0)/C(t_0)$.
\item Set $C_1,C_2$ by Eq.~\eqref{SM:eq:localtwist} and $C_3,C_4$ by Lemma~\ref{SM:lem:macro} and Eq.~\eqref{SM:eq:ratio_bL}.
\item Set $\delta=J_{\min}/64$, $\tau=\beta_0$ by Eq.~\eqref{SM:eq:eps0}, then $A_*\ge\beta_0$ with $12\ln2/A_*\le\delta$, then $L_0$ as in Proposition~\ref{SM:prop:tension_all}, and also $L_0\ge80$ and $\lfloor L_0/1000\rfloor\ge10$ for Secs.~\ref{SM:sec:localcomp} and \ref{SM:sec:geometry}.
\item With $w=6$, $b=4k_R$, $\vartheta$, and $C_{\rm act}$ fixed by the lattice and the range, set $\epsilon_0$ by Eq.~\eqref{SM:eq:eps0}.
\end{enumerate}
The norm constant $C(t_0)$ grows with the slice length and enters only after the geometric conditions of step~2 are fixed; the counting constants $K_{\rm raw},K_{\rm link}$ are independent of the slice length.  The thermal weight $q_0$ and the quantum activity $\bar\epsilon$ are therefore fixed one after the other, each from constants chosen before it.

\subsection{The static theorem}
\label{SM:sec:statictheorem}

\begin{theorem}[static theorem in four dimensions]
\label{SM:thm:static}
With $\beta_0,\epsilon_0,L_0,C$ as above, for all $L\ge L_0$, $\beta\ge\beta_0$, and every Hermitian finite-range $V$ with $\|V\|_\loc\le\epsilon_0$, the measurement of the ideal checks in $\rho\propto e^{-\beta(H_0+V)}$ satisfies
\begin{equation}
\Pr_\rho[\text{a cluster of either type with}\ge u\text{ checks}]\le CL^{20}e^{-4u} .
\end{equation}
\end{theorem}

\begin{proof}
Proposition~\ref{SM:prop:tension_all} supplies Eq.~\eqref{SM:eq:tension} for every valid $\Gamma$ of one type, with $a=C_2\epsilon+2\delta$; Proposition~\ref{SM:prop:marked} then gives Eq.~\eqref{SM:eq:marked2} for every such $\Gamma$, and Proposition~\ref{SM:prop:static4d} sums over clusters.
\end{proof}

The theorem admits arbitrary signs, $Y$ operators, and mixed local matrices in $V$, and non-uniform positive couplings in $H_0$.  For a fixed $\epsilon\le\epsilon_0$ the temperature window is $0<T\le1/\beta_0$, and the threshold is the same at every temperature in it: this is the rectangle of Fig.~2(a).  In the two-dimensional toric code a single thermal point-like excitation is an invalid section on its own, so the factorization into local components of Sec.~\ref{SM:sec:histories} fails at its first step; this is where the proof stops for the two-dimensional toric code.

\section{The memory bound}
\label{SM:sec:memory}

This section proves the memory bound Eq.~(3) from Theorem~\ref{SM:thm:static} and Lemma~\ref{SM:lem:decoder}.  The argument is a stationarity argument on the full logical algebra: a thermally encoded state is close to the Gibbs state in the sense that every local move of the dynamics acts on it, up to the bad set, exactly as it acts on $\rho$, and $\rho$ does not move at all.  The argument is independent of the history expansion, admits time-dependent generators with a common stationary state, and keeps the output on the physical system, so that any later recovery can be appended.  The section closes with the class of dynamics for which the bound applies and a concrete member of it.  Throughout, $\|\cT\|_{1\to1}$ is the induced trace norm of a linear map on operators, and a completely positive trace-preserving map has $\|\cT\|_{1\to1}\le\|\cT\|_\diamond=1$.

\subsection{Thermal encoding and decoder}
\label{SM:sec:encoding}

With $W$, $G$, $B=I-G$ from Sec.~\ref{SM:sec:geometry}, let $p=\Tr(\rho B)$ and
\begin{equation}
\rho_S=\Tr_{\log}(W^\dagger\rho W),\qquad
\cE(X)=W(X\otimes\rho_S)W^\dagger,\qquad
\cD(Y)=\Tr_S(W^\dagger YW),
\label{SM:eq:encdec}
\end{equation}
which is Eq.~(C1).  The \emph{logical algebra} is $\mathfrak A_{\log}=\{W(M\otimes I_S)W^\dagger\}$, the operators that act on the logical factor only in the decoder basis.

\begin{lemma}[encoding and decoder]
\label{SM:lem:encdec}
(i) $\cE$ and $\cD$ are completely positive and trace preserving, and $\cD\cE=\id_d$.  (ii) For every reference system and every operator $\Omega$ on logical$\,\otimes\,$reference, $\|(\cE\otimes\id)(\Omega)\|_1=\|\Omega\|_1$.  (iii) For a logical state $\varphi$ the map
\begin{equation}
\cF_\varphi(Y)=W\bigl[\varphi\otimes\Tr_{\log}(W^\dagger YW)\bigr]W^\dagger
\label{SM:eq:Fomega}
\end{equation}
is completely positive and trace preserving, satisfies $\cF_\varphi(\rho)=\cE(\varphi)$, and has Kraus operators in $\mathfrak A_{\log}$.
\end{lemma}

\begin{proof}
(i) $\cE$ is a tensor product with a state followed by a unitary conjugation, and $\cD$ is a unitary conjugation followed by a partial trace; both are completely positive and trace preserving.  $\cD\cE(X)=\Tr_S(X\otimes\rho_S)=X$.  (ii) $(\cE\otimes\id)(\Omega)$ is the conjugate by the unitary $W\otimes I_{\rm ref}$ of $\Omega\otimes\rho_S$, with $\rho_S$ placed in the syndrome factor on which $W$ acts; the trace norm is invariant under unitary conjugation, and $\|\Omega\otimes\rho_S\|_1=\|\Omega\|_1\|\rho_S\|_1=\|\Omega\|_1$.  (iii) Write $\varphi=\sum_j\lambda_j|e_j\rangle\langle e_j|$ and $\Tr_{\log}(Z)=\sum_\mu(\langle\mu|\otimes I_S)Z(|\mu\rangle\otimes I_S)$ over an orthonormal basis of the logical factor.  Then $\cF_\varphi(Y)=\sum_{j,\mu}K_{j\mu}YK_{j\mu}^\dagger$ with $K_{j\mu}=W(\sqrt{\lambda_j}\,|e_j\rangle\langle\mu|\otimes I_S)W^\dagger\in\mathfrak A_{\log}$, and $\sum_{j\mu}K_{j\mu}^\dagger K_{j\mu}=W(\sum_j\lambda_j\sum_\mu|\mu\rangle\langle e_j|e_j\rangle\langle\mu|\otimes I_S)W^\dagger=I$.  Finally $\Tr_{\log}(W^\dagger\rho W)=\rho_S$ gives $\cF_\varphi(\rho)=\cE(\varphi)$.
\end{proof}

The auxiliary state $\rho_S$ may carry coherence between good and bad patterns, and the decoder traces it out.  The encoding depends on $\beta$ and $V$ through $\rho_S$; the decoder depends on the ideal checks only.

\subsection{Commutation of the replacement map with local maps}
\label{SM:sec:replacement}

A linear map $\cT$ on operators is \emph{supported in a region} $A$ when $\cT=\cT_A\otimes\id_{A^c}$ for a linear map $\cT_A$ on the operators of $A$.  Since the maps $Y_A\mapsto aY_Ab$ span the linear maps on the operators of $A$, every such $\cT$ can be written $\cT(Y)=\sum_\nu A_\nu YB_\nu$ with $A_\nu,B_\nu$ supported in $A$.

\begin{lemma}[commutation of the replacement map]
\label{SM:lem:replacement}
For every logical state $\varphi$, every linear map $\cT$ supported in an allowed region, and every operator $Y$,
\begin{equation}
\cT\cF_\varphi(GYG)=\cF_\varphi \cT(GYG) .
\label{SM:eq:replacement}
\end{equation}
\end{lemma}

\begin{proof}
With the Kraus operators $K_{j\mu}\in\mathfrak A_{\log}$ of Lemma~\ref{SM:lem:encdec}(iii) and $\cT(Y)=\sum_\nu A_\nu YB_\nu$,
$\cT\cF_\varphi(GYG)=\sum_{\nu,j,\mu}A_\nu K_{j\mu}\,GYG\,K_{j\mu}^\dagger B_\nu$.  Lemma~\ref{SM:lem:decoder}, Eq.~\eqref{SM:eq:onesided}, applied to the local operator $A_\nu$ and the logical operator $K_{j\mu}$, gives $A_\nu K_{j\mu}G=K_{j\mu}A_\nu G$; applied to $B_\nu^\dagger$ and $K_{j\mu}$ and then taking adjoints, it gives $GK_{j\mu}^\dagger B_\nu=GB_\nu K_{j\mu}^\dagger$.  Hence every summand equals $K_{j\mu}\,A_\nu GYGB_\nu\,K_{j\mu}^\dagger$, and summing over $\nu$ first gives $\cF_\varphi \cT(GYG)$.
\end{proof}

\begin{lemma}[gentle measurement]
\label{SM:lem:gentle}
For every state $\rho$ and projector $G$ with $p=\Tr(\rho(I-G))$, $\|\rho-G\rho G\|_1\le2\sqrt p$.
\end{lemma}

\begin{proof}
With $B=I-G$, $\rho-G\rho G=B\rho+G\rho B$, so $\|\rho-G\rho G\|_1\le\|B\rho\|_1+\|G\rho B\|_1\le2\|B\rho\|_1$, using $\|G\rho B\|_1\le\|G\|\,\|\rho B\|_1$ and $\|\rho B\|_1=\|B\rho\|_1$.  By the Cauchy--Schwarz inequality for Schatten norms, $\|B\rho\|_1=\|B\rho^{1/2}\rho^{1/2}\|_1\le\|B\rho^{1/2}\|_2\|\rho^{1/2}\|_2=\sqrt{\Tr(\rho^{1/2}B\rho^{1/2})}\cdot1=\sqrt p$.
\end{proof}

\begin{corollary}[commutator bound]
\label{SM:cor:commbound}
For every logical state $\varphi$ and every linear map $\cT$ supported in an allowed region,
\begin{equation}
\|[\cT,\cF_\varphi](\rho)\|_1\le2\|\cT\|_{1\to1}\|\rho-G\rho G\|_1\le4\|\cT\|_{1\to1}\sqrt p .
\label{SM:eq:commbound}
\end{equation}
\end{corollary}

\begin{figure}[t]
\centering
\includegraphics{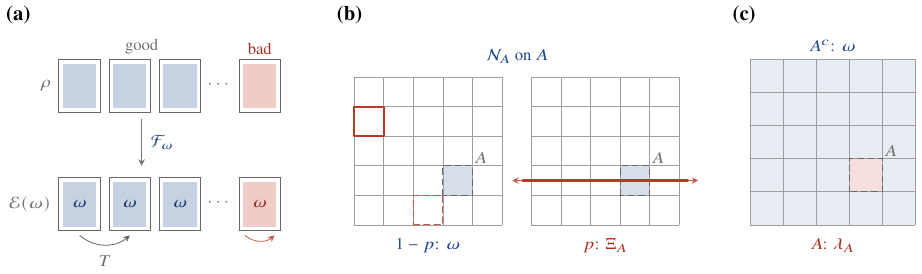}
\caption{Stationarity and local noise.  (a)~In the decoder basis the Gibbs state is a row of logical blocks, one per pattern.  The replacement map $\cF_\varphi$ sets every logical block to $\varphi$ and produces the encoded state; a local map $\cT$ moves patterns and, on good patterns, leaves the logical block untouched, Eq.~\eqref{SM:eq:replacement}, so $\mathcal L_s\mathcal E(\varphi)$ is bounded by the weight $\sqrt p$ of the bad patterns, Eq.~\eqref{SM:eq:rate}.  (b)~Noise $\mathcal N_A$ on an allowed region $A$: on patterns with small loops only (weight $1-p$) the decoder returns $\varphi$; the one failure is a loop threading $A$ and spanning the system (weight $p$), Eq.~\eqref{SM:eq:convex}.  (c)~The same encoding as a secret-sharing scheme: an observer of $A$ sees a fixed state $\lambda_A$ up to $2p$, and the complement recovers $\varphi$ up to $2p$, Eqs.~\eqref{SM:eq:erasure} and \eqref{SM:eq:leakage}.}
\label{SM:fig:S6}
\end{figure}

\begin{proof}
By Lemma~\ref{SM:lem:replacement}, $[\cT,\cF_\varphi](G\rho G)=0$, so $[\cT,\cF_\varphi](\rho)=[\cT,\cF_\varphi](\rho-G\rho G)$.  For every operator $Z$, $\|[\cT,\cF_\varphi](Z)\|_1\le\|\cT\|_{1\to1}\|\cF_\varphi\|_{1\to1}\|Z\|_1+\|\cF_\varphi\|_{1\to1}\|\cT\|_{1\to1}\|Z\|_1=2\|\cT\|_{1\to1}\|Z\|_1$, since $\cF_\varphi$ is completely positive and trace preserving.  Lemma~\ref{SM:lem:gentle} completes the bound [Fig.~\ref{SM:fig:S6}(a)].
\end{proof}

\subsection{Time-dependent dynamics with a common stationary state}
\label{SM:sec:dynamics}

Let $s\mapsto\cL_s$ be a locally integrable family of Lindblad generators, so that the propagator $\Phi(t,s)$, defined by $\partial_t\Phi(t,s)=\cL_t\Phi(t,s)$ and $\Phi(s,s)=\id$, is completely positive and trace preserving for $t\ge s$.  Decompose
\begin{equation}
\cL_s=\sum_jK_{j,s}+\cL_s^{\rm tail},\qquad S(s)=\sum_j\|K_{j,s}\|_{1\to1},\qquad u(s)=\|\cL_s^{\rm tail}\|_{1\to1},\qquad \nu(s)=\|\cL_s(\rho)\|_1 ,
\label{SM:eq:Ldecomp}
\end{equation}
with every $K_{j,s}$ a linear map supported in an allowed region.  Complete positivity and stationarity of $\rho$ are required of the sum alone.

\begin{lemma}[rate bound]
\label{SM:lem:rate}
For every logical state $\varphi$ and every $s$,
\begin{equation}
\|\cL_s\cE(\varphi)\|_1\le b(s):=4S(s)\sqrt p+2u(s)+\nu(s) .
\label{SM:eq:rate}
\end{equation}
\end{lemma}

\begin{proof}
By Lemma~\ref{SM:lem:encdec}(iii), $\cL_s\cE(\varphi)=\cL_s\cF_\varphi(\rho)=[\cL_s,\cF_\varphi](\rho)+\cF_\varphi\cL_s(\rho)$.  The commutator splits along Eq.~\eqref{SM:eq:Ldecomp}: each $\|[K_{j,s},\cF_\varphi](\rho)\|_1\le4\|K_{j,s}\|_{1\to1}\sqrt p$ by Corollary~\ref{SM:cor:commbound}, and $\|[\cL_s^{\rm tail},\cF_\varphi](\rho)\|_1\le2\|\cL_s^{\rm tail}\|_{1\to1}\|\rho\|_1=2u(s)$ by the same two-term estimate.  Finally $\|\cF_\varphi\cL_s(\rho)\|_1\le\|\cL_s(\rho)\|_1=\nu(s)$.
\end{proof}

\begin{lemma}[state bound]
\label{SM:lem:statebound}
With $\mathfrak B(t)=\int_0^tb(s)\,ds$,
\begin{equation}
\sup_\varphi\|\Phi(t,0)\cE(\varphi)-\cE(\varphi)\|_1\le\min\{2,\mathfrak B(t)\} .
\label{SM:eq:statebound}
\end{equation}
\end{lemma}

\begin{proof}
For fixed $X$, $\partial_s[\Phi(t,s)X]=-\Phi(t,s)\cL_sX$, so integrating from $s=0$ to $s=t$ gives Duhamel's formula $\Phi(t,0)X-X=\int_0^t\Phi(t,s)\cL_sX\,ds$.  With $X=\cE(\varphi)$, trace-norm contraction of $\Phi(t,s)$ and Lemma~\ref{SM:lem:rate} give $\|\Phi(t,0)\cE(\varphi)-\cE(\varphi)\|_1\le\int_0^t\|\cL_s\cE(\varphi)\|_1ds\le\mathfrak B(t)$; the difference of two states has trace norm at most $2$.
\end{proof}

\begin{lemma}[diamond-norm lift]
\label{SM:lem:diamondlift}
Let $\Xi$ be a Hermiticity-preserving linear map from the operators of a $d$-dimensional space to operators, with $\|\Xi(\varphi)\|_1\le b$ for every state $\varphi$.  Then $\|\Xi\|_\diamond\le2db$.
\end{lemma}

\begin{proof}
Let $E_{ij}=|i\rangle\langle j|$ be the matrix units of an orthonormal basis.  $E_{ii}$ is a state, so $\|\Xi(E_{ii})\|_1\le b$.  For $i\ne j$ let $\psi_\pm=(|i\rangle\pm|j\rangle)/\sqrt2$ and $\psi_{\pm i}=(|i\rangle\pm i|j\rangle)/\sqrt2$; then $E_{ij}=\tfrac12\bigl(|\psi_+\rangle\langle\psi_+|-|\psi_-\rangle\langle\psi_-|\bigr)+\tfrac i2\bigl(|\psi_{+i}\rangle\langle\psi_{+i}|-|\psi_{-i}\rangle\langle\psi_{-i}|\bigr)$, a combination of four states with coefficients of modulus $\frac12$, so $\|\Xi(E_{ij})\|_1\le2b$.  Let $\Omega\ge0$ be an operator on the space tensored with a reference, and write $\Omega=\sum_{ij}E_{ij}\otimes\Omega_{ij}$ with $\Omega_{ij}=(\langle i|\otimes I)\Omega(|j\rangle\otimes I)$ and $q_i=\Tr\Omega_{ii}$.  Positivity of $\Omega$ gives $\Omega_{ij}=\Omega_{ii}^{1/2}K_{ij}\Omega_{jj}^{1/2}$ with $\|K_{ij}\|\le1$, hence $\|\Omega_{ij}\|_1\le\|\Omega_{ii}^{1/2}\|_2\|\Omega_{jj}^{1/2}\|_2=\sqrt{q_iq_j}$, and
\begin{equation}
\|(\Xi\otimes\id)(\Omega)\|_1\le\sum_{ij}\|\Xi(E_{ij})\|_1\|\Omega_{ij}\|_1\le2b\Bigl(\sum_i\sqrt{q_i}\Bigr)^2\le2bd\sum_iq_i=2bd\,\Tr\Omega .
\end{equation}
A Hermitian $H$ is $H_+-H_-$ with $H_\pm\ge0$ and $\Tr H_++\Tr H_-=\|H\|_1$, so $\|(\Xi\otimes\id)(H)\|_1\le2bd\|H\|_1$, using that $\Xi\otimes\id$ preserves Hermiticity.  For a general $X$, the Hermitian dilation $\tilde X=\begin{pmatrix}0&X\\X^\dagger&0\end{pmatrix}$ on the space tensored with one more qubit satisfies $\|\tilde X\|_1=2\|X\|_1$ and $(\Xi\otimes\id)(\tilde X)=\begin{pmatrix}0&(\Xi\otimes\id)(X)\\(\Xi\otimes\id)(X)^\dagger&0\end{pmatrix}$, whose trace norm is $2\|(\Xi\otimes\id)(X)\|_1$; hence $\|(\Xi\otimes\id)(X)\|_1\le2bd\|X\|_1$ for every $X$ and every reference.
\end{proof}

\begin{proposition}[memory bound]
\label{SM:prop:memory}
For every $t\ge0$,
\begin{equation}
\|\Phi(t,0)\cE-\cE\|_\diamond\le\Delta(t):=\min\{2,\,2d\,\mathfrak B(t)\},\qquad
\|\cD\Phi(t,0)\cE-\id_d\|_\diamond\le\Delta(t) .
\label{SM:eq:diamond}
\end{equation}
This is Eq.~(C2).
\end{proposition}

\begin{proof}
$\Xi=\Phi(t,0)\cE-\cE$ is a difference of completely positive maps, hence Hermiticity preserving, and $\|\Xi(\varphi)\|_1\le\min\{2,\mathfrak B(t)\}$ for every state by Lemma~\ref{SM:lem:statebound}.  Lemma~\ref{SM:lem:diamondlift} gives $\|\Xi\|_\diamond\le2d\,\mathfrak B(t)$, and $\|\Xi\|_\diamond\le\|\Phi(t,0)\cE\|_\diamond+\|\cE\|_\diamond=2$.  For the second bound, $\cD\Phi(t,0)\cE-\id_d=\cD\,\Xi$ by $\cD\cE=\id_d$, and $\|\cD\Xi\|_\diamond\le\|\cD\|_\diamond\|\Xi\|_\diamond=\|\Xi\|_\diamond$.
\end{proof}

The factor $d$ is the logical dimension, $d=64$ in four dimensions, and the bound is linear in $t$.  The first inequality controls the physical output before decoding, which is what Sec.~\ref{SM:sec:operational} needs for recovery after storage.  If the generators preserve $\rho$ exactly, $\nu\equiv0$; if only approximately, the accumulated residual $2d\int_0^t\nu(s)ds$ is the explicit price: a bath whose stationary state differs from $\rho$ at a fixed rate $\nu_*$ adds $2dt\nu_*$, and the exponential lifetime survives as long as this residual stays small over the storage time.

\subsection{The four-dimensional constants}
\label{SM:sec:constants4d}

\begin{proposition}[memory bound in four dimensions]
\label{SM:prop:memory4d}
Let $r=c_rL$ with $c_r=1/8000$, so that every box of radius $r$ lies in an allowed region for $L\ge L_0$, and suppose the dynamics satisfies, for all times,
\begin{equation}
S(s)\le C_SL^4,\qquad u(s)\le C_TL^4e^{-\mu r},\qquad\nu(s)=0 .
\label{SM:eq:dynclass}
\end{equation}
Then for $L\ge L_0$, $\beta\ge\beta_0$, and $\|V\|_\loc\le\epsilon_0$,
\begin{equation}
b(s)\le C_bL^{14}e^{-aL},\qquad a=\min\{1/500,\ \mu c_r\},\qquad
\Delta_L(t)\le\min\{2,\,C_\Delta L^{14}te^{-aL}\},\qquad C_\Delta=2dC_b ,
\label{SM:eq:memory4d}
\end{equation}
which is Eq.~(3) with $c'=14$.
\end{proposition}

\begin{proof}
A pattern lies in the bad set when some cluster has diameter at least $\ell_0$, which requires at least $\ell_0$ checks in that cluster, so Theorem~\ref{SM:thm:static} gives $p=\Tr(\rho B)\le CL^{20}e^{-4\ell_0}$ and $\sqrt p\le\sqrt C\,L^{10}e^{-2\ell_0}$.  With $\ell_0=\lfloor L/1000\rfloor\ge L/1000-1$, $e^{-2\ell_0}\le e^2e^{-L/500}$.  Inserting Eq.~\eqref{SM:eq:dynclass} into Eq.~\eqref{SM:eq:rate},
\begin{equation}
b(s)\le4C_SL^4\sqrt C\,e^2L^{10}e^{-L/500}+2C_TL^4e^{-\mu c_rL}\le C_bL^{14}e^{-aL},\qquad C_b=4e^2\sqrt C\,C_S+2C_T ,
\end{equation}
and Proposition~\ref{SM:prop:memory} with $\mathfrak B(t)\le C_bL^{14}te^{-aL}$ gives the rest.
\end{proof}

For $t\le e^{aL/2}$ the error vanishes as $L\to\infty$.  The exponent $14$ arises as $4+10$: the total strength $L^4$ times $\sqrt p\sim L^{10}e^{-2\ell_0}$.  In the general theorem of Sec.~\ref{SM:sec:general} the same count gives $D(h+3)/2$.

\subsection{The class of dynamics and a Gibbs sampler in it}
\label{SM:sec:dynclass}

The memory bound applies to every family of Lindblad generators with $\rho$ as common stationary state, $O(L^4)$ terms of bounded rate, and spatial tails as in Eq.~\eqref{SM:eq:dynclass}.  The class contains generators without detailed balance and generators that change with time, and each site is driven at a rate of order one.

A concrete member is the Gaussian-filtered exact Gibbs sampler of Chen, Kastoryano, and Gily\'en~\cite{CKG2023}.  For a single-site Hermitian Pauli $A$ and $A_H(t)=e^{iHt}Ae^{-iHt}$, let $\widehat A_H(\varpi)$ be the Fourier transform of $f_\beta(t)A_H(t)$ with $\int|f_\beta|^2=1$, and $0\le\gamma_\beta\le1$ the Gaussian acceptance.  The generator
\begin{equation}
\cL^{(A)}_H(X)=-i[B_H,X]+\mathcal V_H(X)-\tfrac12\{\mathcal V_H^\dagger(I),X\},\qquad
\mathcal V_H(X)=\int\gamma_\beta(\varpi)\widehat A_H(\varpi)X\widehat A_H(\varpi)^\dagger d\varpi ,
\label{SM:eq:CKG}
\end{equation}
with the coherent correction $B_H$ of Ref.~\cite{CKG2023}, preserves $\rho$ exactly.  Three properties of this construction are taken from Ref.~\cite{CKG2023}: (a) the operator Plancherel identity $\int\widehat A_H(\varpi)^\dagger\widehat A_H(\varpi)\,d\varpi=A^\dagger A=I$; (b) the time-domain form of the coherent term, $B_H=\int b_1(t)b_2(t')A_H(\beta(t'-t))^\dagger A_H(-\beta(t'+t))\,dt\,dt'$, with $b_1$ the convolution of $\operatorname{sech}(2\pi t)$ with $\sin(-t)e^{-2t^2}$ and $b_2$ a Gaussian times a phase, so that $b_1b_2$ has finite exponential moments of every order below $2\pi$; (c) exact stationarity of $\rho$ under Eq.~\eqref{SM:eq:CKG}.

\begin{proposition}[uniform rate and quasi-locality]
\label{SM:prop:CKG}
For finite-range $H$ of bounded local strength and $\beta$ in a fixed window $[\beta_0,\beta_1]$, each term \eqref{SM:eq:CKG} has $\|\cL^{(A)}_H\|_{1\to1}\le2+2C_B$, and with $H_r$ the Hamiltonian restricted to a box of radius $r$ around $A$,
\begin{equation}
\|\cL^{(A)}_H-\cL^{(A)}_{H_r}\|_{1\to1}\le C_{\beta_1}e^{-\mu r},
\end{equation}
with constants independent of the number of spins.
\end{proposition}

\begin{proof}
By (a), $0\le\mathcal V_H^\dagger(I)=\int\gamma_\beta\widehat A_H^\dagger\widehat A_H\,d\varpi\le I$, so the dissipative part $X\mapsto\mathcal V_H(X)-\tfrac12\{\mathcal V_H^\dagger(I),X\}$ has induced trace norm at most $2$: $\mathcal V_H$ is completely positive with $\|\mathcal V_H\|_{1\to1}=\|\mathcal V_H^\dagger(I)\|\le1$, and the anticommutator term has norm at most $\|\mathcal V_H^\dagger(I)\|\le1$.  By (b), $\|B_H\|\le C_B=\|b_1\|_1\|b_2\|_1$, and $\|[B_H,\cdot]\|_{1\to1}\le2C_B$.  For the tails, let $(\mathsf W_H\psi)(\varpi)=\sqrt{\gamma_\beta(\varpi)}\,\widehat A_H(\varpi)\psi$ be the Stinespring operator of $\mathcal V_H$, so that $\mathcal V_H(X)=\mathsf W_H^\dagger(X\otimes I)\mathsf W_H$ with $\|\mathsf W_H\|\le1$ by (a), and likewise $\mathsf W_{H_r}$ for $H_r$.  Adding and subtracting $\mathsf W_H^\dagger(X\otimes I)\mathsf W_{H_r}$,
\begin{equation}
\|\mathcal V_H(X)-\mathcal V_{H_r}(X)\|_1\le\|\mathsf W_H-\mathsf W_{H_r}\|\bigl(\|\mathsf W_H\|+\|\mathsf W_{H_r}\|\bigr)\|X\|_1\le2\|\mathsf W_H-\mathsf W_{H_r}\|\,\|X\|_1 ,
\end{equation}
and the compensators differ by $\|\mathcal V_H^\dagger(I)-\mathcal V_{H_r}^\dagger(I)\|\le2\|\mathsf W_H-\mathsf W_{H_r}\|$.  By Plancherel, $\|(\mathsf W_H-\mathsf W_{H_r})\psi\|^2\le\int|f_\beta(u)|^2\|(A_H(u)-A_{H_r}(u))\psi\|^2du$.  A Lieb--Robinson bound~\cite{LiebRobinson1972} on the patch gives $\|A_H(u)-A_{H_r}(u)\|\le C_0e^{-a(r-v|u|)}$ for $|u|\le r/2v$, and the trivial bound $2$ otherwise, so
\begin{equation}
\|\mathsf W_H-\mathsf W_{H_r}\|^2\le C_0^2e^{-ar}\int_{|u|\le r/2v}|f_\beta(u)|^2du+4\int_{|u|>r/2v}|f_\beta(u)|^2du\le Ce^{-\mu_1r},
\end{equation}
because $f_\beta$ is a Gaussian of width proportional to $\beta\le\beta_1$, whose tail beyond $r/2v$ is bounded by $e^{-\mu_1r}$ with $\mu_1$ depending on $\beta_1$ and $v$.  For the coherent term, split the double integral in (b) at $|t|+|t'|=r/(2v\beta_1)$: inside, both time arguments are at most $r/2v$ in modulus, and the patch estimate bounds the difference of the integrands by $2C_0e^{-ar/2}|b_1(t)b_2(t')|$; outside, each integrand has norm at most $|b_1(t)b_2(t')|$, and the finite exponential moments of $b_1b_2$ bound the tail by $Ce^{-\mu_2r/\beta_1}$.  Hence $\|B_H-B_{H_r}\|\le Ce^{-\mu r}$ and $\|[B_H-B_{H_r},\cdot]\|_{1\to1}\le2Ce^{-\mu r}$.  All constants are set by the kernels, the Lieb--Robinson velocity, and the range.
\end{proof}

Summing one such term per site and Pauli gives $O(L^4)$ total strength and $O(L^4)e^{-\mu r}$ truncation error, so Eq.~\eqref{SM:eq:dynclass} holds; adding the Hamiltonian evolution $-i[H,\cdot]$ preserves $\rho$ and has a finite-range decomposition.  Time-dependent schedules $\cL_s=\sum_ac_a(s)\cL^{(a)}-if(s)[H,\cdot]$ with bounded, possibly site-dependent weights $c_a(s)\ge c_->0$ belong to the class as well.  Including all single-site $X,Y,Z$ jumps makes the finite-system Gibbs state the unique stationary state and excludes purely imaginary eigenvalues of the generator, by the standard argument that every fixed operator of the adjoint commutes with all jumps; the semigroup therefore thermalizes, and the memory bound describes slow logical relaxation inside a dynamics that eventually reaches $\rho$.  The jump intensity has a positive lower bound uniform in $L$ on each temperature window, by the operator-valued Plancherel identity $\int\varpi^2\widehat A_H^\dagger\widehat A_H\,d\varpi\preceq(\beta_0^{-2}+\|[H,A]\|^2)I$, which keeps at least half of the Fourier mass at $|\varpi|\le\Omega=\sqrt2(\beta_0^{-2}+\|[H,A]\|^2)^{1/2}$.

The static threshold $\epsilon_0$ is uniform in temperature; the constants $C_\Delta,a$ of the dynamical bound inherit the temperature dependence of the tail constants $\mu,C_T$ of the specific bath.  For the sampler above these constants are uniform on every fixed window $[\beta_0,\beta_1]$, and the bound of Eq.~\eqref{SM:eq:memory4d} holds on each such window with constants independent of $L$.

\section{Operational consequences}
\label{SM:sec:operational}

Everything in this section follows from the one-sided identity Eq.~\eqref{SM:eq:onesided}, the encoding Eq.~\eqref{SM:eq:encdec}, and the two numbers $p$ and $\Delta(t)$.  The identity gives an exact convex decomposition of the logical channel induced by any noise on an allowed region, Eq.~(7), with bad-branch weight $p$ instead of the $\sqrt p$ of the stationarity argument.  From it follow erasure recovery with the ideal decoder, a leakage bound, a long-lived coherent family of encoded states, and a mixing obstruction.  Throughout, the adjoint $\cT^\dagger$ of a linear map on operators is taken with respect to the Hilbert--Schmidt product, $\Tr(X^\dagger\cT(Y))=\Tr(\cT^\dagger(X)^\dagger Y)$; the adjoint of a completely positive trace-preserving map is completely positive and unital, and $\cD^\dagger(M)=\bar M$ for every logical operator, since $\Tr\bigl(M^\dagger\Tr_S(W^\dagger YW)\bigr)=\Tr\bigl((M^\dagger\otimes I)W^\dagger YW\bigr)=\Tr\bigl((W(M\otimes I)W^\dagger)^\dagger Y\bigr)$.

\subsection{Convex decomposition of local noise}
\label{SM:sec:convex}

With allowed regions as defined in Sec.~\ref{SM:sec:geometry}, let $G_S^\perp=I-G_S$ and, if $p>0$,
\begin{equation}
\rho_{S,B}=G_S^\perp\rho_SG_S^\perp/p,\qquad\cE_B(X)=W(X\otimes\rho_{S,B})W^\dagger .
\label{SM:eq:badencoding}
\end{equation}

\begin{lemma}[bad-branch encoding]
\label{SM:lem:badenc}
$p=\Tr(\rho_SG_S^\perp)$, $\rho_{S,B}$ is a state, $\cE_B$ is completely positive and trace preserving with $\cD\cE_B=\id_d$, and $B\cE(X)B=p\,\cE_B(X)$ for every $X$.
\end{lemma}

\begin{proof}
$p=\Tr(\rho B)=\Tr\bigl(W^\dagger\rho W(I\otimes G_S^\perp)\bigr)=\Tr(\rho_SG_S^\perp)$ by the definition of $\rho_S$ as a partial trace.  $G_S^\perp\rho_SG_S^\perp\ge0$ has trace $p$, so $\rho_{S,B}$ is a state, and $\cE_B$ has the form of $\cE$ with $\rho_S$ replaced by $\rho_{S,B}$, so Lemma~\ref{SM:lem:encdec}(i) applies.  Finally $B=W(I\otimes G_S^\perp)W^\dagger$ gives $B\cE(X)B=W\bigl(X\otimes G_S^\perp\rho_SG_S^\perp\bigr)W^\dagger=p\,\cE_B(X)$.
\end{proof}

\begin{proposition}[local noise]
\label{SM:prop:noise}
For every completely positive trace-preserving $\cN_A$ supported on an allowed region $A$,
\begin{equation}
\cD\cN_A\cE=(1-p)\,\id_d+p\,\Xi_A,\qquad\Xi_A=\cD\cN_A\cE_B ,
\label{SM:eq:convex}
\end{equation}
with $\Xi_A$ a channel; hence $\|\cD\cN_A\cE-\id_d\|_\diamond\le2p$.  For $p=0$ the right side is $\id_d$.
\end{proposition}

\begin{proof}
Let $O_\alpha$ be Kraus operators of $\cN_A$, supported in $A$, with $\sum_\alpha O_\alpha^\dagger O_\alpha=I$, so that $\cN_A^\dagger(Y)=\sum_\alpha O_\alpha^\dagger YO_\alpha$.  For a logical operator $\bar M$, Eq.~\eqref{SM:eq:onesided} with $O=O_\alpha$ gives $\bar MO_\alpha G=O_\alpha\bar MG$, hence
\begin{equation}
\cN_A^\dagger(\bar M)\,G=\sum_\alpha O_\alpha^\dagger\bar MO_\alpha G=\sum_\alpha O_\alpha^\dagger O_\alpha\bar MG=\bar MG .
\end{equation}
Applying this to $\bar M^\dagger=\overline{M^\dagger}$ and taking adjoints gives $G\,\cN_A^\dagger(\bar M)=G\bar M$, because $\cN_A^\dagger$ preserves adjoints.  So $X=\cN_A^\dagger(\bar M)-\bar M$ satisfies $XG=GX=0$, and with $I=G+B$, $X=(G+B)X(G+B)=BXB$.  Let $\cQ=\cD\cN_A-\cD$, a map from physical to logical operators; then $\cQ^\dagger(M)=\cN_A^\dagger(\cD^\dagger(M))-\cD^\dagger(M)=\cN_A^\dagger(\bar M)-\bar M=B[\cN_A^\dagger(\bar M)-\bar M]B$, and for every $Y$ and $M$
\begin{equation}
\Tr\bigl(M^\dagger\cQ(Y)\bigr)=\Tr\bigl(\cQ^\dagger(M)^\dagger Y\bigr)=\Tr\bigl(B[\cN_A^\dagger(\bar M)-\bar M]^\dagger BY\bigr)=\Tr\bigl(\cQ^\dagger(M)^\dagger BYB\bigr)=\Tr\bigl(M^\dagger\cQ(BYB)\bigr),
\end{equation}
that is, $\cQ(Y)=\cQ(BYB)$: the blocks $GYG$, $GYB$, and $BYG$ contribute nothing, which is the origin of the factor $p$ in place of $\sqrt p$.  Therefore, by Lemma~\ref{SM:lem:badenc} and $\cD\cE=\cD\cE_B=\id_d$,
\begin{equation}
\cD\cN_A\cE(X)=X+\cQ(\cE(X))=X+\cQ\bigl(B\cE(X)B\bigr)=X+p\,\cQ(\cE_B(X))=X+p\bigl[\cD\cN_A\cE_B(X)-X\bigr],
\end{equation}
which is Eq.~\eqref{SM:eq:convex}.  $\Xi_A$ is a composition of completely positive trace-preserving maps, and the diamond distance of two channels is at most $2$, so $\|\cD\cN_A\cE-\id_d\|_\diamond=p\|\Xi_A-\id_d\|_\diamond\le2p$.  For $p=0$, $B\cE(X)B=0$ and $\cQ(\cE(X))=0$.
\end{proof}

Equation~\eqref{SM:eq:convex} is Eq.~(7) [Fig.~\ref{SM:fig:S6}(b)].  It is an exact decomposition of the logical channel: the noise on one region may be arbitrarily strong, only its support is restricted, and the bad branch carries exactly the weight $p$.

\subsection{Erasure recovery and leakage}
\label{SM:sec:erasure}

\begin{proposition}[erasure recovery]
\label{SM:prop:erasure}
Let $A$ be an allowed region, $\sigma_A$ any fixed state on $A$, $\cJ_{A,\sigma}(Y_{A^c})=\sigma_A\otimes Y_{A^c}$, and $\cR_A=\cD\cJ_{A,\sigma}$.  Then
\begin{equation}
\|\cR_A\Tr_A\cE-\id_d\|_\diamond\le2p .
\label{SM:eq:erasure}
\end{equation}
\end{proposition}

\begin{proof}
$\cN_A=\cJ_{A,\sigma}\Tr_A$ is a completely positive trace-preserving map supported on $A$, the replacement of the state of $A$ by $\sigma_A$, and $\cR_A\Tr_A\cE=\cD\cN_A\cE$; apply Proposition~\ref{SM:prop:noise}.
\end{proof}

The recovery uses the code, the region, and the chosen $\sigma_A$ only, and it is polynomial-time whenever the decoder is.

Let $\Psi_A=\Tr_{A^c}\cE$ be what an observer of $A$ sees, $\lambda_A=\Psi_A(I_d/d)$, and $\cC_{\lambda_A}(X)=\Tr(X)\lambda_A$ the constant channel; let $\Psi_{A,B}=\Tr_{A^c}\cE_B$ and $\lambda_{A,B}=\Psi_{A,B}(I_d/d)$.  Let $\cP(Y)=W\bigl(I_d/d\otimes\Tr_{\log}W^\dagger YW\bigr)W^\dagger$ be the logical depolarizing twirl.

\begin{lemma}[twirl identities]
\label{SM:lem:twirl}
(i) $\cP$ is completely positive, trace preserving, self-adjoint, and an average over the $d^2$ logical Weyl unitaries $\bar U_{jk}$: $\cP(Y)=d^{-2}\sum_{jk}\bar U_{jk}Y\bar U_{jk}^\dagger$.  (ii) For every operator $O_A$ supported in an allowed region $A$, $[O_A-\cP(O_A)]G=G[O_A-\cP(O_A)]=0$.  (iii) For every operator $O_A$ on $A$, extended by $I_{A^c}$, $\cE^\dagger(\cP(O_A))=\Tr(\lambda_AO_A)\,I_d$ and $\cE_B^\dagger(\cP(O_A))=\Tr(\lambda_{A,B}O_A)\,I_d$.
\end{lemma}

\begin{proof}
(i) The Weyl unitaries $U_{jk}=X^jZ^k$ on $\cH_{\log}$ satisfy $d^{-2}\sum_{jk}U_{jk}MU_{jk}^\dagger=\Tr(M)I_d/d$ for every $M$, hence $d^{-2}\sum_{jk}(U_{jk}\otimes I)Z(U_{jk}^\dagger\otimes I)=I_d/d\otimes\Tr_{\log}Z$ for every operator $Z$ on $\cH_{\log}\otimes\cH_S$, by linearity from $Z=M\otimes T$.  With $\bar U_{jk}=W(U_{jk}\otimes I)W^\dagger$ and $Z=W^\dagger YW$ this reads $d^{-2}\sum_{jk}\bar U_{jk}Y\bar U_{jk}^\dagger=\cP(Y)$.  A unitary average is completely positive and trace preserving, and it is self-adjoint because the set of unitaries is closed under adjoints.  (ii) By Eq.~\eqref{SM:eq:onesided} with $O=O_A$ and the logical operator $\bar U_{jk}^\dagger$, $O_A\bar U_{jk}^\dagger G=\bar U_{jk}^\dagger O_AG$; multiplying by $\bar U_{jk}$ on the left and averaging, $\cP(O_A)G=d^{-2}\sum_{jk}\bar U_{jk}O_A\bar U_{jk}^\dagger G=d^{-2}\sum_{jk}\bar U_{jk}\bar U_{jk}^\dagger O_AG=O_AG$, where $G$ commutes with every logical operator (Lemma~\ref{SM:lem:decoder}).  The identity $G\cP(O_A)=GO_A$ follows by taking adjoints, since $\cP$ preserves adjoints.  (iii) $W^\dagger\cP(Y)W=I_d/d\otimes T$ with $T=\Tr_{\log}(W^\dagger YW)$, and $\cE^\dagger(Z)=\Tr_S[(I\otimes\rho_S)W^\dagger ZW]$ is the adjoint of $X\mapsto W(X\otimes\rho_S)W^\dagger$, so $\cE^\dagger(\cP(Y))=d^{-1}\Tr(\rho_ST)\,I_d=d^{-1}\Tr\bigl((I\otimes\rho_S)W^\dagger YW\bigr)I_d=\Tr\bigl(\cE(I_d/d)\,Y\bigr)I_d$; with $Y=O_A\otimes I_{A^c}$ this is $\Tr(\lambda_AO_A)I_d$.  The same computation with $\rho_{S,B}$ gives the statement for $\cE_B$.
\end{proof}

\begin{proposition}[leakage]
\label{SM:prop:leakage}
For every allowed region $A$,
\begin{equation}
\Psi_A-\cC_{\lambda_A}=p\,(\Psi_{A,B}-\cC_{\lambda_{A,B}}),\qquad
\|\Psi_A-\cC_{\lambda_A}\|_\diamond\le2p,\qquad
\|\Psi_A(\varphi_1)-\Psi_A(\varphi_2)\|_1\le p\,\|\varphi_1-\varphi_2\|_1
\label{SM:eq:leakage}
\end{equation}
for all logical states $\varphi_1,\varphi_2$.
\end{proposition}

\begin{proof}
Let $O_A$ be an operator on $A$, extended by $I_{A^c}$, and $X=O_A-\cP(O_A)$.  By Lemma~\ref{SM:lem:twirl}(ii), $XG=GX=0$, so $X=BXB$, and
\begin{equation}
\cE^\dagger(X)=\Tr_S\bigl[(I\otimes\rho_S)W^\dagger BXBW\bigr]=\Tr_S\bigl[(I\otimes G_S^\perp\rho_SG_S^\perp)W^\dagger XW\bigr]=p\,\cE_B^\dagger(X),
\end{equation}
using $W^\dagger BW=I\otimes G_S^\perp$.  By Lemma~\ref{SM:lem:twirl}(iii), $\cE^\dagger(O_A)-\cC_{\lambda_A}^\dagger(O_A)=\cE^\dagger(X)$ and $\cE_B^\dagger(O_A)-\cC_{\lambda_{A,B}}^\dagger(O_A)=\cE_B^\dagger(X)$, where $\cC_\lambda^\dagger(O_A)=\Tr(\lambda O_A)I_d$.  Hence $(\Psi_A-\cC_{\lambda_A})^\dagger=p\,(\Psi_{A,B}-\cC_{\lambda_{A,B}})^\dagger$ on operators of $A$, and taking adjoints gives the first identity.  The diamond norm of the difference of two channels is at most $2$, which gives the second bound.  For the third, $\cC_{\lambda_{A,B}}(\varphi_1-\varphi_2)=\Tr(\varphi_1-\varphi_2)\lambda_{A,B}=0$, so $\Psi_A(\varphi_1)-\Psi_A(\varphi_2)=p\,\Psi_{A,B}(\varphi_1-\varphi_2)$, and $\Psi_{A,B}$ is a trace-norm contraction.
\end{proof}

The diamond-norm form controls the joint output of $A$ with any reference entangled to the input.  Together, Eqs.~\eqref{SM:eq:erasure} and \eqref{SM:eq:leakage} say that the complement of an allowed region is an authorized set and the region itself is an unauthorized set, in the sense of quantum secret sharing~\cite{CleveGottesmanLo1999}, with access structure fixed by geometry [Fig.~\ref{SM:fig:S6}(c)].

\begin{proposition}[recovery and leakage after storage]
\label{SM:prop:afterstorage}
For every dynamics of Sec.~\ref{SM:sec:memory} and every allowed region $A$,
\begin{equation}
\|\cR_A\Tr_A\Phi(t,0)\cE-\id_d\|_\diamond\le\min\{2,\Delta(t)+2p\},\qquad
\|\Tr_{A^c}\Phi(t,0)\cE-\cC_{\lambda_A}\|_\diamond\le\min\{2,\Delta(t)+2p\} .
\label{SM:eq:afterstorage}
\end{equation}
In four dimensions both are at most $C_\Delta L^{14}te^{-aL}+2C_pL^{20}e^{-4\ell_0}$.
\end{proposition}

\begin{proof}
Insert $\cE$: $\cR_A\Tr_A\Phi(t,0)\cE-\id_d=\cR_A\Tr_A\bigl[\Phi(t,0)\cE-\cE\bigr]+\bigl[\cR_A\Tr_A\cE-\id_d\bigr]$.  The diamond norm is submultiplicative and equals $1$ on completely positive trace-preserving maps, so the first term is at most $\|\Phi(t,0)\cE-\cE\|_\diamond\le\Delta(t)$ by Proposition~\ref{SM:prop:memory}, and the second is at most $2p$ by Proposition~\ref{SM:prop:erasure}.  The second bound follows in the same way from Proposition~\ref{SM:prop:leakage}, and both differences of channels have diamond norm at most $2$.  The four-dimensional constants are those of Proposition~\ref{SM:prop:memory4d} and Theorem~\ref{SM:thm:static}.
\end{proof}

The erasure happens once, at readout; the region is lost after the evolution, and the Hamiltonian during storage is the full one.

\subsection{The coherent family of encoded states}
\label{SM:sec:coherent}

Let $\cS_{\beta,V}=\{\cE(\varphi):\varphi\ \text{a logical density matrix}\}$.

\begin{proposition}[coherent family]
\label{SM:prop:coherent}
(i) $\cS_{\beta,V}$ is affinely isomorphic to the full $d$-dimensional state space, $\|\cE(\varphi_1)-\cE(\varphi_2)\|_1=\|\varphi_1-\varphi_2\|_1$; its real dimension is $d^2-1=4095$ in four dimensions.  (ii) For every Hermitian logical operator $X$ and every $s$, $\|\cL_s\cE(X)\|_1\le b(s)\|X\|_1$.  (iii) For a maximally entangled state $|\Phi_d\rangle$ of the logical system with a reference,
\begin{equation}
\langle\Phi_d|\bigl(\cD\Phi(t,0)\cE\otimes\id\bigr)(|\Phi_d\rangle\langle\Phi_d|)|\Phi_d\rangle\ge1-\Delta(t)/2 .
\label{SM:eq:entfidelity}
\end{equation}
\end{proposition}

\begin{proof}
(i) is Lemma~\ref{SM:lem:encdec}(ii) applied to differences of states; $\cE$ is injective and affine.  (ii) Write $X=X_+-X_-$ with $X_\pm\ge0$ and $\Tr X_++\Tr X_-=\|X\|_1$; for $X_\pm\ne0$, $\cE(X_\pm)=\Tr(X_\pm)\,\cE(X_\pm/\Tr X_\pm)$ with $X_\pm/\Tr X_\pm$ a state, so Lemma~\ref{SM:lem:rate} gives $\|\cL_s\cE(X)\|_1\le b(s)(\Tr X_++\Tr X_-)$.  (iii) For a pure state $\Phi$ and any state $\sigma$, $\langle\Phi|\sigma|\Phi\rangle\ge1-\tfrac12\|\sigma-|\Phi\rangle\langle\Phi|\|_1$, and $\|((\cD\Phi(t,0)\cE)\otimes\id)(|\Phi_d\rangle\langle\Phi_d|)-|\Phi_d\rangle\langle\Phi_d|\|_1\le\|\cD\Phi(t,0)\cE-\id_d\|_\diamond\le\Delta(t)$ by Proposition~\ref{SM:prop:memory}.
\end{proof}

By (ii) the whole encoded operator space is an approximate null space of the dynamics, and Eq.~\eqref{SM:eq:diamond} bounds the drift of every member uniformly; (iii) is an operational test of coherence.  Long-lived refers to the drift of states, Eq.~\eqref{SM:eq:diamond}.  Drift and spectral gap are different quantities: on a one-way chain of $m+1$ classical states $0\to1\to\cdots\to m$ with unit rates and $m$ absorbing, the uniform distribution $\xi$ on the first $m$ states has $\|\cL\xi\|_1=2/m$, since only the flux out of state $0$ and into state $m$ is unbalanced, while every nonzero eigenvalue of the generator equals $-1$.

\subsection{Lower bound on the mixing time}
\label{SM:sec:sampling}

Let $m_\rho(t)=\sup_\xi\frac12\|\Phi(t,0)(\xi)-\rho\|_1$ over all initial states $\xi$; it is non-increasing in $t$, because $\rho$ is stationary and the propagator contracts the trace norm.

\begin{proposition}[worst-case mixing bound]
\label{SM:prop:mixing}
For all $t\ge0$,
\begin{equation}
m_\rho(t)\ge\max\{0,\ 1-1/d-\mathfrak B(t)/2\} .
\label{SM:eq:mixing}
\end{equation}
\end{proposition}

\begin{proof}
For any logical orthonormal basis $\{|j\rangle\}$, $\sum_j\langle j|\cD(\rho)|j\rangle=1$, so some $j_*$ has $\langle j_*|\cD(\rho)|j_*\rangle\le1/d$.  Let $\varphi_*=|j_*\rangle\langle j_*|$, $\xi_*=\cE(\varphi_*)$, and $E_*=\cD^\dagger(\varphi_*)=\bar\varphi_*$, a projector on the physical space.  Then $\Tr(E_*\rho)=\langle j_*|\cD(\rho)|j_*\rangle\le1/d$, while
\begin{equation}
\Tr\bigl(E_*\Phi(t,0)\xi_*\bigr)=\langle j_*|\cD\Phi(t,0)\cE(\varphi_*)|j_*\rangle\ge1-\tfrac12\|\cD\Phi(t,0)\cE(\varphi_*)-\varphi_*\|_1\ge1-\tfrac12\|\Phi(t,0)\cE(\varphi_*)-\cE(\varphi_*)\|_1\ge1-\mathfrak B(t)/2 ,
\end{equation}
by the pure-state inequality of Proposition~\ref{SM:prop:coherent}(iii), the contraction $\|\cD(Z)\|_1\le\|Z\|_1$, $\cD\cE=\id_d$, and Lemma~\ref{SM:lem:statebound}.  Since $0\le E_*\le I$, $\tfrac12\|\Phi(t,0)\xi_*-\rho\|_1\ge\Tr\bigl(E_*(\Phi(t,0)\xi_*-\rho)\bigr)\ge1-\mathfrak B(t)/2-1/d$.
\end{proof}

\begin{corollary}[mixing time]
\label{SM:cor:tmix}
Let $\bar b$ be a uniform bound on $b(s)$ and $0<\eta<1-1/d$.  The worst-case mixing time $t_{\rm mix}(\eta)=\inf\{t:m_\rho(t)\le\eta\}$ obeys $t_{\rm mix}(\eta)\ge2(1-1/d-\eta)/\bar b$, and in four dimensions
\begin{equation}
t_{\rm mix}(1/4)\ge\frac{47}{32C_b}L^{-14}e^{aL},
\label{SM:eq:tmix}
\end{equation}
which exceeds $e^{cn^{1/4}}$ for a constant $c>0$ and all large $L$, with $n=6L^4$ the number of qubits.
\end{corollary}

\begin{proof}
$m_\rho(t)\le\eta$ requires $1-1/d-\mathfrak B(t)/2\le\eta$ by Eq.~\eqref{SM:eq:mixing}, and $\mathfrak B(t)\le\bar bt$.  In four dimensions $d=64$, $2(1-1/64-1/4)=47/32$, and $\bar b=C_bL^{14}e^{-aL}$ by Proposition~\ref{SM:prop:memory4d}; with $L=(n/6)^{1/4}$ the polynomial prefactor is absorbed into the exponent for large $L$.
\end{proof}

In terms of accumulated activity $\cA(t)=\int_0^tS$ and tail $\cU(t)=\int_0^tu$, with $\nu\equiv0$, reaching $m_\rho\le\eta$ requires $2\sqrt p\,\cA(t)+\cU(t)\ge1-1/d-\eta$, a requirement on accumulated activity that is independent of the clock rate.  The obstruction concerns quasi-local Gibbs-preserving dynamics at low temperature, including time-dependent schedules, started from a thermally encoded logical basis state; at high temperature such dynamics mix rapidly~\cite{Bakshi2024,Kochanowski2024,DobrushinShlosman2026}.

\section{The topological Peierls condition and the model family}
\label{SM:sec:general}

The four-dimensional proof uses the lattice only through a short list of properties.  This section states them as conditions on a family of stabilizer codes, proves the theorem of Appendix~A by substituting the general constants into Secs.~\ref{SM:sec:window}--\ref{SM:sec:memory}, step by step and lemma by lemma, verifies the conditions for the generalized toric codes of Appendix~A, and proves that the class is closed under stacking, pinned ancillas, and finite-depth local unitaries.

\subsection{Conditions (A)--(D)}
\label{SM:sec:conditions}

\emph{(A) Geometry.}  Qubits and checks sit on a $D$-dimensional periodic lattice of linear size $L$ with a bounded number of labels per unit cell; distances are the ambient distances up to a fixed bi-Lipschitz equivalence, and balls of radius $R$ contain at most $C_{\rm ball}(R+1)^D$ cells.  The number of qubits is $n_L\le C_{\rm vol}L^D$ and the number of checks of each type is at most $C_{\rm vol}L^D$.  Checks are commuting non-scalar Hermitian Pauli operators of weight at most $w$, at most $b_0$ of them touch any qubit, the group they generate excludes $-I$, and $H_0=-\sum_aJ_aS_a$ with $0<J_{\min}\le J_a\le J_{\max}$.  The code space has dimension $d_L=2^{k_L}$ with $2\le d_L\le d_{\max}$.  Every connected set of diameter $r<c_{\rm box}L$ lies in a non-winding box of diameter at most $C_{\rm box}(r+1)$, where a non-winding box is one that lifts uniquely to $\Z^D$.  The perturbation is any Hermitian $V$ of fixed range $R$ with $\|V\|_\loc=\epsilon$; $k_R$ bounds the number of qubits of a term and $b=b_0k_R$ the number of checks of one type that fail to commute with a term.

\emph{(B) Clusters, local erasure, and global constraints.}  Checks come in $s_*$ types $\alpha$, each with a fixed adjacency of range at most $r_{\rm adj}$ and degree at most $\deg_\alpha$ that groups the violated checks of that type into clusters.  A pattern is valid if and only if each of its type components is valid, and a type-$\alpha$ pattern is valid when it is the pattern of a Pauli operator that creates no excitation of any other type.  (B1)~Every cluster $C$ of diameter $r<c_{\rm rep}L$ that occurs in a valid pattern is $\syn(F_C)$ for a Pauli operator $F_C$ supported within distance $C_{\rm rep}(r+1)$ of $C$ that creates no excitation of any other type; $F_C$ is chosen as a function of $C$ alone.  (B2)~Each type carries a charge $q_\alpha:\mathcal C_\alpha\to\F_2^{h_\alpha}$ on clusters such that valid type-$\alpha$ patterns have zero total charge and every union of non-touching clusters with zero total charge is a valid type-$\alpha$ pattern; $h=\max_\alpha h_\alpha$.

\emph{(C) Local generation.}  Every Pauli operator supported in a non-winding box of diameter $r<c_{\rm gen}L$ that commutes with all checks is $\zeta\prod_{a\in Y}S_a$ with the supports of the checks in $Y$ within distance $C_{\rm gen}(r+1)$ of the box.

\emph{(D) Dynamics.}  A family of Lindblad generators with $\rho_{\beta,V}$ as common stationary state and, for truncation radius $r$, $\cL=\sum_jK_{j,r}+\cL_r^{\rm tail}$ with $\sum_j\|K_{j,r}\|_{1\to1}\le C_SL^D$ and $\|\cL_r^{\rm tail}\|_{1\to1}\le C_TL^De^{-\mu r}$, the $K_{j,r}$ supported in boxes of diameter at most $C_{\rm dyn}r$.

Conditions (B1) and (C) are the two properties of the main text, (B2) is the bookkeeping of global constraints, and (B1), (B2), (C) are conditions (i), (ii), (iii) of Appendix~A.  (B1) and (C) are independent: the two-dimensional toric code satisfies (C) and violates (B1).  All constants named in (A)--(D) are the \emph{structural constants}.

\subsection{The structural stability theorem}
\label{SM:sec:generaltheorem}

\begin{theorem}[structural stability]
\label{SM:thm:general}
For every family satisfying (A)--(C) and every range of $V$, there are $\epsilon_0,\beta_0,a,C,L_0>0$ depending on the structural constants only, such that for $L\ge L_0$, $\beta\ge\beta_0$, $\|V\|_\loc\le\epsilon_0$:
(a)~$\Pr_{\rho}[\text{a cluster of some type with}\ge u\text{ checks}]\le CL^{D(h+1)}e^{-4u}$;
(b)~the decoder $\cD_L$ built from the checks and the erasure rule (B1), and a linear thermal encoding $\cE_{\beta,V}$ with $\cD_L\cE_{\beta,V}=\id_{d_L}$, satisfy under (D)
\begin{equation}
\|\cD_Le^{t\cL}\cE_{\beta,V}-\id_{d_L}\|_\diamond\le CL^{D(h+3)/2}\,t\,e^{-aL} .
\end{equation}
\end{theorem}

\begin{proof}
The four-dimensional proof is rerun with the general constants.  Each step names the lemmas of Secs.~\ref{SM:sec:geometry}--\ref{SM:sec:memory} it reuses, the four-dimensional inputs they used, and the replacement supplied by (A)--(D); Table~\ref{SM:tab:substitution} summarizes the replacements.

\emph{Step 1: sectors and histories (Sec.~\ref{SM:sec:histories}).}  Lemmas~\ref{SM:lem:sectors}--\ref{SM:lem:sectortrace} are statements about a stabilizer group of rank $n_L-k_L$ not containing $-I$: the sectors have rank $d_L$, valid patterns are the syndromes, and the sector trace of a zero-syndrome Pauli operator is $\zeta\chi_S(s)$ for a check product times a phase and $0$ for the $4^{k_L}-1$ nontrivial logical classes.  Lemmas~\ref{SM:lem:locality}--\ref{SM:lem:expansion} use only the finite range of $V$, the bounded weight and degree of the checks, and $J_{\max}$, through $b_R$, $c_R$, and the dependency graph; they hold verbatim with $d$ replaced by $d_L$.  Lemma~\ref{SM:lem:components} uses rule (2) with the type adjacency of (B) in place of the line graph; its proof is unchanged, and clusters are now clusters of each type.  Lemma~\ref{SM:lem:signflip} is Pauli algebra.

\emph{Step 2: local components (Lemma~\ref{SM:lem:local}).}  By (A), ${\rm sp}(C)$, of diameter at most $C_gm(C)$, lies in a non-winding box of diameter at most $C_{\rm box}(C_gm(C)+1)$ when $C_gm(C)<c_{\rm box}L$; enlarge it by the collar $\max\{C_{\rm rep},C_{\rm gen}\}(C_{\rm box}(C_gm(C)+1)+1)$ to obtain $Y_C$, with $\diam Y_C\le c_1m(C)$ for a structural $c_1$, and choose $c_0$ so that $c_1m_0$ is below $c_{\rm box}L$, $c_{\rm rep}L$, and $c_{\rm gen}L$.  Part (i): each cluster of a section is a cluster of the valid pattern $s_k$ of diameter below $c_{\rm rep}L$, so (B1) supplies $F_C$, supported in $Y_C$, with syndrome equal to that cluster and no excitation of other types; their product is $R_C(k)$, and the section is valid as the syndrome of $R_C(k)$.  Part (ii): $Q_C$ has empty syndrome and support in a non-winding box of diameter below $c_{\rm gen}L$, so (C) gives $Q_C=\zeta_CS_C$ with generating checks in $Y_C$.  Part (iii) is unchanged.

\emph{Step 3: signs (Secs.~\ref{SM:sec:pairwise}--\ref{SM:sec:emptybox}).}  Lemmas~\ref{SM:lem:netcommute}--\ref{SM:lem:cut} use Pauli commutation and the characters of Lemma~\ref{SM:lem:characters} only.  Lemma~\ref{SM:lem:emptybox} uses the repair $R_C$ and the generation of $S_C$ inside $Y_C$, which Step~2 provides; Eq.~\eqref{SM:eq:linklocal} follows with $c_3=c_1+2$.  Mutual statistics in any dimension enter through the same character formula.

\emph{Step 4: counting and the gas (Secs.~\ref{SM:sec:gasdef}--\ref{SM:sec:return}).}  $\Delta_{\mathfrak G}$ is bounded by (A) and (B).  The ball count becomes $C_{\rm ball}(R+1)^{D+1}$, so $\Upsilon(q)$ is replaced by $\Upsilon_D(q)=C_*\sum_m(m+1)^{D+1}(K_{\rm link}eq)^m$ and $q_0$ is chosen by Eq.~\eqref{SM:eq:q0} with $\Upsilon_D$; Lemmas~\ref{SM:lem:countcomp}--\ref{SM:lem:countlink} give structural $K_{\rm raw},K_{\rm link}$.  Lemmas~\ref{SM:lem:actbound}--\ref{SM:lem:background} are unchanged, since they use only Eq.~\eqref{SM:eq:incompsum}, which now reads $\sum_{\gamma'\not\sim\gamma}\rho_{\gamma'}e^{m(\gamma')}\le m(\gamma)\Upsilon_D(q_0)$.  Lemma~\ref{SM:lem:macro} and Proposition~\ref{SM:prop:ratio_bL} hold with $C_{\rm env}NL^D$ nodes per slice and $\beta\le A_*L^D$.

\emph{Step 5: bridge (Sec.~\ref{SM:sec:bridge}).}  $\ln\mathsf D=n_L\ln2\le C_{\rm vol}L^D\ln2$ and $\beta_*=A_*L^D$ with $2C_{\rm vol}\ln2/A_*\le\delta$; Proposition~\ref{SM:prop:tension_all} then gives Eq.~\eqref{SM:eq:tension} with $C_{\rm tw}=C_2$ for every valid type-$\alpha$ pattern $\Gamma$, with $F$ as in Step~6.

\emph{Step 6: fixed window and the static bound (Sec.~\ref{SM:sec:window}).}  Lemma~\ref{SM:lem:frozen} uses the weight $w$ and the bound $b$ from (A) and an erasing operator $F$ for $\Gamma$ creating no other type, which (B) supplies for every valid type-$\alpha$ pattern.  Lemmas~\ref{SM:lem:duhamel}--\ref{SM:lem:high} and Proposition~\ref{SM:prop:marked} use $J_{\min},J_{\max},w,b$ only, with $\vartheta$ and $C_{\rm act}$ defined as before.  Grouping: by (B2), as in Lemma~\ref{SM:lem:grouping}, every cluster $C_0$ of type $\alpha$ in a valid pattern belongs to a union of at most $h_\alpha+1\le h+1$ clusters of that type with zero total charge, which is valid.  Counting: connected sets of $m$ checks containing a given check number at most $K_\alpha^m$ with $K_\alpha=\max\{2,\deg_\alpha\}^2$, by the closed-walk argument of Lemma~\ref{SM:lem:clustercount}; in Eq.~\eqref{SM:eq:eps0} replace $196$ by $K_{\max}=\max_\alpha K_\alpha$ and $392$ by $2K_{\max}$.  Proposition~\ref{SM:prop:static4d} then gives, with $M_\alpha\le C_{\rm vol}L^D$ checks of type $\alpha$,
\begin{equation}
\Pr_\rho[\text{a cluster of some type with}\ge u\text{ checks}]\le\sum_\alpha\frac{2q^u}{1-q}\sum_{r=1}^{h_\alpha+1}M_\alpha^r\Bigl(\frac q{1-q}\Bigr)^{r-1}\le CL^{D(h+1)}e^{-4u},
\end{equation}
which is (a).

\emph{Step 7: decoder (Sec.~\ref{SM:sec:decoder}).}  With $\ell_0=c_{\rm dec}L$, the decoder erases, type by type in a fixed order, every cluster of diameter below $8\ell_0$ by its operator $F_C$ of (B1), and treats all other patterns by global elimination; good patterns are those all of whose clusters have diameter below $\ell_0$, and an allowed region is one whose union with the collar $C_{\rm rep}(8\ell_0+1)+r_{\rm adj}$ lies in a non-winding box of diameter below $c_{\rm gen}L$.  Lemma~\ref{SM:lem:W} holds verbatim.  In the proof of Lemma~\ref{SM:lem:decoder}, Step~1 used that the clusters changed by $O$ lie within a collar of the support of $O$ of thickness $\ell_0$ plus the adjacency range, that unchanged clusters carry identical erasing operators, which is the clause ``$F_C$ is a function of $C$ alone'' of (B1), and that the residual zero-syndrome Pauli operator, supported in a box of diameter $O(\ell_0)<c_{\rm gen}L$, is a check product by (C); Step~2 is unchanged.  A cluster of diameter at least $\ell_0$ has at least $\ell_0/r_{\rm adj}$ checks, so by (a) the bad set has weight $p\le CL^{D(h+1)}e^{-4\ell_0/r_{\rm adj}}=CL^{D(h+1)}e^{-c'L}$.

\emph{Step 8: memory (Sec.~\ref{SM:sec:memory}).}  Lemmas~\ref{SM:lem:encdec}--\ref{SM:lem:diamondlift} and Proposition~\ref{SM:prop:memory} are dimension-free, with $d_L$ in place of $d$.  Under (D) with $r=c_rL$ chosen so that boxes of diameter $C_{\rm dyn}r$ are allowed regions, Lemma~\ref{SM:lem:rate} gives $b(s)\le4C_SL^D\sqrt p+2C_TL^De^{-\mu c_rL}$, and $\sqrt p\le CL^{D(h+1)/2}e^{-c'L/2}$ from Step~7, so $b(s)\le CL^{D+D(h+1)/2}e^{-aL}=CL^{D(h+3)/2}e^{-aL}$ with $a=\min\{c'/2,\mu c_r\}$; Proposition~\ref{SM:prop:memory} with the factor $2d_L\le2d_{\max}$ gives (b).
\end{proof}

The temperature window is $0<T\le1/\beta_0$ with $\epsilon_0$ uniform in it.  The common $\epsilon_0,\beta_0$ depend on the fixed family; they are uniform in $L$, $\beta$, and $V$.  The operational statements of Sec.~\ref{SM:sec:operational} hold throughout the class, with $p\le C_pL^{D(h+1)}e^{-4\ell_0}$ and $\Delta_L(t)\le CL^{D(h+3)/2}te^{-aL}$.

\begin{table}[t]
\caption{Four-dimensional inputs and their general replacements.}
\label{SM:tab:substitution}
\renewcommand{\baselinestretch}{1}\small
\begin{ruledtabular}
\begin{tabular}{p{2.6cm}p{5.0cm}p{7.4cm}}
Step & Four-dimensional datum & General replacement \\
\colrule
fixed window & $w=6$, $b=4k_R$, line-graph degree $14$ & $w,b,J_{\min},J_{\max}$ from (A); $K_\alpha=\max\{2,\deg_\alpha\}^2$; erasing $F$ from (B) \\
sections of a component & closed 1-chains filled by $H_Q$ & cluster-by-cluster erasure (B1) \\
net Pauli of a component & Lemma~\ref{SM:lem:generation} & local generation (C) \\
signs & loops and membranes & Pauli bilinear characters, repair and generation boxes \\
counting & five-dimensional balls, $NL^4$ roots & $(D+1)$-dimensional balls, $O(NL^D)$ roots, $\Upsilon_D(q)$ \\
entropy bridge & $\ln\mathsf D=6L^4\ln2$ & $n_L\ln2\le C_{\rm vol}L^D\ln2$ \\
grouping & $h=4$, at most five clusters & $h_\alpha\le h$, at most $h+1$ clusters \\
decoder & fillings of Eq.~\eqref{SM:eq:filling} & erasing operators $F_C$ of (B1), residual by (C) \\
memory & $d=64$ & $d_L\le d_{\max}$, one-sided identity from (A)--(C) \\
\end{tabular}
\end{ruledtabular}
\end{table}

\subsection{Generalized toric codes}
\label{SM:sec:family}

\begin{figure}[t]
\centering
\includegraphics{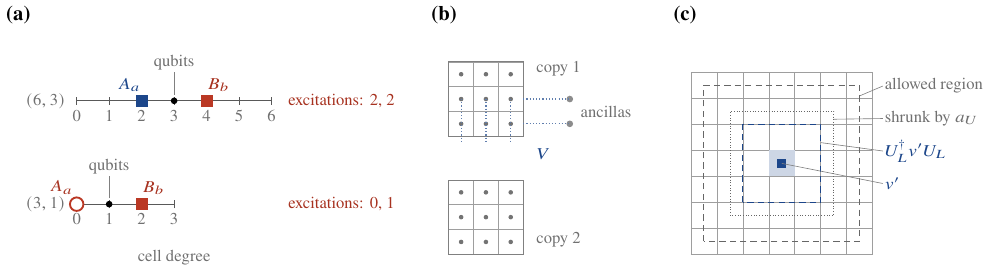}
\caption{The model family and its closure.  (a)~Cell degrees of the $p$-cell toric code: $X$-checks $A_a$ on $(p-1)$-cells, qubits on $p$-cells, $Z$-checks $B_b$ on $(p+1)$-cells; violated $A$-checks form closed $(p-1)$-chains and violated $B$-checks closed dual $(D-p-1)$-chains.  For $(6,3)$ both degrees equal $2$ and both excitation types are closed surfaces; for $(3,1)$ one degree is $0$ and that excitation type is point-like.  (b)~Stacks and pinned ancillas, Proposition~\ref{SM:prop:stacking}: the perturbation may couple different copies and the ancillas to the code.  (c)~A finite-depth local unitary $U_L$ thickens every term of $V'$ by the light-cone radius $a_U$, so $\|V\|_{\rm loc}\le K_U\|V'\|_{\rm loc}$, and allowed regions shrink by $a_U$, Proposition~\ref{SM:prop:unitary}.}
\label{SM:fig:S7}
\end{figure}

Fix $D\ge4$ and $2\le p\le D-2$.  Qubits sit on the $p$-cells of $T_L^D$; for every $(p-1)$-cell $a$ and $(p+1)$-cell $b$,
\begin{equation}
A_a=\prod_{f\supset a}X_f,\qquad B_b=\prod_{f\subset\partial b}Z_f,\qquad H_0=-\sum_aJ_aA_a-\sum_bK_bB_b .
\label{SM:eq:gentoric}
\end{equation}
Table~\ref{SM:tab:family} lists the first members, and Fig.~\ref{SM:fig:S7}(a) shows the cell degrees for $(6,3)$ and $(3,1)$; dual pairs such as $(5,2)$ and $(5,3)$ share one mechanism.

\begin{table}[t]
\caption{Generalized toric codes in the class.  Excitation degrees are the chain degrees of the two excitation types, $p-1$ for violated $A$-checks on the primal lattice and $D-p-1$ for violated $B$-checks on the dual lattice.  $h_{D,p}=\max\{\binom D{p-1},\binom D{p+1}\}$ is the number of global constraints per type.}
\label{SM:tab:family}
\renewcommand{\baselinestretch}{1}\small
\begin{ruledtabular}
\begin{tabular}{cccccc}
$(D,p)$ & qubits on & excitation degrees & $k$ & $h_{D,p}$ & exponents $D(h+1)$, $D(h+3)/2$ \\
\colrule
$(4,2)$ & plaquettes & $1,1$ & $6$ & $4$ & $20$, $14$ \\
$(5,2)$ & plaquettes & $1,2$ & $10$ & $10$ & $55$, $32.5$ \\
$(5,3)$ & 3-cells & $2,1$ & $10$ & $10$ & $55$, $32.5$ \\
$(6,2)$ & plaquettes & $1,3$ & $15$ & $20$ & $126$, $69$ \\
$(6,3)$ & 3-cells & $2,2$ & $20$ & $15$ & $96$, $54$ \\
$(6,4)$ & 4-cells & $3,1$ & $15$ & $20$ & $126$, $69$ \\
\end{tabular}
\end{ruledtabular}
\end{table}

\begin{corollary}
\label{SM:cor:family}
For every fixed $(D,p)$ with $D\ge4$, $2\le p\le D-2$, the family \eqref{SM:eq:gentoric} satisfies (A)--(C) with $h=h_{D,p}$, so Theorem~\ref{SM:thm:general} holds with exponents $D(h_{D,p}+1)$ and $D(h_{D,p}+3)/2$ and protects all $\binom Dp$ logical qubits.
\end{corollary}

\begin{proof}
\emph{(A).}  There are $\binom DpL^D$ qubits, $\binom D{p-1}L^D$ checks $A_a$, and $\binom D{p+1}L^D$ checks $B_b$.  A $(p-1)$-cell lies in the $p$-cells obtained by adding one of the remaining $D-p+1$ directions in either orientation, so $A_a$ has weight $2(D-p+1)$; a $(p+1)$-cell has $2(p+1)$ faces, so $B_b$ has weight $2(p+1)$; a $p$-cell has $2p$ faces and lies in $2(D-p)$ cells of dimension $p+1$.  Two checks of opposite kind share the $p$-cells that contain $a$ and lie in $\partial b$, which number two if $a$ is a face of a face of $b$ and zero otherwise, by $\partial_p\partial_{p+1}=0$; so all checks commute, and every product of checks is $X(\delta y)Z(\partial u)$ with coefficient $+1$, never $-I$.  The number of logical qubits is $k=\dim H_p(T^D;\F_2)=\binom Dp$, by the K\"unneth decomposition of the torus into $D$ circles.  Non-winding boxes and the ball count are those of the cubical lattice.

\emph{(B1).}  The pattern of a $Z$-type operator with support $z$ is $\partial_pz$, a primal $(p-1)$-chain; that of an $X$-type operator is, under cellular duality, the boundary of a dual $(D-p)$-chain, a dual $(D-p-1)$-chain; both are closed.  For a chain degree $q\ge1$ define the adjacency of $q$-cells by sharing a $(q-1)$-face.  A valid type pattern $s$ has $\partial_qs=0$; every $(q-1)$-face touching $s$ lies in a single cluster, so the boundary cancellations of $\partial_qs$ happen within clusters, and every cluster is a closed $q$-chain, branched or not.  Lemma~\ref{SM:lem:contraction} holds in every dimension and degree, so a cluster $C$ of diameter below $L/8$ satisfies $\partial_{q+1}(H_QC)=C$ inside its bounding box $Q$ (the dual box for the dual type); the operator $Z(H_QC)$, respectively $X$ of the dual filling, has syndrome $C$ and creates no excitation of the other type, since $Z$-type operators commute with the $B$-checks and $X$-type operators with the $A$-checks.  It depends on $C$ alone.  Both degrees $p-1$ and $D-p-1$ are at least $1$ exactly when $2\le p\le D-2$.

\emph{(B2).}  A closed $q$-chain bounds if and only if its class in $H_q(T^D;\F_2)=\F_2^{\binom Dq}$ vanishes.  Explicit charges are the parities $\lambda_I(z)=\sum_{(x,I):\,x_i=L-1\ \forall i\in I}z_{(x,I)}$ over direction sets $I$ of size $q$: they vanish on the boundary of every $(q+1)$-cell, whose two faces in a given direction set cross the hyperplanes together or not at all, and they form the Kronecker pairing with the coordinate $q$-tori, which are a basis of $H_q$.  Hence a union of non-touching closed clusters with all charges zero bounds, and is the pattern of a $Z$-type or $X$-type operator creating no other type.  So $h_A=\binom D{p-1}$, $h_B=\binom D{D-p-1}=\binom D{p+1}$, and the two types are grouped separately.

\emph{(C).}  For a Pauli operator $\zeta X(x)Z(z)$ in a non-winding box commuting with all checks, $z$ is a closed $p$-chain, filled by a $(p+1)$-chain inside the box by Lemma~\ref{SM:lem:contraction}, so $Z(z)$ is a product of the $B_b$ of that chain; $*x$ is a closed dual $(D-p)$-chain filled inside the dual box, so $X(x)$ is a product of the $A_a$ of the corresponding edges, within two layers of the box.

\emph{Counting.}  A $q$-cell has $2q$ faces of dimension $q-1$, each contained in $2(D-q+1)$ cells of dimension $q$, so the adjacency degree is at most $\deg_q=2q[2(D-q+1)-1]$ and $K_q=\max\{2,\deg_q\}^2$; for $D=4$, $q=1$ this returns $\deg_1=14$ and $K=196$.
\end{proof}

For $p=1$ or $p=D-1$ one excitation type is a $0$-chain, which $\Pi_Q$ maps to the corner of the box rather than to zero, so an isolated point cannot be erased inside any box; (B1) fails for that type.  Point-like excitations are created in pairs and can only be moved, which is the two- and three-dimensional toric code.  No integer $p$ satisfies $2\le p\le D-2$ for $D\le3$, so these codes lie outside the class.  For $(6,3)$ both excitation types are closed surfaces, so the condition concerns erasability of membranes as well as loops.  The lifetime exponent $e^{-aL}$ holds for every member; for membrane-like excitations it is a lower bound.

\subsection{Closure of the class}
\label{SM:sec:closure}

\begin{proposition}[stacking and pinned ancillas]
\label{SM:prop:stacking}
A fixed finite stack of codes satisfying (A)--(C) in the same dimension, with a bounded number of ancilla qubits per unit cell added through single-qubit checks $Z_x$ with couplings $J_x$ in a fixed positive interval, satisfies (A)--(C).
\end{proposition}

\begin{proof}
Give the checks of each copy and the ancilla checks their own types, keeping the adjacency of each copy.  (A): weights, degrees, and cell counts are the maxima and sums over the copies, the checks of different copies act on disjoint qubits and commute, the stabilizer group is the direct product and excludes $-I$, and the logical dimension is the product of the logical dimensions.  (B1): a cluster belongs to one copy and is erased by that copy's operator $F_C$, which acts on that copy only and creates no excitation of any other type; an isolated ancilla excitation at $x$ is erased by $X_x$.  (B2): each copy keeps its charges; the ancilla type has trivial charge space, since every set of ancilla excitations is the syndrome of a product of $X_x$.  (C): a Pauli operator on the stack is a tensor product of Pauli operators on the copies and on the ancillas; commuting with all checks, each factor commutes with the checks of its copy and is a product of nearby checks by (C) for that copy, and an ancilla factor commuting with $Z_x$ is $I$ or $Z_x$ up to phase.
\end{proof}

The perturbation may then couple different copies and the ancillas to the code arbitrarily [Fig.~\ref{SM:fig:S7}(b)], within $\|V\|_\loc\le\epsilon_0$; the lifetime is a single statement for the whole stack.

\begin{proposition}[finite-depth local unitaries]
\label{SM:prop:unitary}
Let $U_L$ be a local unitary circuit of fixed depth and gate range, with light-cone radius $a_U$ independent of $L$, and $H'=U_LH_0U_L^\dagger+V'$.  Then the thermal state of $H'$, the encoding $\cE'=\cU_L\cE$, and the decoder $\cD'=\cD\cU_L^{-1}$, with $\cU_L(X)=U_LXU_L^\dagger$, satisfy Theorem~\ref{SM:thm:general} with $\epsilon_0$ replaced by $\epsilon_0/K_U$, $K_U=\sup_x|\{y:\dist(x,y)\le a_U\}|$, and allowed regions shrunk by $a_U$.
\end{proposition}

\begin{proof}
Set $V=U_L^\dagger V'U_L$, so that $H'=U_L(H_0+V)U_L^\dagger$.  Each term $U_L^\dagger v'_iU_L$ keeps its norm, and its support is the light cone of $\supp v'_i$, which thickens it by $a_U$ [Fig.~\ref{SM:fig:S7}(c)]; a qubit lies in the light cones of at most $K_U$ supports containing a given qubit, so $\|V\|_\loc\le K_U\|V'\|_\loc$, and $\|V\|_\loc\le\epsilon_0$ whenever $\|V'\|_\loc\le\epsilon_0/K_U$.  The thermal state of $H'$ is $U_L\rho U_L^\dagger$ with $\rho$ the thermal state of $H_0+V$, so $\cE'$ is a thermal encoding for $H'$ and $\cD'\cE'=\cD\cE=\id$.  A dynamics $\cL'$ of the class (D) for $H'$ conjugates to $\cL=\cU_L^{-1}\cL'\cU_L$, which has $\rho$ as stationary state, the same $1\to1$ norms, and supports thickened by $a_U$; so $\cL$ belongs to the class (D) for $H_0+V$ with allowed regions shrunk by $a_U$.  Finally $\cD'e^{t\cL'}\cE'=\cD\cU_L^{-1}\cU_Le^{t\cL}\cU_L^{-1}\cU_L\cE=\cD e^{t\cL}\cE$, and Theorem~\ref{SM:thm:general} applies.
\end{proof}

The reference Hamiltonian $U_LH_0U_L^\dagger$ has commuting terms that are in general not Pauli, since $U_L$ need not be Clifford.  The class therefore contains models with non-Pauli checks, with the known circuit entering the decoder.  The operational statements of Sec.~\ref{SM:sec:operational} carry over with the same conjugation.

\section{Exactly solvable single-plaquette sector and the crossover temperature}
\label{SM:sec:checks}

The crossover temperature $T_\times$ of the main text, below which the excitations that the checks detect are virtual rather than thermal, can be isolated in one exactly solvable sector.

\begin{lemma}[single-plaquette sector]
\label{SM:lem:pair}
Let $H=H_0-\epsilon X_p$ be the uniform four-dimensional code with couplings $J$ and a field on one plaquette $p$, and let $\Delta=8J$.  For every code state $|\bar\mu\rangle$ the two-dimensional space spanned by $|\bar\mu\rangle$ and $X_p|\bar\mu\rangle$ is invariant under $H$, on which, in this basis and up to the constant $E_{\rm ref}$,
\begin{equation}
H_{\rm pair}=\begin{pmatrix}0&-\epsilon\\-\epsilon&\Delta\end{pmatrix} .
\label{SM:eq:Hpair}
\end{equation}
With $E_\epsilon=\sqrt{\Delta^2/4+\epsilon^2}$, the probability that the ideal checks show the excitation in the Gibbs state of this sector is
\begin{equation}
p_{\rm pair}(\beta,\epsilon)=\frac12\Bigl(1-\frac{\Delta}{2E_\epsilon}\tanh\beta E_\epsilon\Bigr)=\frac{\epsilon^2}{2E_\epsilon(E_\epsilon+\Delta/2)}+\frac{\Delta}{2E_\epsilon}\,\frac{e^{-2\beta E_\epsilon}}{1+e^{-2\beta E_\epsilon}},
\qquad p_{\rm pair}(\infty,\epsilon)=\frac{\epsilon^2}{\Delta^2}+O(\epsilon^4/\Delta^4) .
\label{SM:eq:pair}
\end{equation}
\end{lemma}

\begin{proof}
$X_p$ anticommutes with the four edge checks of $p$ and commutes with all other checks, so $X_p|\bar\mu\rangle$ lies in the sector of those four checks, an eigenvector of $H_0$ with energy $E_{\rm ref}+4\cdot2J=E_{\rm ref}+\Delta$, while $H_0|\bar\mu\rangle=E_{\rm ref}|\bar\mu\rangle$; and $-\epsilon X_p$ exchanges the two vectors since $X_p^2=I$.  This gives Eq.~\eqref{SM:eq:Hpair}.  Write $H_{\rm pair}=(\Delta/2)I+\mathsf H$ with $\mathsf H=\begin{pmatrix}-\Delta/2&-\epsilon\\-\epsilon&\Delta/2\end{pmatrix}$, $\mathsf H^2=E_\epsilon^2I$, so $e^{-\beta\mathsf H}=\cosh(\beta E_\epsilon)I-(\sinh(\beta E_\epsilon)/E_\epsilon)\mathsf H$ and
\begin{equation}
p_{\rm pair}=\frac{\langle2|e^{-\beta H_{\rm pair}}|2\rangle}{\Tr e^{-\beta H_{\rm pair}}}=\frac{\cosh\beta E_\epsilon-(\Delta/2E_\epsilon)\sinh\beta E_\epsilon}{2\cosh\beta E_\epsilon}=\frac12\Bigl(1-\frac{\Delta}{2E_\epsilon}\tanh\beta E_\epsilon\Bigr) .
\end{equation}
The decomposition follows from $\tfrac12(1-\tanh x)=e^{-2x}/(1+e^{-2x})$: $\tfrac12(1-\tfrac{\Delta}{2E_\epsilon}\tanh\beta E_\epsilon)=\tfrac12(1-\tfrac{\Delta}{2E_\epsilon})+\tfrac{\Delta}{2E_\epsilon}\cdot\tfrac12(1-\tanh\beta E_\epsilon)$ and $\tfrac12(1-\tfrac{\Delta}{2E_\epsilon})=(E_\epsilon-\Delta/2)/(2E_\epsilon)=\epsilon^2/(2E_\epsilon(E_\epsilon+\Delta/2))$.  At $\beta=\infty$, $E_\epsilon=(\Delta/2)\sqrt{1+4\epsilon^2/\Delta^2}=\Delta/2+\epsilon^2/\Delta+O(\epsilon^4/\Delta^3)$, so $\epsilon^2/(2E_\epsilon(E_\epsilon+\Delta/2))=\epsilon^2/\Delta^2+O(\epsilon^4/\Delta^4)$.
\end{proof}

\begin{figure}[t]
\centering
\includegraphics{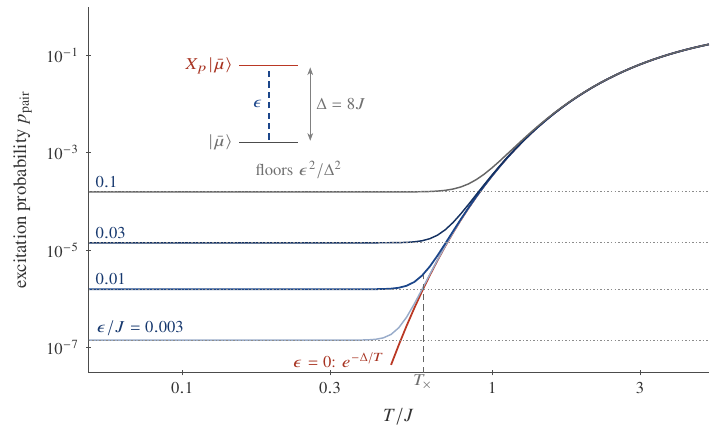}
\caption{The exactly solvable single-plaquette sector, Eq.~\eqref{SM:eq:pair}.  Probability that the ideal checks show the excitation, as a function of temperature, for plaquette fields $\epsilon/J=0$, $0.003$, $0.01$, $0.03$, $0.1$ with $\Delta=8J$.  The thermal curve $e^{-\Delta/T}/(1+e^{-\Delta/T})$ drops without bound; each field adds a temperature-independent term $\epsilon^2/\Delta^2$ (dotted), which the thermal curve crosses at $T_\times$ of the main text, marked for $\epsilon=0.01J$.  Inset: the two-level sector, the code state and the single flipped plaquette at energy $\Delta$, coupled by $\epsilon$.}
\label{SM:fig:S9}
\end{figure}

The first term of Eq.~\eqref{SM:eq:pair} is the bubble weight $(\epsilon/\Delta)^2$ of Sec.~\ref{SM:sec:histories}, the second the thermal density $e^{-\Delta/T}$ up to the factor $\Delta/2E_\epsilon$, and the two are equal at $e^{-\Delta/T}=\epsilon^2/\Delta^2$ to leading order, that is, at $T_\times=\Delta/[2\ln(\Delta/\epsilon)]$ (Fig.~\ref{SM:fig:S9}).  Below $T_\times$ the thermal term keeps falling while the first term stays, so the excitations detected at low temperature are quantum.  The sector is conditional on one plaquette and isolates the mechanism of Fig.~1(c).

\section{Comparison with earlier methods}
\label{SM:sec:comparison}

Several lines of work bound thermal states or memories, and each supplies one ingredient used here.  Table~\ref{SM:tab:comparison} summarizes them; the paragraphs that follow state what is taken from each and what is new.

\begin{table}[t]
\caption{Earlier methods for thermal stability and their reference objects.}
\label{SM:tab:comparison}
\renewcommand{\baselinestretch}{1}\footnotesize
\begin{ruledtabular}
\begin{tabular}{p{3.6cm}p{3.0cm}p{3.6cm}p{4.6cm}}
Method & Reference state & Regime & Ingredient used here \\
\colrule
Quantum Pirogov--Sinai~\cite{Ginibre1969,BKU1996,DFF1996} & product ground states & $T>0$, quantum perturbation of classical models & contour counting, free-energy comparison of phases \\
Zero-temperature cluster expansion~\cite{Klich2010} & one ground state & $T=0$, abelian perturbations & two-layer counting of linked clusters \\
Stationarity criterion~\cite{CLBT2010} & commuting code, Gibbs fixed point & $T>0$, $V=0$ & thermal encoding, approximate fixed point \\
Sector mixing~\cite{BGL2025} & commuting code & $T>0$, $V=0$ & the complementary statement: fast relaxation within a sector \\
Circuit-based phases~\cite{MKS2026,Sang2025} & local channels between states & $T>0$, conditional on clustering & target definition of a thermal phase \\
Bottleneck~\cite{Rakovszky2026,GamarnikKianiZlokapa2024} & extensive energy barrier & $T>0$, small perturbation & mixing lower bound from a rare set \\
Gauge averaging~\cite{Thorngren2026} & classical gauge theory & $T>0$, classical perturbation & average over the stabilizer group \\
Statistical-mechanics mapping~\cite{LiODeaKhemani2025} & decoding success of a classical model & $T>0$ classical, $T=0$ quantum & continuum limit of the imaginary-time picture \\
\end{tabular}
\end{ruledtabular}
\end{table}

\emph{Quantum Pirogov--Sinai theory} expands a quantum perturbation of a classical spin model around its product ground states and matches contours between phases~\cite{Ginibre1969,BKU1996,DFF1996}.  Its fixed-window contour counting and its comparison of restricted partition functions are the pattern for Sec.~\ref{SM:sec:window}.  The history expansion of Secs.~\ref{SM:sec:histories}--\ref{SM:sec:gas} is the version of the theory in which the reference objects are patterns: the thermal state is a sum over entangled sectors labeled by patterns, the comparison is between two Hamiltonians that differ by the erasure of one loop, Eq.~\eqref{SM:eq:KKF}, and the signs of mutual statistics are computed exactly.

\emph{Klich's expansion}~\cite{Klich2010} controls the ground-state energy of a perturbed stabilizer Hamiltonian by a cluster expansion in which linked clusters are counted with a two-layer encoding, and observes that linking of clusters obstructs the factorization of vacuum expectations.  Section~\ref{SM:sec:gas} uses the same counting idea and resolves the obstruction: linked components are grouped, their signs are evaluated by Eq.~\eqref{SM:eq:sigma}, and Lemma~\ref{SM:lem:emptybox} shows that the signs act at short range.  The new element is the thermal circle: a bubble linked with a sheet that winds around it, and the extension of the empty-box argument to the whole macroscopic background with its logical trace kept intact.

\emph{The stationarity criterion of Chesi, Loss, Bravyi, and Terhal}~\cite{CLBT2010} proves memory times from a Gibbs fixed point and protected observables that commute with the Hamiltonian and with a safe projector.  Section~\ref{SM:sec:memory} is its fully quantum form for noncommuting $H$: the replacement map $\cF_\varphi$ replaces protected observables, the one-sided identity replaces commutation, the entire logical algebra and any reference system are kept, and the generators may depend on time.  Both arguments use the Gibbs fixed point alone.

\emph{Rapid mixing within a logical sector}~\cite{BGL2025} and the \emph{bottleneck theorems}~\cite{Rakovszky2026,GamarnikKianiZlokapa2024} describe the dynamics of commuting codes from the opposite ends: fast relaxation inside a sector, and slow passage between sectors when an extensive energy barrier separates them.  The mixing obstruction of Sec.~\ref{SM:sec:operational} applies to generic noncommuting perturbations of the four-dimensional and generalized codes, where the barrier is linear in $L$ and the sectors are labeled by the ideal checks, which the perturbation mixes.

\emph{Circuit-based definitions of thermal phases}~\cite{MKS2026,Sang2025} place two states in one phase when local channels map each to the other, and relate self-correction to stable clustering of correlations.  Decoding and re-encoding with the decoder of Sec.~\ref{SM:sec:operational} performs this transformation globally for every state in the stable region; a local form of it would identify the whole stable region as a thermal phase, as discussed at the end of the main text.

\emph{Gauge averaging}~\cite{Thorngren2026} proves robustness of the classical Wegner memory by averaging a classical perturbation over the gauge group; the fixed-slice representation of Sec.~\ref{SM:sec:histories}, which keeps every pattern between slices exact, plays the same role for the quantum problem.  The classical memory there is protected by thermal fluctuations; the protection here is energetic, and the window is uniform down to $T=0$.  \emph{The statistical-mechanics mapping}~\cite{LiODeaKhemani2025} relates decoding thresholds to the order of an auxiliary classical model, including the continuous-imaginary-time limit of a quantum Hamiltonian; the fixed-slice representation of Sec.~\ref{SM:sec:histories} is the discrete form of that limit, with the slice length fixed so that the activities depend on $\epsilon$ and $t_0$ alone.

\section{Higher-form symmetries and Table~I}
\label{SM:sec:symmetry}

The two local properties of the main text are statements about the symmetries of the ideal code.  This section records the dictionary: the ideal code is a $\Z_2$ $p$-form gauge theory with two higher-form symmetries, the first property says that the excitations, which sit on the boundaries of open symmetry generators, are extended, and the second says that the charged operators of each symmetry are nonlocal.  In this language Table~I is a classification by the degree $q$ of the symmetries, and the theorem states that for $1\le q\le D-2$ the finite-temperature breaking of a $\Z_2$ higher-form symmetry survives generic explicit breaking.  The two statements proved here, the area law for point-like excitations and the perimeter law for extended ones, concern the ideal code and use its classical excitation gas.

\subsection{Conventions}
\label{SM:sec:conventions}

\begin{figure}[t]
\centering
\includegraphics{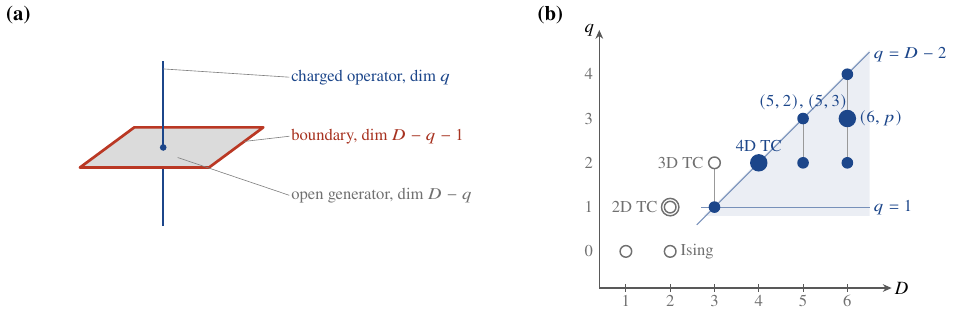}
\caption{Higher-form symmetries and Table~I.  (a)~A $q$-form symmetry in $D$ spatial dimensions, drawn for $q=1$, $D=3$: an open generator on a $(D-q)$-dimensional surface, its boundary of dimension $D-q-1$, where the excitations sit, and a charged operator of dimension $q$ piercing it.  (b)~The degrees $(q_X,q_Z)=(p,D-p)$ of the two symmetries of the $(D,p)$ toric codes, with the one- and two-dimensional Ising models at $q=0$; filled markers lie in the window $1\le q\le D-2$ and open markers outside it.  Spontaneous breaking survives at $T>0$ for $q\le D-2$ and resists explicit breaking for $q\ge1$; a code whose two symmetries both lie in the shaded window is a quantum memory, which is the condition $2\le p\le D-2$.  Pairs joined by a line are the two symmetries of one code; a double marker is a code whose two symmetries have equal degree.}
\label{SM:fig:S10}
\end{figure}

In $D$ spatial dimensions a $q$-form symmetry~\cite{GKSW2015,McGreevy2023} is generated by operators supported on closed $(D-q)$-dimensional surfaces; its charged operators are $q$-dimensional; and the boundary of an open generator, called a symmetry defect in that literature, is $(D-q-1)$-dimensional [Fig.~\ref{SM:fig:S10}(a)]; in the code the excitations sit on these boundaries.  A $0$-form symmetry is an ordinary global symmetry, with codimension-one domain walls and local charged operators.

\subsection{The two symmetries of the $p$-cell toric code}
\label{SM:sec:twosym}

For qubits on $p$-cells with checks as in Eq.~\eqref{SM:eq:gentoric}, the products $\prod Z_f$ over closed $p$-cycles commute with $H_0$ and generate a symmetry of degree $q_Z=D-p$; an open $p$-chain ends on a $(p-1)$-chain of violated $A$-checks, of dimension $D-q_Z-1$, and its charged operators are the $X$-type membranes on dual $(D-p)$-cycles.  Likewise $\prod X_f$ over closed dual $(D-p)$-cycles generate a symmetry of degree $q_X=p$, whose open generators end on $(D-p-1)$-chains of violated $B$-checks and whose charged operators are the $Z$-type membranes.  The logical operators of the code are the charged operators of the two symmetries, and the two excitation types sit on the boundaries of their open generators.  Each symmetry is broken in the ground space, which is the statement that the toric code is a spontaneously broken higher-form symmetry~\cite{KapustinThorngren2017,McGreevy2023}.

\subsection{Finite-temperature breaking for $q\le D-2$}
\label{SM:sec:finiteT}

The order parameter of a $q$-form symmetry is a charged operator on a closed $q$-cycle.  Take the $X$-type symmetry of the $p$-cell code, $q=p$, and a contractible $p$-cycle $\partial m$ bounding a $(p+1)$-chain $m$, and let $W_Z(\partial m)=\prod_{f\in\partial m}Z_f$.

\begin{lemma}[lattice Stokes identity and linking]
\label{SM:lem:stokes}
(i) $W_Z(\partial m)=\prod_{b\in m}B_b$, so in every state $\langle W_Z(\partial m)\rangle=\langle(-1)^{N_m}\rangle$, where $N_m$ is the number of violated $B$-checks in $m$.  (ii) In the ideal code the $B$-type excitations form closed dual $(D-p-1)$-chains $s$, and for every $s$, $(-1)^{|s\cap m|}=\prod_\gamma(-1)^{|\gamma\cap m|}$ over the clusters $\gamma$ of $s$.  (iii) If a cluster $\gamma$ of $s$ lies in a non-winding box $Q$, then $(-1)^{|\gamma\cap m|}=(-1)^{|\Sigma\cap\partial m|}$ for the dual $(D-p)$-chain $\Sigma=H_{Q^*}\gamma$ bounded by $\gamma$ inside $Q^*$; in particular $(-1)^{|\gamma\cap m|}=1$ unless $\partial m$ meets $Q^*$, which is within distance $C(|\gamma|+1)$ of $\gamma$.
\end{lemma}

\begin{proof}
(i) $\prod_{b\in m}B_b=\prod_{b\in m}\prod_{f\subset\partial b}Z_f=\prod_fZ_f^{\,n_f}$ with $n_f$ the number of cells of $m$ having $f$ in their boundary; $n_f$ is odd exactly for $f\in\partial m$.  Since $B_b$ has eigenvalue $(-1)^{[b\ \text{violated}]}$, the product has eigenvalue $(-1)^{N_m}$ on every joint eigenstate of the checks, and the thermal state is a mixture of such states.  (ii) By the computation of (B1) in the proof of Corollary~\ref{SM:cor:family}, the $B$-type excitations of a Pauli operator form a closed dual $(D-p-1)$-chain, and a pattern is the disjoint union of its clusters.  (iii) The dual complex of $T^D_L$ is a cubical complex, so Lemma~\ref{SM:lem:contraction} applies in $Q^*$: $\gamma=\partial\Sigma$ with $\Sigma$ inside $Q^*$.  Under the duality, a dual $(D-p)$-chain $\Sigma$ corresponds to a primal $p$-cochain and a dual $(D-p-1)$-chain to a primal $(p+1)$-cochain, with $\partial$ on the dual side corresponding to the coboundary $\delta$ on the primal side; the $\F_2$ pairing $\langle\cdot,\cdot\rangle$ between $k$-cochains and $k$-chains, which counts common cells, satisfies $\langle\delta u,m\rangle=\langle u,\partial m\rangle$ by the definition of $\delta$.  Hence $|\gamma\cap m|=\langle\delta\Sigma,m\rangle=\langle\Sigma,\partial m\rangle=|\Sigma\cap\partial m|$ modulo $2$.  If $\partial m$ does not meet $Q^*$, the right side is $0$; and $Q^*$, the dual box of the bounding box of the connected set $\gamma$, has diameter at most $C(|\gamma|+1)$.
\end{proof}

The expectation of the order parameter is therefore the parity of the number of excitations linked with it.  In the ideal code the two types are independent: a pattern is valid when each type bounds, the energy is the sum of the two types, and $(-1)^{N_m}$ depends on the $B$-type alone, so
\begin{equation}
\langle W_Z(\partial m)\rangle_{\rm ideal}=\frac{\sum_{s}e^{-\beta E(s)}(-1)^{|s\cap m|}}{\sum_{s}e^{-\beta E(s)}}=\frac{Z_B^{(m)}}{Z_B},
\label{SM:eq:Wratio}
\end{equation}
the sums over the valid $B$-type patterns.

\begin{lemma}[area law for point-like excitations]
\label{SM:lem:arealaw}
Let a type of excitations consist of points of equal energy $2J$ at $M$ sites, every even subset being a valid pattern (the two-dimensional toric code, and the point-like type of the three-dimensional code).  Then, with $\varsigma=e^{-2\beta J}$ and $\varrho_T=\varsigma/(1+\varsigma)$,
\begin{equation}
\langle(-1)^{N_m}\rangle_{\rm ideal}=\frac{(1-\varsigma)^{|m|}(1+\varsigma)^{M-|m|}+(1+\varsigma)^{|m|}(1-\varsigma)^{M-|m|}}{(1+\varsigma)^M+(1-\varsigma)^M}
\le2\,(1-2\varrho_T)^{\min\{|m|,M-|m|\}} .
\label{SM:eq:arealaw}
\end{equation}
\end{lemma}

\begin{proof}
Summing $\prod_bw^{n_b}(-1)^{n_b[b\in m]}$ over all subsets and projecting onto even subsets with $\tfrac12[1+\prod_b(-1)^{n_b}]$ gives the numerator and, without the sign, the denominator.  The bound follows from $(1-\varsigma)/(1+\varsigma)=1-2\varrho_T$, each of the two terms being at most $(1+\varsigma)^M$ times $(1-2\varrho_T)$ to the power $|m|$ or $M-|m|$, and $(1+\varsigma)^M\le(1+\varsigma)^M+(1-\varsigma)^M$.
\end{proof}

For $|m|\le M/2$ the order parameter decays as $(1-2\varrho_T)^{|m|}$, with the volume it bounds, at every $T>0$: the breaking is lost.  For extended excitations the decay is with the boundary only.

\begin{proposition}[perimeter law for extended excitations]
\label{SM:prop:perimeter}
Let the $B$-type excitations of the ideal code be closed dual chains of degree at least $1$ (the four-dimensional code and every member of Corollary~\ref{SM:cor:family}), with adjacency degree at most $\deg$ and $K=\max\{2,\deg\}^2$, and let $\varsigma=e^{-2\beta J_{\min}}$.  There are $\beta_1$ and $L_1$, depending on the lattice only, such that for $\beta\ge\beta_1$, $L\ge L_1$, and every contractible $\partial m$,
\begin{equation}
\langle W_Z(\partial m)\rangle_{\rm ideal}\ \ge\ \exp\bigl[-c(T)\,|\partial m|-4\delta_L\bigr],\qquad
c(T)=2C\sum_{\ell\ge1}(\ell+1)^{D}(Ke\varsigma)^\ell\xrightarrow[T\to0]{}0,\quad\delta_L=C'L^Dr^{m_0},\ r=Ke\varsigma<1 .
\label{SM:eq:perimeterlaw}
\end{equation}
\end{proposition}

\begin{proof}
Both sums in Eq.~\eqref{SM:eq:Wratio} are hard-core gases of closed dual chains: a valid pattern is a family of pairwise non-touching clusters, every cluster of diameter below $L/8$ bounds inside a non-winding box by Lemma~\ref{SM:lem:contraction}, and a union of bounding clusters bounds.  Call a cluster \emph{local} if it has fewer than $m_0$ cells and \emph{macroscopic} otherwise, and let $\Lambda_{\rm cl}$ be the finite set of local clusters, with the incompatibility relation ``share a cell or touch'', cost $m(\gamma)=|\gamma|$, activities $z_\gamma=\prod_{b\in\gamma}e^{-2\beta J_b}$ for $Z_B$ and $z^{(m)}_\gamma=z_\gamma(-1)^{|\gamma\cap m|}$ for $Z_B^{(m)}$, and radii $\rho_\gamma=\varsigma^{|\gamma|}$.  A family of local clusters is a valid pattern whenever its members are pairwise compatible, so the local parts of $Z_B$ and $Z_B^{(m)}$ are the polymer partition functions $Z_{\Lambda_{\rm cl}}$ with these two activity families.

\emph{Hypotheses of Lemma~\ref{SM:lem:deletion}.}  A cluster $\gamma'$ incompatible with $\gamma$ contains a cell adjacent to, or equal to, a cell of $\gamma$; there are at most $\deg+1$ such cells per cell of $\gamma$ and at most $K^{\ell}$ clusters of size $\ell$ through a given cell (the closed-walk count of Lemma~\ref{SM:lem:clustercount}).  Hence $\sum_{\gamma'\not\sim\gamma}\rho_{\gamma'}e^{m(\gamma')}\le|\gamma|(\deg+1)\sum_{\ell\ge1}(Ke\varsigma)^\ell\le|\gamma|/16$ and $qe^{1/8}\le\frac12$ once $\varsigma\le \varsigma_1$, which defines $\beta_1$.

\emph{Local comparison.}  The activities differ only on clusters linked with $\partial m$, $(-1)^{|\gamma\cap m|}=-1$, and by Lemma~\ref{SM:lem:stokes}(iii) such a cluster has a cell within distance $C(|\gamma|+1)$ of some cell of $\partial m$.  Lemma~\ref{SM:lem:interpolation} gives
\begin{equation}
\bigl|\ln|Z^{(m)}_{\Lambda_{\rm cl}}|-\ln|Z_{\Lambda_{\rm cl}}|\bigr|\le\sum_{\gamma\ \text{linked}}2\varsigma^{|\gamma|}e^{|\gamma|}\le|\partial m|\sum_{\ell\ge1}C(\ell+1)^{D}K^\ell\,2(qe)^\ell=c(T)|\partial m| ,
\end{equation}
counting, for each cell of $\partial m$ and each size $\ell$, the cells within distance $C(\ell+1)$ and the clusters through each of them.  Along the real interpolation path $z(t)=(1-t)z+tz^{(m)}$ the partition function is real and nonvanishing by Lemma~\ref{SM:lem:deletion}, and $Z_{\Lambda_{\rm cl}}>0$, so $Z^{(m)}_{\Lambda_{\rm cl}}>0$ as well.

\emph{Macroscopic clusters.}  As in Lemmas~\ref{SM:lem:background} and \ref{SM:lem:macro}, group the valid patterns by their family of macroscopic clusters.  The local gas in the presence of a macroscopic family is the gas on $\Lambda_{\rm cl}$ with the activities of the clusters touching the family set to zero, which by Lemma~\ref{SM:lem:interpolation} changes $\ln|Z|$ by at most $\varkappa_{\rm cl}$ times the number of macroscopic cells, with $\varkappa_{\rm cl}=2(\deg+1)\sum_{\ell\ge1}(Ke\varsigma)^\ell\le\frac18$; the bare weight of a macroscopic family is at most $\varsigma$ to the power of its number of cells, and $(-1)^{|\gamma\cap m|}$ does not change moduli.  Summing over families as in Lemma~\ref{SM:lem:macro}, with at most $C_{\rm vol}L^D$ cells and $K^\ell$ clusters of size $\ell$ through a cell, $|Z_B/Z_{\Lambda_{\rm cl}}-1|\le e^{\delta_L}-1$ with $\delta_L=2C_{\rm vol}L^Dr^{m_0}$, and the same for $Z_B^{(m)}$; for $L\ge L_1$, $\delta_L\le\frac14$ and each ratio has $|\ln|\cdot||\le2\delta_L$.  The same argument shows $Z_B^{(m)}$ has the sign of $Z^{(m)}_{\Lambda_{\rm cl}}$, hence is positive.

Combining, $\ln\langle W_Z(\partial m)\rangle=\ln(Z_B^{(m)}/Z_B)\ge-c(T)|\partial m|-4\delta_L$.
\end{proof}

The same argument with primal and dual exchanged treats the $Z$-type symmetry.  Hence a discrete $q$-form symmetry stays spontaneously broken at $T>0$, in the ideal code, exactly when
\begin{equation}
q\le D-2 ,
\label{SM:eq:qD2}
\end{equation}
the condition obtained in Refs.~\cite{GKSW2015,McGreevy2023} by reducing the thermal theory on its time circle, where a $q$-form symmetry of the quantum theory yields a $q$-form symmetry of the $D$-dimensional classical theory, and in Ref.~\cite{Lake2018} for continuous symmetries.  At $T=0$ the Euclidean theory has dimension $D+1$ and the same count gives $q\le D-1$.  This is why the two-dimensional toric code, $q=1=D-1$, is ordered in its ground space and disordered at every $T>0$; why the three-dimensional code keeps its $q=1$ symmetry broken and loses its $q=2$ symmetry; and why the ideal four-dimensional code has topological order at $T>0$~\cite{Hastings2011}, both of its symmetries having $q=2=D-2$.  For the code, Eq.~\eqref{SM:eq:qD2} is the first property: an extended closed excitation is the boundary of a local membrane and is erased by flipping it, Eq.~\eqref{SM:eq:filling}, while a point-like excitation is created in pairs and can only be moved.

\subsection{Robustness to explicit breaking for $q\ge1$}
\label{SM:sec:explicit}

A generic local perturbation breaks both symmetries of the ideal code explicitly: a single $Z_f$ anticommutes with every $X$-type generator through $f$, and once it is added to the Hamiltonian the generators are no longer conserved.  For $q\ge1$ the order parameter is invisible to every local term.

\begin{lemma}[local indistinguishability of the code states]
\label{SM:lem:indist}
Let $A$ be a region inside a non-winding box.  Then $\Tr_{A^c}|\bar\psi\rangle\langle\bar\psi|$ is the same for every code state $|\bar\psi\rangle$; equivalently, $\Pi_\emptyset O\Pi_\emptyset=c(O)\Pi_\emptyset$ for every operator $O$ supported in $A$.
\end{lemma}

\begin{proof}
Expand $O$ in the Pauli operators supported in $A$.  A Pauli operator $P$ that anticommutes with some check maps the code space into the sector $\Pi_{\syn(P)}\ne\Pi_\emptyset$ (Lemma~\ref{SM:lem:sectors}(iii)), so $\Pi_\emptyset P\Pi_\emptyset=0$; one that commutes with every check is a check product times a phase by Lemma~\ref{SM:lem:generation}, and a check product is the identity on the code space, so $\Pi_\emptyset P\Pi_\emptyset=\zeta\Pi_\emptyset$.  Hence $\langle\bar\psi|O|\bar\psi\rangle=c(O)$ for every code state, which is the statement about reduced states.
\end{proof}

The order parameter is a charged operator on a closed $q$-cycle, nonlocal for $q\ge1$, and the states it distinguishes, the logical states of the code, are indistinguishable on every region inside a non-winding box.  A local term therefore cannot select among them at first order, and the rigorous form of the statement that the degeneracy and the perimeter law of the dressed charged operators survive is the stability of topological order at zero temperature~\cite{HastingsWen2005,BHM2010} and, at finite temperature, Theorem~\ref{SM:thm:general}; this is the robustness of emergent higher-form symmetries stated in Ref.~\cite{McGreevy2023}.  For $q=0$ the order parameter is local, the magnetization of the Ising model, and a field couples to it directly: the stored bit is erased.  In the code the dividing line is the second property: every local operator commuting with the checks is a product of nearby checks, Lemma~\ref{SM:lem:generation}, so the stored information is invisible to local operators, which is the lattice form of ``no local order parameter''.

\subsection{Classification by symmetry degree}
\label{SM:sec:classification}

The two properties therefore translate into the two inequalities $1\le q\le D-2$, required of every symmetry of the code.  For the $p$-cell toric code the degrees are $(q_X,q_Z)=(p,D-p)$, and both lie in the window exactly when $2\le p\le D-2$, the condition of Corollary~\ref{SM:cor:family}.  Table~\ref{SM:tab:qclass} lists the examples of Table~I by symmetry degree, and Fig.~\ref{SM:fig:S10}(b) places them in the $(D,q)$ plane; it is the same table with rows $q\le D-2$ and columns $q\ge1$.

\begin{table}[t]
\caption{Table~I of the main text as a classification by symmetry degree.}
\label{SM:tab:qclass}
\renewcommand{\baselinestretch}{1}\small
\begin{ruledtabular}
\begin{tabular}{p{3.7cm}ccp{3.0cm}cp{3.4cm}}
Model & $(q_X,q_Z)$ or $q$ & $D-2$ & $q\le D-2$ & $q\ge1$ & Outcome \\
\colrule
1D Ising & $0$ & $-1$ & no & no & no memory \\
2D Ising & $0$ & $0$ & yes & no & thermal bit, erased by a field \\
2D toric code & $(1,1)$ & $0$ & no, no & yes & no memory at $T>0$ \\
3D toric code & $(1,2)$ & $1$ & yes, no & yes & classical bit (Wilson loop, Wegner) \\
4D toric code ($p=2$) & $(2,2)$ & $2$ & yes, yes & yes & quantum memory \\
$(D,p)$ toric code & $(p,D-p)$ & $D-2$ & iff $2\le p\le D-2$ & yes & quantum memory, $k=\binom Dp$ \\
\end{tabular}
\end{ruledtabular}
\end{table}

\subsection{Restatement of the theorem}
\label{SM:sec:symtheorem}

Theorem~\ref{SM:thm:general} with Corollary~\ref{SM:cor:family} states: for $\Z_2$ higher-form symmetries of degree $1\le q\le D-2$ realized by a Pauli stabilizer code or its finite-depth image, the finite-temperature spontaneous breaking survives generic explicit breaking by finite-range local perturbations, with a threshold uniform down to $T=0$; when both symmetries of a $\Z_2$ $p$-form gauge theory lie in this window, its low-temperature thermal state is a robust quantum memory.  This is the finite-temperature counterpart of the zero-temperature robustness of emergent higher-form symmetries~\cite{HastingsWen2005,BHM2010,McGreevy2023}.  The theorem covers $\Z_2$ Pauli codes and their finite-depth local-unitary images; continuous and non-abelian higher-form symmetries are the next targets.  Related uses of this language for fracton phases and for a Landau--Ginzburg description of one-form symmetries appear in Refs.~\cite{QiRadzihovskyHermele2021,IqbalMcGreevy2022}.

\section{Extensions within the hypotheses}
\label{SM:sec:extensions}

The proof uses the lattice through (A)--(C) only.  Two extensions follow by verifying (A)--(C); the last subsection identifies the condition that remains for codes without a lattice.

\subsection{Other cellulations and manifolds}
\label{SM:sec:cellulations}

\begin{proposition}[cellulated manifolds]
\label{SM:prop:cellulation}
Let $(M_L)$ be a family of cellulations of a fixed closed $D$-manifold $M$ with bounded local geometry: bounded cell degrees, cells of bounded size, and the \emph{uniform local acyclicity} property that every closed $q$-chain, $q\ge1$, of the primal or the dual complex inside a ball of radius $r<c_{\rm ac}L$ is the boundary of a chain inside the ball of radius $C_{\rm ac}(r+1)$, by a rule determined by the chain alone.  Then the $p$-cell code on $M_L$, $2\le p\le D-2$, satisfies (A)--(C) with
\begin{equation}
h=\max\bigl\{\dim H_{p-1}(M;\F_2),\ \dim H_{D-p-1}(M;\F_2)\bigr\},\qquad k=\dim H_p(M;\F_2),
\end{equation}
so Theorem~\ref{SM:thm:general} holds with exponents $D(h+1)$ and $D(h+3)/2$.
\end{proposition}

\begin{proof}
(A) is the bounded local geometry, with non-winding boxes replaced by balls of radius below $c_{\rm ac}L$ and $k=\dim H_p(M;\F_2)$ the number of logical qubits, as for any CSS code built from a chain complex.  (B1): as in Corollary~\ref{SM:cor:family}, the $B$-type excitations of a $Z$-type operator form a closed dual $(D-p-1)$-chain and the $A$-type excitations of an $X$-type operator a closed primal $(p-1)$-chain, every cluster of a valid pattern is closed, and uniform local acyclicity in the dual, respectively primal, complex supplies a filling inside the ball of radius $C_{\rm ac}(r+1)$ determined by the cluster alone; the corresponding $Z$-type or $X$-type operator creates the cluster and no excitation of the other type.  (B2): take the charge of a closed $(p-1)$-chain to be its class in $H_{p-1}(M;\F_2)$ and that of a closed dual $(D-p-1)$-chain its class in $H_{D-p-1}(M;\F_2)$; valid patterns are boundaries and have zero class, and a union of non-touching closed chains with zero total class is a boundary, hence the pattern of a $Z$-type or $X$-type operator creating no other type.  (C): a Pauli operator $\zeta X(x)Z(z)$ in a ball commuting with all checks has $z$ a closed $p$-chain and $*x$ a closed dual $(D-p)$-chain, both filled inside the enlarged ball by uniform local acyclicity, so $Z(z)$ is a product of $B$-checks and $X(x)$ a product of $A$-checks of the fillings, within distance $C_{\rm ac}(r+1)$ of the ball.  The counting constants are determined by the cell degrees.
\end{proof}

The cubical contraction $H_Q$ is uniform local acyclicity for the torus, with the box in place of the ball.  Verifying it is the one step specific to a given cellulation.

\subsection{$\Z_{\mathsf N}$ qudit codes}
\label{SM:sec:ZN}

For qudits of dimension $\mathsf N$ with generalized Pauli operators $X,Z$, $ZX=\zeta_{\mathsf N}XZ$, $\zeta_{\mathsf N}=e^{2\pi i/\mathsf N}$, the $p$-cell code has checks $A_a=\prod X_f^{\pm1}$ and $B_b=\prod Z_f^{\pm1}$ with the orientation signs of the cellular boundary, and $H_0=-\sum_aJ_a(S_a+S_a^\dagger)/2$.

\begin{proposition}[$\Z_{\mathsf N}$ codes]
\label{SM:prop:ZN}
Theorem~\ref{SM:thm:general} holds for the $\Z_{\mathsf N}$ $p$-cell codes, $2\le p\le D-2$, with $2J_{\min}$ replaced by $J_{\min}(1-\cos2\pi/\mathsf N)$ in every energy comparison, the counting constants multiplied by $\mathsf N-1$, $\ln\mathsf D=n\ln \mathsf N$, and $d_L=\mathsf N^{\binom Dp}$.
\end{proposition}

\begin{proof}
The proof of Theorem~\ref{SM:thm:general} is rerun with the following replacements, each of which affects the named lemmas only in the way stated.

\emph{Energies and patterns.}  The joint eigenvalues of a check are $\zeta_{\mathsf N}^j$, $j\in\Z_{\mathsf N}$, and an excitation of charge $j\ne0$ costs $J_a(1-\cos2\pi j/\mathsf N)\ge J_a(1-\cos2\pi/\mathsf N)$, which replaces $2J_{\min}$ in Lemma~\ref{SM:lem:energy}, Lemma~\ref{SM:lem:local}(iii), and Eq.~\eqref{SM:eq:t0eps}; a pattern is a $\Z_{\mathsf N}$-valued function on the checks, the syndrome of $X(x)Z(z)$ is the pair $(\partial z,\partial(*x))$ of $\Z_{\mathsf N}$-chains, and valid patterns are the bounding ones.

\emph{Contraction.}  Over $\Z_{\mathsf N}$ the cellular boundary carries signs, and the interval maps of Eq.~\eqref{SM:eq:intervalmaps} satisfy $\partial H_I+H_I\partial=\id-\pi_I$: $\partial H_I(\mathbf v_j)=\sum_{k=0}^{j-1}(\mathbf v_{k+1}-\mathbf v_k)=\mathbf v_j-\mathbf v_0$ and $H_I\partial(\mathbf e_j)=H_I(\mathbf v_{j+1}-\mathbf v_j)=\mathbf e_j$.  With the Koszul sign convention for tensor products, $\partial(x\otimes y)=\partial x\otimes y+(-1)^{|x|}x\otimes\partial y$ and $(\id\otimes H_I)(x\otimes y)=(-1)^{|x|}x\otimes H_Iy$, the cross terms in $\partial H^{(j)}+H^{(j)}\partial$ cancel as in Lemma~\ref{SM:lem:contraction}, the terms with $i>j$ acquiring opposite signs from passing $\partial_i$ across the degree-one map $H_{I_j}$, and the sum telescopes to $\id-\Pi_Q$.  Hence Eq.~\eqref{SM:eq:filling} holds over $\Z_{\mathsf N}$, and (B1), (C) follow as in Corollary~\ref{SM:cor:family}; (B2) holds with charges in $H_q(T^D;\Z_{\mathsf N})=\Z_{\mathsf N}^{\binom Dq}$, detected by the signed crossing counts, with the grouping of Lemma~\ref{SM:lem:grouping} unchanged.

\emph{Sectors and characters.}  Lemmas~\ref{SM:lem:sectors}--\ref{SM:lem:sectortrace} hold for the abelian group generated by the checks, with characters valued in the $\mathsf N$th roots of unity: $\chi_S(s)$ is the eigenvalue of the check product $S$ on the sector $s$, multiplicative in $s$ and in $S$, and $RSR^\dagger=\chi_S(\syn R)S$ for every generalized Pauli operator $R$, since $R$ conjugates each check to $\zeta_{\mathsf N}^{j}$ times itself with $j$ the charge of $\syn R$ at that check.

\emph{Counting.}  Each excitation cell carries one of $\mathsf N-1$ charges, which multiplies $K_{\rm raw}$, $K_{\rm link}$, and the cluster-counting constants by $\mathsf N-1$; the Pauli alphabet on $m$ qudits has $\mathsf N^{2m}$ letters, which enters $K_{\rm raw}$ through $\mathsf N^{2c_R}$ in place of $4^{c_R}$.

\emph{Signs.}  Two generalized Pauli operators satisfy $xy=\zeta_{\mathsf N}^{\kappa(x,y)}yx$ with $\kappa\in\Z_{\mathsf N}$, so Lemma~\ref{SM:lem:regroup} gives the regrouping phase $\prod_{\alpha<\beta}\zeta_{\mathsf N}^{\kappa(C_\alpha,C_\beta)}$ with $\kappa(C,D)\in\Z_{\mathsf N}$ the sum of the commutation exponents over the inverted pairs; the independence of the extraction order holds because the net operators commute, and Eq.~\eqref{SM:eq:sigma} becomes $\sigma(C,D)=\zeta_{\mathsf N}^{\kappa(C,D)}\chi_{S_C}(s_D(0))\chi_{S_D}(s_C(0))$, a root of unity.  The proof of Lemma~\ref{SM:lem:emptybox} uses only commutation of operators with disjoint supports and the identity $R_CS_DR_C^\dagger=\chi_{S_D}(s_C(0))S_D$, both of which hold, so $\sigma(C,D)=1$ under its hypotheses.  Linked clusters are defined by $\sigma\ne1$, the activities are complex, and Lemmas~\ref{SM:lem:deletion}--\ref{SM:lem:background} admit complex activities as they stand.

\emph{Bridge and decoder.}  $\ln\mathsf D=n\ln \mathsf N$ in Lemma~\ref{SM:lem:bridge}, which changes $A_*$ by the factor $\ln \mathsf N/\ln2$.  Lemma~\ref{SM:lem:decoder} uses only the logical algebra, the sector structure, and (C); the operational results of Sec.~\ref{SM:sec:operational} use the one-sided identity and the encoding, with $d_L=\mathsf N^{\binom Dp}$.
\end{proof}

The conclusion is the same theorem for $\Z_{\mathsf N}$ $p$-form gauge theories, with the symmetry dictionary of Sec.~\ref{SM:sec:symmetry} read with $\Z_{\mathsf N}$ in place of $\Z_2$.

\subsection{Codes without a lattice}
\label{SM:sec:ldpc}

For a quantum low-density parity-check code on a bounded-degree graph, three parts of the proof apply as they stand: the Peierls counting of Proposition~\ref{SM:prop:static4d} needs only bounded degree, the empty-box lemma needs only metric balls, and the entropy bridge needs only the number of qubits.  The part that uses the lattice is (B2): the global grouping needs a number $h$ of global constraints that stays bounded as $n$ grows, and it produces the prefactor $n^{h+1}$ in Theorem~\ref{SM:thm:general}(a).  In good codes the number of logical qubits and the rank of the check redundancies grow with $n$, so $h$ grows with $n$.  A counting that uses local redundancy in place of the global grouping is the step that extends the theorem to constrained LDPC codes with local erasure and local generation, the setting of the finite-temperature memories of Ref.~\cite{HongGuoLucas2025} and of the LDPC codes studied as phases of matter in Refs.~\cite{RakovszkyKhemani2023,RakovszkyKhemani2024,Rakovszky2026}.

\end{document}